\documentclass[debug, preprint, twocolumn]{rmaa}
\usepackage{natbib}

\usepackage{paralist}
\usepackage{pdfpages}
\usepackage{psfrag,color}

\usepackage{amsmath}
\usepackage[utf8]{inputenc}

\usepackage{spverbatim}

\usepackage{color}
\definecolor{URLCOLOR}{rgb}{0.1,0.1,0.7} 
\definecolor{CHANGECOLOR}{rgb}{0.0,0.0,0.0} 

\usepackage{./aas_macros}
\def\emmmhz{{10~MHz}} 
\def\emmcm{{29.98 m}} 

\def\egamry{{$7.354\times 10^6 \Ryd$}} 
\def\egamrymev{{$100 \MeV$}} 

\newcommand{\Cloudy}{\textsc{Cloudy}}

\font\manual=manfnt at 7pt \def\dbend{\hbox{\raise0.9ex\hbox{\manual\char127\hspace{0.6em}}}}

\providecommand{\e}[1]{\ensuremath{\times 10^{#1}}}

\newcounter{INTERNALionstage}

\newcommand{\Av}{A$_{\rm V}$}

\def\gtsim{\mathrel{\hbox{\rlap{\hbox{\lower4pt\hbox{$\sim$}}}\hbox{$>$}}}}
\def\lesssim{\mathrel{\hbox{\rlap{\hbox{\lower4pt\hbox{$\sim$}}}\hbox{$<$}}}}

\def\cm{{\rm\thinspace cm}}

\def\erg{{\rm\thinspace erg}}

\def\km{{\rm\thinspace km}}

\def\MeV{{\rm\thinspace MeV}}

\def\micron{\hbox{$\mu$m}}

\def\ps{{\rm\thinspace s^{-1}}}
\def\pcc{{\rm\thinspace cm^{-3}}}
\def\Ryd{{\rm\thinspace Ryd}}
\def\s{{\rm\thinspace s}}

\def\kmps{\mbox{$\km\ps\,$}}

\def\ps{\mbox{$\s^{-1}\,$}}
\def\pscm{\mbox{$\cm^{-2}\,$}}

\def\pscm{\mbox{$\cm^{-2}\,$}}

\def\hi{\mbox{{\rm H~{\sc i}}}}
\def\hii{\mbox{{\rm H~{\sc ii}}}}
\def\hei{\mbox{{\rm He~{\sc i}}}}
\def\heii{\mbox{{\rm He~{\sc ii}}}}

\def\oii{\mbox{{\rm O~{\sc ii}}}}
\def\oiii{\mbox{{\rm O~{\sc iii}}}}

\def\sii{\mbox{{\rm S~{\sc ii}}}}

\def\htwo{\mbox{{\rm H}$_2$}}
\def\hone{\mbox{{\rm H}$^0$}}
\def\h0{\mbox{{\rm H}$^0$}}
\def\hplus{\mbox{{\rm H}$^+$}}

\def\hO{\mbox{{\rm H}$^0$}}

\DeclareMathAlphabet{\vib}{OML}{cmm}{m}{it}
\newcommand*{\satellite}[1]{\textit{#1}}
\newcommand*{\xmm}{\satellite{XMM-Newton}}
\newcommand*{\chandra}{\satellite{Chandra}}

\usepackage[draft]{hyperref}
\hypersetup{
    bookmarks=true, 
    unicode=false,  
    pdftoolbar=true,   
    pdfmenubar=true,   
    pdffitwindow=true,      
    pdftitle={The 2017 Release of \Cloudy},    
    pdfauthor={G. J. Ferland et al.},     
    pdfsubject={Astrophysics},   
    pdfnewwindow=true,      
    pdfkeywords={keywords}, 
    colorlinks=true,   
    linkcolor=black, 
    citecolor=black, 
    filecolor=black, 
    urlcolor=URLCOLOR   
} 
\newcommand\URL[1]{\href{http://#1}{\nolinkurl{#1}}}

\title{The 2017 Release of \Cloudy{}}
\shorttitle{The 2017 Release of \Cloudy{}}

\author{
G.~J. Ferland\altaffilmark{1}, 
M.~Chatzikos\altaffilmark{1}, 
F.~Guzm\'{a}n\altaffilmark{1}, 
M.~L. Lykins\altaffilmark{1}, 
P.~A.~M. van Hoof\altaffilmark{2},
R.~J.~R. Williams\altaffilmark{3},
\\
N.~P. Abel\altaffilmark{4},
N.~R.~Badnell\altaffilmark{5},
F. P. Keenan\altaffilmark{6},
R.~L. Porter\altaffilmark{7}, 
P.~C. Stancil\altaffilmark{7}
}
\shortauthor{Ferland et al.}

\altaffiltext{1}{University of Kentucky, Lexington, USA}

\altaffiltext{2}{Royal Observatory of Belgium}

\altaffiltext{3}{AWE plc, UK}

\altaffiltext{4}{University of Cincinnati, USA}

\altaffiltext{5}{University of Strathclyde, Glasgow, UK}

\altaffiltext{6} {Queen's University Belfast, Belfast, UK}

\altaffiltext{7}{University of Georgia, USA}

\fulladdresses{
\item Nicholas P. Abel: Department of Physics, University of Cincinnati, 4200 Clermont College Drive, Batavia, Ohio, 
45103, USA (npabel2@gmail.com).
\item N.~R.~Badnell: Department of Physics, University of Strathclyde, Glasgow, G4 0NG, UK (badnell@phys.strath.ac.uk)
\item M.~Chatzikos, F.~Guzm\'{a}n, Gary J. Ferland and Matthew L. Lykins: Department of Physics and Astronomy, University of Kentucky, 
Lexington, KY 40506, USA (gary@pa.uky.edu; mchatzikos@gmail.com; 
francisco.guzman@uky.edu; mllyki3@uky.edu).
\item Francis P. Keenan:Astrophysics Research Centre, School of Mathematics and Physics, Queen's University Belfast, Belfast BT7 1NN, Northern Ireland, U.K. (f.keenan@qub.ac.uk)
\item Ryan L. Porter and Phillip Stancil: Department of Physics and Astronomy and Center for Simulational Physics, The University of Georgia, Athens, Georgia 30602-2451, USA (ryanlporter@gmail.com, stancil@physast.uga.edu).
\item Peter A.~M. van Hoof: Royal Observatory of Belgium, Ringlaan 3, 1180 Brussels,
Belgium (p.vanhoof@oma.be).
\item Robin J.~R. Williams: AWE plc, Aldermaston, Reading RG7 4PR, UK (robin.williams@awe.co.uk)
}

\abstract{
We describe the 2017 release of the spectral synthesis code \Cloudy{},
summarizing the many improvements to the scope and accuracy of the
physics which have been made since the previous release.
Exporting the atomic data into external data files has enabled many new large
datasets to be incorporated into the code.
The use of the complete datasets is not realistic for most calculations,
so we describe the limited subset of data used by default, which predicts 
significantly more lines than the previous release of \Cloudy{}.
This version is nevertheless faster than the previous release, as a result of code optimizations. 
We give examples of the accuracy limits using small models, and the performance
requirements of large complete models.
We summarize several advances in the H- and He-like iso-electronic sequences
and use our complete collisional-radiative models to establish the 
densities where the coronal and local thermodynamic equilibrium approximations work. 
}

\resumen{
Se presenta la versi\'on 2017 del c\'odigo de s\'{\i}ntesis espectral \Cloudy{}
junto con las mejoras de precisi\'on y tratamiento de la f\'{\i}sica
desde la version previa.
La exportaci\'on de los datos at\'omicos hacia bases de datos externas ha
permitido la incorporaci\'on de grandes conjuntos de nuevos datos.
El uso completo de estos no es pr\'actico para la mayor\'ia de
las simulaciones, describi\'endose el subconjunto de datos 
usado por defecto y que predice un n\'umero
considerable mayor de l\'{\i}neas que en la versi\'on previa. No obstante, la versi\'on actual
es m\'as r\'apida debido a la optimizaci\'on del c\'odigo.
Se dan ejemplos de los l\'{\i}mites de precisi\'on de los modelos peque\~nos y
de los requisitos de rendimiento para los modelos
completos. Se sintetizan avances en los modelos para iones con secuencias
isoelectr\'onicas de H y He y el uso de modelos
colisionales-radiativos completos en la determinaci\'on de las densidades donde
las aproximaciones coronal y de equilibrio termodin\'amico local funcionan.

}

\addkeyword{atomic processes}
\addkeyword{galaxies: active}
\addkeyword{methods: numerical}
\addkeyword{molecular processes}
\addkeyword{radiation mechanisms}

\reviewarticle

\SetFirstPage{1}
\SetYear{2017}

\begin{document}

\RescaleLengths{1.1}

\maketitle
\clearpage
{\small
\reviewtoc
  \bigskip
   \bigskip
}

\section{Introduction}
\label{sec:intro}

We introduce the next major version of \Cloudy{}, version C17.
\Cloudy{} is a non-local thermodynamic equilibrium (NLTE) spectral synthesis and
 plasma simulation code
designed to simulate astrophysical environments and predict their spectra.
The previous version of \Cloudy{}, C13, is described in \citet{CloudyReview13},
hereafter referred to as C13, while the last
major review before C13 was \citet{Ferland1998}.
These give an overview of the scope and goals for our simulation code.
The basic physics is described in \citet{AGN3}, hereafter referred to as AGN3.
\citet{Ferland2003b} goes into some atomic and plasma physics questions
with examples of \Cloudy{} applications to photoionized clouds.

A great effort since C13 has gone into moving \\Cloudy{}'s atomic and molecular 
data into external databases.
These external databases make it possible to compute intensities of a great many
 emission lines.
A theme running through this review is the tradeoff between the increased accuracy
that comes from including larger and more complete models, and the associated increase
in time and memory. 
For this reason using the full databases is usually not practical.
Command-line user options control the size of the various model atoms, ions, and molecules.
With all databases fully used the number of lines is increased by well over an order of magnitude,
and the default setup predicts significantly more lines than the previous release.
Despite the increased number of lines, in its default state
C17 is actually faster than C13
because of a good number of optimizations introduced to the code base.

The next sections describe how we incorporate a number of external databases to compute
large and complex models of ionic and molecular emission.
We then discuss how we determine the ionization and emission of the gas,
and its range of validity.  
Other major changes to the physics and functionality of the code are also reviewed.
The external data, with its common user interface and underlying software, make it simple to 
report such quantities as the column density or population in a particular level of a species, or its spectrum.
We give examples of using \Cloudy{} to compute both equilibrium 
\citep{Lykins.M12Radiative-cooling-in-collisionally-and-photo, 2014MNRAS.440.3100W}
and non-equilibrium \citep{Chatzikos2015} cooling.
The former occurs if the system has not changed over timescales longer than those
required for atomic processes to reach steady state.
The latter occurs mainly at temperatures at or below the $10^5$~K peak in the cooling function,
if the temperature changes too rapidly for the system to come into ionization equilibrium.
Another change includes options to remove isotropic continuum sources, such as the CMB,
from the spectral predictions. 

We do not review those parts of the code, its documentation, user support sites, or web access
that have not changed since the release of C13.  The C13 review paper remains the
primary documentation for those parts of this release.

\section{Databases}
\label{sec:databases}

\subsection{The move to external databases}

The greatest effort since C13 has gone into a massive reorganization of our treatment of 
ions, atoms, or molecules, which we collectively refer to as ``species''.
\Cloudy{} originally added species models  
on a case-by-case basis, with the data mixed in with the source.
This was a significant maintenance problem since only people with a working
knowledge of C++ could update the data.
We have largely moved the data into external databases.
As much as possible we treat all species with a common code base.
New models are added to our Stout database  \citep{LykinsStout15}, 
which was designed to present data in formats as close
as possible to the original data sources, for ease of maintenance. 

These databases make it possible to create very extensive models of line emission that
include a very large number of levels, although this comes at the cost of longer computing times.
They also provide the flexibility of including far fewer levels, with faster execution time, but with a less
realistic representation of the physics.
This compromise between speed and precision will be a theme running throughout this review.

There are five distinct databases used to model spectral lines.
These are outlined here and in more detail in later sections.

\subsubsection{The H-like and He-like iso-electronic sequences}

We treat atomic one and two electron systems (except H$^-$) with full collisional radiative models,
referred to as CRM (see the review by \citet{FerlandWilliams16}).
These models are described in greater detail in Section \ref{sec:iso-sequences} below.
The models are complete, are capable of making highly accurate predictions
of emission, and go to the interstellar medium (ISM), Local Thermodynamic Equilibrium (LTE)  
and Strict Thermodynamic Equilibrium (STE) limits when appropriate.
As described in Section \ref{sec:TwoLevelApproximation} below, 
our models of other ions are not as complete.

\subsubsection{The \htwo{} molecule}

This, the most common molecule in the universe, is treated as an extensive model
introduced in Gargi Shaw's thesis \citep{2005ApJ...624..794S}.
Improvements are described in greater detail in Section \ref{sec:GrainMolec} below.
  
 \subsubsection{Stout, CHIANTI, and LAMDA models}
 
 We treat emission from atoms, ions, and molecules (other than those described in the previous sections) 
 using the Stout, CHIANTI, and LAMDA databases.
 These use a common codebase and are controlled in very similar ways.
 The H-like and He-like iso sequences, and the \htwo{} model, were created
 as separate projects and are controlled by a separate set of commands.
  
For molecules, we use 
the Leiden Atomic and Molecular
Database ``LAMDA''\footnote{\url{http://www.strw.leidenuniv.nl/~moldata/}} 
\citep{Schoier.F05An-atomic-and-molecular-database-for-analysis}.
Section \ref{sec:LAMDA} below gives more details.
For some ions, we use version 7.1.4 of the 
CHIANTI\footnote{\url{http://www.chiantidatabase.org/}}   database
\citep{1997A&AS..125..149D,2012ApJ...744...99L},
as described by \citet{Lykins.M12Radiative-cooling-in-collisionally-and-photo}.
 
 We add new species to our Stout database \citep{LykinsStout15}.
 The data format is designed to be as close as possible to the presentation
 tables in the original publications.
 This makes Stout easy to maintain and update.
 We use NIST energy levels where possible.  
   
 The original publications defining the LAMDA and CHIANTI databases should be consulted
 to find the original references for individual data sources.
 Our Stout database is constantly updated.
 Appendix \ref{sec:AppendixStoutData}  gives a summary and references for the data it uses.

There are many species for which
 NIST gives level energies and transition probabilities but no collision data are available.
 For these we use  NIST data  with collision rates from the g-bar approximation \citep{Burgess1992}.
We refer to these as ``baseline'' models in Appendix \ref{sec:AppendixStoutData}.
 \citet{LykinsStout15} gives further details.
   
Baseline  model wavelengths should be accurate, and the transition probabilities are matched to
 the energy levels, but the g-bar collision strengths are highly approximate.
 High-quality collision data {\it are} available for most astrophysically important species,
 as shown in Appendix \ref{sec:AppendixStoutData}, so baseline models are mainly used for
 species that are not commonly observed.
 
There are several considerations to keep in mind if a baseline species is important in a particular application.
First, the collision rates are highly approximate, so at low densities the line intensities will be too.
If the density if high enough for the levels to be in LTE the predictions will be fine.
However, with some effort, the predictions could be improved.  First, the OPEN-ADAS data 
collection\footnote{\url{http://open.adas.ac.uk/}} does include 
plane-wave Born or distorted-wave collision rates.  These are better than g-bar
but the data set is not matched to NIST energies.  This matching can be done with some effort,
as has been done for Fe~II \citep{Verner1999} and Si~II \citep{Laha2016}.
Alternatively, members of the atomic physics community could be asked to 
produce close-coupling collision rates for the astrophysical application.
 
 \Cloudy{} has long included a large and very complete model of Fe~II emission
 which was developed as part of Katya Verner's thesis
 \citep{Verner1999}.  Modern atomic calculations now routinely provide
 datasets of similar or larger size, so the current version of \Cloudy{} can
 create complex emission models of any species with sufficient data.
The  \citet{Verner1999} Fe~II model is now fully integrated into the Stout database.

\subsection{Species and their names}

\Cloudy\ simulates gas ranging from fully ionized to molecular.
Nomenclature varies considerably between chemical, atomic, and plasma physics.
We have adopted a naming convention that tries to find a middle ground between
these different fields.

A particular atom, ion, or molecule is referred to as a ``species''.
A species is a baryon,
and this release of \Cloudy{} has 625 species.
Examples are CO, \htwo, H$^+$, and Fe$^{22+}$.
Species are treated using a common approach, as much as possible.
Our naming convention melds a bit of each of these fields
because a single set of rules must apply to all species.

\begin{itemize}

 \item
 Species labels are case sensitive, to distinguish between the
 molecule ``CO'' and the atom ``Co''.
 
 \item
 At present we do not use ``\verb|_|'' to indicate subscripts, or ``\verb|^|''
 to indicate  charge.

 \item
 Molecules are written pretty much as they appear in texts.
 \htwo, CO, and H$^-$ would be written as ``H2'', ``CO'', and ``H-''.

 \item 
 Atoms are the element symbol by itself.
 Examples are ``H'' or ``He''
and \emph{not} the atomic physics notation
 H$^0$ or He$^0$.
 
 \item 
 Ions are given by ``+'' followed by the net charge.
 Examples are ``He+2'' or ``Fe+22''
 and \emph{not} the correct atomic physics notation,
 He$^{2+}$ or Fe$^{22+}$.
 The latter would clash with notation for molecular ions.
 ``C2+'' indicates C$_2^+$ in our notation.
 
\item We specify isotopes using ``\^{}'' and the atomic weight
  placed before the atom to which it refers.  For example,
  ``\^{}13CO'' is the carbon monoxide isotopologue $\rm ^{13}CO$.

 \end{itemize}

Appendix \ref{sec:AppendixSpecies} lists our species together with the database
used to treat them.

\subsection{Working with spectral lines}

These species may emit  a collection of photons which we refer to as
a spectrum, although the species and spectrum may be labeled differently.  
We follow the spectroscopic convention that a spectral line is identified by a 
 label and a wavelength.
The next sections discuss how each is specified.

\subsubsection{Specifying spectral lines}

 We follow a modified atomic physics notation for the spectrum.
 In atomic physics,
 H~I, He~II, and C~IV are the spectra emitted by H$^0$, He$^+$, and C$^{3+}$.
 ``H~I'' indicates a collection of photons while H$^0$ is a baryon.  
  In our notation, we  replace the Roman numeral with an integer so
we refer to the spectrum as ``{\tt H  1}'' and the baryon as ``{\tt H}'' in the output.
For example, H~I $\lambda 4861$\AA, 
 He~II  $\lambda 4686$\AA,
 and the C~IV  $\lambda 1549$\AA{} doublet
  would appear in the output as
H~~1 4861.36, 
 He~2  4686.01,
 and Blnd  1549.00
 (blends are discussed below).

Chemistry does not suffer from this distinction between baryons and spectra
so the species label is also the spectroscopic ID.
Some examples of molecular lines in the output might include
H2O 538.142m,
HNC  1652.90m,  
HCS+ 1755.88m,
CO   2600.05m, 
CO   1300.05m, 
\^{}13CO 906.599m.
In this context the ``m'' indicates microns rather than our default 
angstrom unit.
 
To summarize, atomic hydrogen would be referenced as ``H'' while the L$\alpha$
line would be ``H~~1''.  The distinction is important because,
depending on whether it is formed by collisional excitation or recombination, 
L$\alpha$ can trace either H$^0$ or H$^+$.

We continue to follow the spectroscopic convention of denoting a line by its species label and wavelength.
This has the problem that several lines in a rich spectrum may have the save wavelength, at
least to our quoted precision.  An example is the strongest molecular hydrogen line that
can be measured from the ground.
The \htwo{} 1-0 S(1) transition has a wavelength of 2.121 \micron.
However there are a number of \htwo{} lines with nearly this wavelength;
3-2 S(23)	2.11751\micron, 1-0 S(1)	2.12125\micron, and 3-2 S(4)	2.12739\micron.
We ameliorate this confusion by reporting the wavelengths with six significant figures
in this version.
However, this method of identifying lines is fragile and it is still possible that the code will find a line
with the specified wavelength that is not the intended target.

\subsubsection{Line blends}

We introduce the concept of ``line blends'' in this version.  
These have the label ``Blnd'' in the main output, and a simplified wavelength.
An example is {\tt Blnd      2.12100m}, which is the sum of the three \htwo{} lines
mentioned above.
Operationally, a spectrometer measures the total flux through one spectral resolution element
and it is frequently not possible to identify individual contributors to what appears
as a single spectral feature. 
The {\tt Blnd} output option makes it possible for \Cloudy{} to report what is measured.

There are other cases where spectroscopists report the total intensity of a multiplet
even when individual members can be measured.  Two examples are
the [\oii{}] $\lambda \lambda 3726, 3729$ and [\sii{}] $\lambda \lambda 6731, 6716$ doublets.
We report the total multiplet intensity as {\tt Blnd 3727} and {\tt Blnd 6720}.
Such multiplet sums had been added to \Cloudy{} on an ad hoc basis in previous versions,
often with the label ``TOTL''.  
The ``Blnd'' entry makes the notation consistent across the code and allows it to be included in the
reporting framework described in Section \ref{sec:SaveLineLabels}.

\subsubsection{Air vs vacuum wavelengths}
The convention in spectroscopy, dating back to 19$^{\rm th}$ century experimental atomic physics,
is to quote line wavelengths in vacuum for  $\lambda < 2000$\AA\ and  air wavelengths
for $\lambda \ge 2000$\AA.
\Cloudy{} has long followed this convention.

There is an increasing trend to use vacuum for all wavelengths, e.g. due to satellite missions and
the Sloan project\footnote{\url{http://www.sdss.org/dr12/spectro/spectro_basics/}}.
We provide a command,
{\tt print line vacuum},
to use vacuum wavelengths throughout.
The continuum reported by the family of {\tt save continuum} commands,
used in several of the examples presented below,
is always reported in vacuum wavelengths to avoid 
a discontinuity at 2000\AA.

\subsection{Which database for which species?}

The H-like and He-like iso-electronic sequences are always included, although
the default number of levels is a compromise between speed and precision.
This is discussed in Section \ref{sec:iso-sequences}.
It is not possible to substitute other models, for instance, CHIANTI, for these species.
These iso-sequences are integrated into the ionization-balance solver so
are needed for it to function.

The large \htwo{} model is not used by default.  It is enabled with
the command {\tt database H2}.
In this comprehensive model,
radiative/collisional processes are coupled to the
dissociation/formation mechanisms and resulting chemistry.

The remaining databases, Stout, CHIANTI, and LAMDA, have different emphases.
LAMDA has a focus on molecules and PDRs, while CHIANTI is optimized for solar physics
and models in collisional ionization equilibrium.
Nonetheless, there are some species that are present in more than one database.

Each database has its own ``{\tt masterlist}'' file that
specifies which of its models to use.
The {\tt masterlist} file follows the naming convention used within its database.
For CHIANTI and Stout, the internal structure of C$^{3+}$, which produces C~IV emission,
is called ``c\_4''.
The water molecule in LAMDA is referenced as ``H2O''.
If a particular species is specified in more than one {\tt masterlist} file we will
use Stout if it exists, then CHIANTI, followed by LAMDA.  

 A small part of the default Stout masterlist file is shown here:
 
{\footnotesize
\begin{verbatim}
#c mn_23  
#n mn_3  
#n mn_4  
mn_5  
mn_6  
#c mn_8  
#c mn_9  
# 50 levels for N I to do continuum pumping discussed in
# >>refer	 Ferland et al., 2012, ApJ 757, 79
n_1  50
#c n_2  
#c n_3  
#c n_4  
n_5  
na_1  
na_2  
#c na_3  
#c na_4  
\end{verbatim}}
\noindent
This file lives in the {\tt cloudy/data/stout/masterlist} directory.  Similar files are located in the
{\tt cloudy/data/chianti/masterlist} and {\tt cloudy/data/lamda/masterlist} directories.
Each line in the file has a species label and those beginning with ``\#'' are available but are commented out
(we use the ``\#'' character to indicate comment lines across our data files).
This shows that we use Stout models of Mn~V, Mn~VI, N~I, Na~I, and Na~II.  We might 
use CHIANTI data, or ignore, species that are commented out.

It is easy for the user to use species from different databases
by editing the {\tt masterlist} files.
But there are consequences of using a non-default database.
The biggest is that different databases will often have different versions of the level energies.
The line wavelengths may change because we derive the wavelength from the level energies.
We use both the line label and the extended precision form of the wavelength to match lines.
This may break if the wavelengths change significantly.

We do support changing the databases but in release versions of the code have
created MD5 checksums for all of the data included in the download.  A caution
will be reported if the non-default data files are used.
This is intended to remind the user that our default data files have been changed in some way.

\subsection{How complete a model should be done?}

The default setup for the iso-electronic sequences is described below.
When our \htwo{} model is selected, the full dataset is used.

The Stout, CHIANTI, and LAMDA databases have similar user interfaces.
The default number of levels is described in \citet{Lykins.M12Radiative-cooling-in-collisionally-and-photo}.
For a particular species, the temperature of maximum ion abundance 
is hotter in a collisionally-ionized gas
than in the photoionized case.
Because of this higher kinetic temperature, 
more levels will be energetically accessible in the collisional case.
By default, we use 50 levels for the collisional and 15 levels for the photoionization case.

The number of levels can be adjusted to suit particular needs.
There are several ways to do this.
The command {\tt database species "name" levels xxx} 
will change the number of levels for a particular species.  The command
 {\tt database CHIANTI  levels maximum}  will make all of the CHIANTI
 models as large as possible.  Similar commands also work for the Stout and LAMDA databases.
 Finally, the minimum number of levels for a species can be specified in its {\tt masterlist} file
 by entering a number after the species label.
 For instance, faint optical [N~I] lines in H~II regions are mainly excited by continuum fluorescence
 \citep{FerlandNIpump12}.
 This physics requires that the lowest fifty levels of N~I be included.  This was done in C13
 by explicitly including those levels in the C++ source.  
 In this version we specify this minimum number of
 levels in the Stout {\tt masterlist} file.
 The example portion of the Stout {\tt masterlist} file given above includes this particular case.
 
\begin{figure}
\centering
\includegraphics[width=\linewidth]{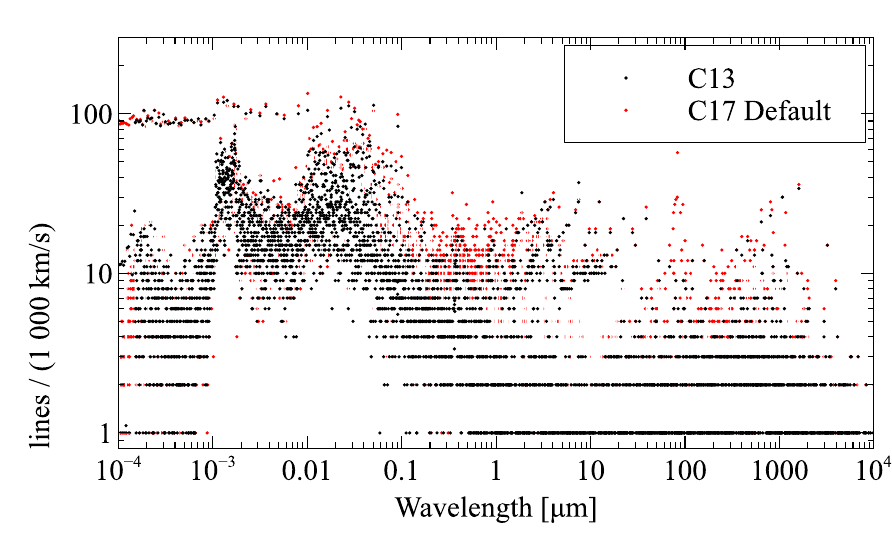}
\caption[Number of lines, default and maximum databases]{
\label{fig:LineDensity1317}
This compares the number of lines that fall into 1000 \kmps velocity bins
in C13 (black) and the default C17 setup (red).
}
\end{figure}

With these databases we predict, by default, more lines in this version of 
\Cloudy{} than with C13.
This actually takes less computer time  because of memory and other optimizations described below.
Figure \ref{fig:LineDensity1317} compares the density of lines per 1000 \kmps velocity interval for C13,
given as the black dots, and
our default setup for C17, the red dots.  
Most spectral regions now have more lines, often by up to 50\%.

\begin{figure}
\centering
\includegraphics[width=\linewidth]{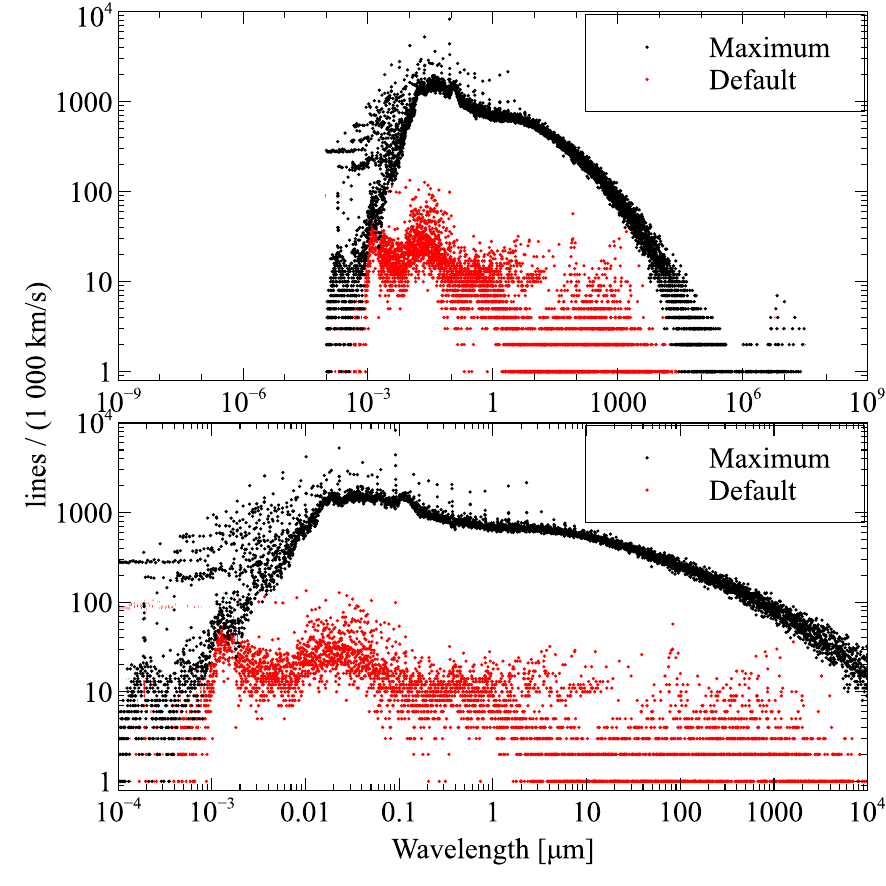}
\caption[Number of lines, default and maximum databases]{
\label{fig:LineDensity}
This shows the number of lines that fall into 1000 km/s velocity bins,
across the spectrum.
The red points for  default setup
and the black points give the number of lines the are predicted when
the databases are made as large as possible.
The upper panel shows the full spectral range considered by the
code, while the lower panel shows the peak of the line density.}
\end{figure}

Figure \ref{fig:LineDensity} compares the density of lines per velocity in our default
setup versus a calculation with the databases made as large as possible.
The upper panel presents the ``big picture'', the line density over the full spectral range we cover,
a frequency
of \emmmhz\ $(\lambda \approx$ \emmcm\ -- this is approximately the lowest
frequency observable through the ionosphere) and an energy
of $h\nu = $\egamry\ $(\approx$ \egamrymev).
Few lines are present shortward of $10^{-4} \micron = 1$\AA\ or
longward of 3\e{6} \micron = 3m.
The lower panel zooms into the mid-spectral region with significant numbers of lines.
The  red points indicate our default C17 setup.
The upper envelope of black points results from making the
databases as large as possible.

We will give example input scripts across this document to show how various Figures were produced.
The large-database calculation in Figure \ref{fig:LineDensity} was created with the following input deck:

{\footnotesize
\begin{verbatim}
table AGN
ionization parameter -2
stop zone 1
constant temperature 1e4 K
hden 0
database H2
database H-like levels resolved 5
database He-like levels resolved 5
database H-like levels collapsed 200
database He-like levels collapsed 200
database CHIANTI levels maximum
database stout levels maximum
database LAMDA levels maximum
save continuum units microns "mesh.con"
\end{verbatim}}

All commands are fully documented in ``Hazy 1'', \Cloudy's documentation,
which is part of the download.
Most commands are unchanged from C13.
The spectral energy distribution (SED) is our generic Active Galactic Nucleus (AGN) 
continuum, with an ionization parameter of $U = 10^{-2}$.
The geometry is a single zone with a hydrogen density of $1 \pcc$ and a gas kinetic temperature 
set to 10$^4$~K.  
These have to be specified to get the code to run and were chosen to do the simplest 
and fastest calculation.
As described here, the {\tt database} commands are new in C17 and control
the behavior of the databases.

The number of lines per spectral bin is one of the items in the file produced by the
{\tt save continuum} command.
The width of each continuum bin, or, equivalently, the spectral resolution, can be adjusted in two ways.
The default continuum mesh can be changed by a uniform scale factor with the 
{\tt set continuum resolution ...} command.
This method is used in several simulations presented below to increase the spectral resolution to highlight
particular issues in the spectrum.
Alternatively, the initialization file {\tt continuum\_mesh.ini} in the {\tt data} directory can be changed to 
alter the default continuum mesh. 
This second method was used to make the C13 and C17 continuum meshes
the same, to allow the comparison in  Figure \ref{fig:LineDensity1317}.

Figure \ref{fig:LineDensity} shows that it is possible to predict a very large number of lines,
but this comes at great cost.  
The default C17 calculation took 4.3 s on an Intel Core I7 processor while the calculation
using the large databases took 1822~s, roughly half an hour.
It would not now be feasible to use the full databases in a realistic calculation in which
the temperature solver is used and the cloud has a significant column density so that
optical depths are important.

\subsection{Generating reports}

\subsubsection{database print}

The command {\tt database print} generates a report listing all species. 
The following would generate a report for \Cloudy{} in its default setup:

{\footnotesize
\begin{verbatim}
test
database print
\end{verbatim}}
\noindent
This command was used to generate Tables \ref{tab:HLevels} and \ref{tab:HeLevels} 
giving the default setup for the one- and two-electron iso-sequences.
A small portion of the report for the Stout, CHIANTI, and LAMDA databases follows:

{\tiny
\begin{verbatim}
Using LAMDA model SO with 70 levels of 91 available.
Using LAMDA model SiC2 with 40 levels of 40 available.
Using LAMDA model CS with 31 levels of 31 available.
Using LAMDA model C2H with 70 levels of 102 available.
Using LAMDA model OH+ with 49 levels of 49 available.
Using STOUT spectrum Al 1 (species: Al) with 15 levels of 187 available.
Using STOUT spectrum Al 3 (species: Al+2) with 15 levels of 83 available.
Using STOUT spectrum Al 4 (species: Al+3) with 15 levels of 115 available.
Using STOUT spectrum Zn 2 (species: Zn+) with 15 levels of 27 available.
Using STOUT spectrum Zn 4 (species: Zn+3) with 2 levels of 2 available.
Using CHIANTI spectrum Al 2 (species: Al+) with 15 experimental energy levels of 20 available.
Using CHIANTI spectrum Al 5 (species: Al+4) with 3 experimental energy levels of 3 available.
Using CHIANTI spectrum Al 7 (species: Al+6) with 15 experimental energy levels of 15 available.
Using CHIANTI spectrum Al 8 (species: Al+7) with 15 experimental energy levels of 20 available.
Using CHIANTI spectrum Al 9 (species: Al+8) with 15 experimental energy levels of 54 available.
\end{verbatim}}

Each line of output gives the database name, the spectroscopic designation, the species designation,
the number of levels used, and the total number available.
With CHIANTI there is the further option to use all levels, or only those
with experimental (measured) energies.

\subsubsection{save species labels all}

The {\tt save species labels all} 
command will produce a file containing the full list of species labels.
One can generate this list by running the following input deck:
\begin{verbatim}
test
save species labels "test.slab" all 
\end{verbatim}

\noindent
Table \ref{tab:SaveSpeciesLabel} shows a small part of the resulting output.
There are several important points in this Table.
First, several species do not list a database.
The cases of ``H+'', ``He+2'', and ``C+6'' are bare nuclei and have no electronic levels, while
the negative hydrogen ion``H-'' and the molecules  ``HeH+'', ``C2+'' and ``CN+'' do have
internal levels in nature, but we currently do not have models of these systems.
The remainder are treated with one of the databases described above.
Although many of these species have no internal structure, other species properties,
especially the column density, are computed and reported.

Note a likely source of confusion.  As described above, ``C+2'' is doubly ionized carbon,
while ``C2+'' is an ion of molecular carbon.

\begin{table}[t!]
\centering
\small
\caption{Save species label  example}
\label{tab:SaveSpeciesLabel} 
\tablecols{2}
\setlength\tabnotewidth{\linewidth}
\setlength\tabcolsep{2\tabcolsep}
\begin{tabular}{@{} cr@{\hspace*{2em}} cr @{\quad\quad}}
\toprule
Species label&Database\\
H&H-like \\
H+&\\
H-&\\
He&He-like\\
He+&H-like \\
He+2&\\
HeH+&\\
C&Stout\\
C+&Stout\\
C+2&Stout\\
C+3&CHIANTI\\
C+4&He-like\\
C+5&H-like \\
C+6&\\
C2+&\\
C2H&LAMDA\\
NH3&LAMDA\\
CN&LAMDA\\
CN+&\\
HCN&LAMDA\\
\bottomrule \\
\end{tabular}
\end{table}

\subsubsection{print citation}

The {\tt print citation} command reports the ADS links to papers defining the databases
we use.
We encourage users to cite the original source of any data that played an important role in an investigation.
 This will support and encourage atomic, molecular, and chemical physicists 
 to continue their valuable work.

\subsubsection{save species commands}

It is easy to report internal properties of a species,
such as the column density or population of a particular level.
The following is an example of a {\tt save species} command used
 to report the column densities 
of several species and the visual extinction:
\begin{verbatim}
save species column densities "test.col"
"e-"
"CO[2]"
"C[1:5]"
"H2"
"H"
"H+"
"*AV"
end of species
\end{verbatim}

\noindent
The total column density of electrons, \htwo{}, \h0, and \hplus{} would be reported,
along with the population in the $J=1$ level of CO, and the first five levels of C$^0$.
Note that the level index within the ``[xx:yy]'' counts from a lowest level of 1.

\subsubsection{Save line labels}
\label{sec:SaveLineLabels}

The {\tt save line labels} command creates a file listing all spectral-line labels and 
wavelengths in the same format as they
appear in the main output's emission-line list. 
This is a useful way to obtain a list of lines to use
when looking for a specific line. The file is tab-delimited, 
with the first column giving the line's
index within the large stack of spectral lines, 
the second giving the character string that identifies
the line in the output, and the third giving the line's wavelength in any of several units. 
Each entry ends with a description of the spectral line. 
Lines derived from databases (CHIANTI, Stout, LAMDA) are
followed by a comment that contains the database of origin 
and the indices of the energy levels,
as listed in the original data.

An example of some of its output follows:

{\tiny
\begin{verbatim}
     4	Inci            0 		total luminosity in incident continuum
     5	TotH            0 		total heating, all forms, information since individuals added later 
     6	TotC            0 		total cooling, all forms, information since individuals added later 
  1259	H  1      911.759A		radiative recombination continuum
  1260	H  1      3646.00A		radiative recombination continuum
  1261	H  1      3646.00A		radiative recombination continuum
  1262	H  1      8203.58A		radiative recombination continuum
  3552	H  1      1215.68A		H-like, 1 3,   1^2S -   2^2P
  3557	H  1      1025.73A		H-like, 1 5,   1^2S -   3^2P
  3562	H  1      972.543A		H-like, 1 8,   1^2S -   4^2P
  5328	Ca B      1640.00A		case a or case b from Hummer & Storey tables
  5329	Ca B      1215.23A		case a or case b from Hummer & Storey tables
 73487	CO        2600.05m		LAMDA, 1 2
 73492	CO        1300.05m		LAMDA, 2 3
 73497	CO        866.727m		LAMDA, 3 4
 85082	C  3      1908.73A		Stout, 1 3
 85087	C  3      1906.68A		Stout, 1 4
 85092	C  3      977.020A		Stout, 1 5
180217	Al 8      5.82933m		CHIANTI, 1 2
180222	Al 8      2139.33A		CHIANTI, 1 4
180227	Al 8      381.132A		CHIANTI, 1 8
312763	H2        1.13242m		diatoms lines
312768	H2        1.26316m		diatoms lines
312854	Blnd      2798.00A		Blend: "Mg 2      2795.53A"+"Mg 2      2802.71A"
312855	Blnd      615.000A		Blend: "Mg10      609.794A"+"Mg10      624.943A"
\end{verbatim}}

\section{The ionization equilibrium}

\begin{figure}
\centering
\includegraphics[width=\linewidth]{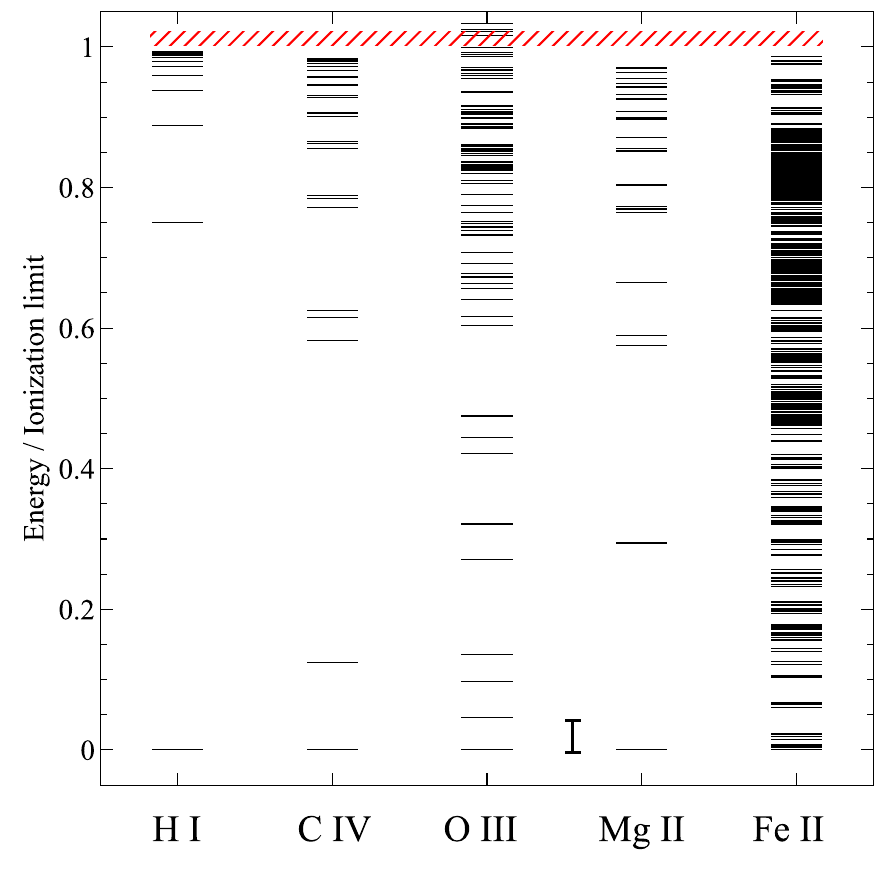}
\caption[Energy levels for some elements]{
\label{fig:EnergyLevels}
Experimental energy levels \citep{NIST_ASD} for some species present in an ionized gas.
The energies are given relative to the ionization potential (IP).  Of these ions,
only O~III and Mg~II have data for autoionizing levels,
shown as the levels above the ionization limit indicated by the red hashed box.
The autoionizing levels of Mg~II are not
visible since they are far above the ionization limit.
The vertical bar in the middle, corresponding to E/IP = 0.05, is a typical gas
kinetic energy in a photoionized plasma and is shown to indicate
which levels are energetically accessible from the ground state. 
}
\end{figure}

Our goal is to compute the ionization for a very wide range of densities and temperatures,
as shown in Figures 17 and 18 of C13, for the first thirty elements and all of their 
ions.\footnote{Those figures
were created using the output of {\tt grid\_extreme} from our the test suite.}
This development is complete for the H- and He-like iso-electronic sequences,
but is still in progress for many-electron systems.
This is challenging because of limitations in the available atomic data, computer hardware, and human effort.
We currently use a hybrid scheme, outlined below, which is motivated by the astrophysical problems
we wish to solve.
Figure \ref{fig:EnergyLevels} shows energy levels for five typical species,
to serve as an example of some of the issues involved in the following discussion.
 
 There are two simple limits for the solution of the ionization equilibrium.
 At the low densities found in the interstellar medium (ISM),  most ions are in the ground state and
 the ionization rate per unit volume is proportional to the atom density
 multiplied either by the electron density, in the case of collisional ionization, or the
 ionizing photon flux, in the case of photoionization.
 The recombination rate per unit volume is proportional to the product of the ion
 and electron density.  The density dependence drops out for the case of collisional ionization
 (see Equation \ref{eq:CollTwoLevel} below).
 In contrast, at  very high densities, such as the lower regions of a stellar atmosphere,
  the ionization is given by the Saha-Boltzmann equation and is 
 inversely proportional to the electron density (Equation \ref{eq:SahaBoltzmann} below).
 \Cloudy{} spans both regions so neither approximation can be made
 and a full solution of the NLTE equations should be performed to obtain the ionization balance and level populations.
 
In a collisional-radiative model (CRM), the name given to the full NLTE treatment in plasma physics, 
the level populations are determined 
self-consistently with the ionization.  That is, the populations in  bound levels and
the continuum above them in Figure \ref{fig:EnergyLevels} are solved as a coupled system of equations.
We use this approach for the H-like and He-like iso-sequences, as described in C13 and
the following sections, and shown in Figures 17 and 18 of C13.
A modified two-level approximation, described  below, is used for other species.
This hybrid approach  is motivated by the physics of the systems shown 
in Figure \ref{fig:EnergyLevels}.

\subsection{The H- and He-like iso-electronic sequences}
\label{sec:iso-sequences}

Energy levels for  \hone{}  are shown in
Figure{} \ref{fig:EnergyLevels} on the left.
This structure is valid for all ions of the H-like isoelectronic sequence.
For  \hone{}, the 
 first excited $n=2$ level occurs at an energy $E \approx (1 - 1/n^2) \approx 0.75$ of the ionization limit.
This lowest excited level is much closer to the continuum above it than to the ground state below it.
At the low kinetic temperatures found in photoionization equilibrium 
there should be little collisional coupling between excited and ground states because of
this large energy separation, although very highly excited levels are strongly 
coupled to the continuum above.
As a result, the very highly excited ``Rydberg'' levels are populated 
following recombination from the ion above it, rather
than collisional excitation from the ground state, in photoionization equilibrium.
Most optical and infrared emission is produced following recombination in this case.
The energies of the $n$-shells
are also roughly valid for the He-like sequence.  These two-electron systems
behave in a similar manner; optical and IR lines form by recombination.

We treat these two iso-electronic systems with a full CRM because
of the strong coupling of excited level populations to the continuum (C13).
A large coupled system of equations is solved to determine both the level populations
and the ionization, so the two are entirely self-consistent.
Predictions over a wide range of density and temperature are shown in figures 17 and 18 of C13.
We developed a unified model for both the H-like and He-like isoelectronic sequences,
that extends from H to Zn, as described by
\citet{2005ApJ...628..541B, 2007ApJ...657..327P, 2007ApJ...664..586P, 2012MNRAS.425L..28P}
and \citet{ 2009ApJ...691.1712L},
As shown in C13, and previously by \citet{Ferland1988},
our model of the ionization and chemistry of hydrogen 
does go to the correct limits at high (LTE) and low (ISM) densities,
and to the strict thermodynamic equilibrium (STE) limit
when exposed to a true blackbody.
This is only possible when the ionization and level populations are self-consistently
determined by solving the full collisional-radiative problem.

\subsubsection{Recent Developments}

The H- and He-like isoelectronic-sequences, 
coupled with the cosmic abundances of the elements, 
cause their spectra to fall into two very different regimes.  
For brevity, we refer to these as iso-sequences in the following.
Hydrogen and helium have low nuclear charge $Z$ and so have low ionization potentials, 
$IP \sim Z^2$.
As a result,  \hi{},  \hei{}, and  \heii{}
emission is produced in gas with nebular temperatures, $\sim 10^4$~K, and occurs mainly in
the optical and infrared.
A goal of the current development is the prediction of highly accurate line emissivities
as a step towards measuring the primordial helium abundance \citep{2007ApJ...657..327P}.

The next most abundant elements, starting with carbon ($Z=6$), have high
ionization potentials, $IP \geq 6^2$~Ryd, so are produced in very hot gas,
$T \geq 10^6$~K, and emit
in the X-rays \citep{2007ApJ...664..586P, 2006PASP..118..920P}.
Heavy elements of these iso-sequences fall into very different spectral regimes than 
hydrogen and helium, 
probe gas
with very different physical conditions, and so are found in distinctly different environments.

The high precision needed for primordial helium measurements means that the atomic data
must be quite accurate.
We are revisiting this problem.
The original papers on  \hi{} and  \hei{} emission
\citep{1971MNRAS.153..471B, 1972MNRAS.157..211B, 1999ApJ...514..307B,1987MNRAS.224..801H} all
used the \citet{1964MNRAS.127..165P} theory of $l$-changing collisions.
\citet{2001JPhB...34L...1V, 2001PhRvA..63c2701V} and \citet{ 2012ApJ...747...56V}
present an improved theory for these collisions, which predict rate coefficients
that are $\sim$6 times smaller.
We have used these newer rates in most of our published work
\citep{2005ApJ...628..541B, 2007ApJ...657..327P, 2007ApJ...664..586P, 2012MNRAS.425L..28P, 2009ApJ...691.1712L}, and in C13.

There are good reasons,
outlined in \citet{Guzman.I.2016}, to prefer the \citet{1964MNRAS.127..165P} theory.
\citet{Guzman.II.2017} extend this treatment to He-like systems, in which
the low-$l$ S, P, and D states are not energy-degenerate, so an extra 
  cut-off energy term is applied to the probability integral as in
  \citet{1964MNRAS.127..165P}. \citet{Guzman.II.2017} also improve the
\citet{1964MNRAS.127..165P}  approximations to deal with low 
densities and high
temperatures, where the original formulae could produce negative values. 
They call this the modified \citet[PS-M]{1964MNRAS.127..165P} approach.
\citet{2017JPhB...50k5201W} extend the semi-classical theory of \citet{2001JPhB...34L...1V} to provide thermodynamically-consistent
$l$-changing rates, which are found to agree quite closely with the results of the other approaches.

\begin{figure}
\centering
\includegraphics[clip,width=\linewidth]{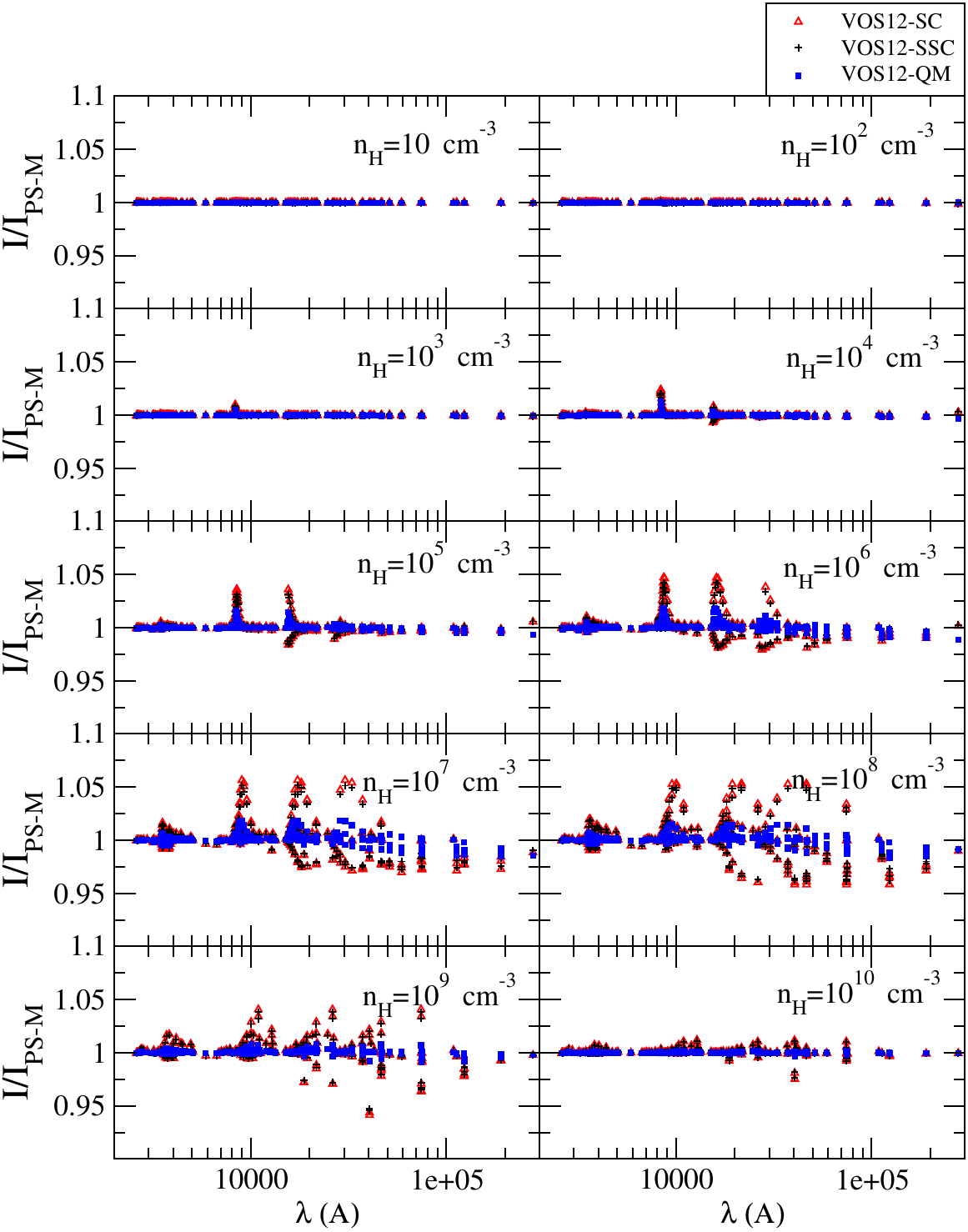}
\caption[l-changing lines]{
  \label{fig:helike}
  Ratios of He I lines using the different datasets
  with respect to PS-M, see text for details.
  Figure from \citep{Guzman.II.2017}.
}
\end{figure}

These differences affect the
predictions. In Figure \ref{fig:helike}, the line emissivities from the different approaches are divided by
the emissivities obtained using PS-M.
Here, a single layer of gas has been considered and emissivities from recombination lines calculated.
A cosmic helium abundance, He/H =0.1, was  assumed. The cloud is radiated by a
monochromatic radiation field using \Cloudy's {\tt laser} command.
It was centered at 2 Ryd, with an ionization
parameter of U = 0.1 \citep{AGN3}. The state of the emitting gas in these conditions
resembles that in  \hii{} regions, observationally relevant for the determination of He abundances
\citep{Izotov2014}. The monochromatic radiation field was used to prevent internal excitations that
would be produced by a broadband incident radiation field.
 A constant gas kinetic temperature of 1\e{4}~K is
 assumed. We assume `Case B' \citep{Baker1938}, where
Lyman lines with upper shell $n>2$ are assumed to scatter often enough to be degraded into Balmer lines and Ly$\alpha$. 
The hydrogen density is varied over a wide range and the electron density is calculated self-consistently.
The latter is approximately 10\% greater than the hydrogen density since He is singly ionized. 
The atomic data used for He and H emission, except for $l$-changing collisions, 
are the `standard' set of data that has been described in previous work \citep{Porter2005}.

In Figure \ref{fig:helike}, the most accurate quantum mechanical treatment,
VOS12-QM \citep{2001JPhB...34L...1V, 2001PhRvA..63c2701V, 2012ApJ...747...56V},
agrees closely with PS-M.
The semiclassical treatments, VOS12-SC \citep{2001JPhB...34L...1V, 2001PhRvA..63c2701V}, and
the further simplified approach,  VOS12-SSC \citep{2012ApJ...747...56V}, produce differences up to 10\%
in the predicted line intensities.
The input to generate Figure \ref{fig:helike} for the quantum mechanical case can be as follows\footnote{The commands must be written in one line.
Here, ``\textbackslash'' is used to break the command in two lines for
presentation purposes.}:

\begin{verbatim}
laser 2
ionization parameter -1
hden 4  
element helium abundance -1
init file "hheonly.ini"
stop zone 1
set dr 0
database he-like resolved levels 30
database he-like collapsed levels 170
database he-like collisions l-mixing S62 \
      no degeneracy thermal VOS12 quantal
constant temperature 4
case b no photoionzation no Pdest 
no scattering escape
no induced processes
iterate
normalise to "He 1" 4471.49A
monitor line luminosity "He 1" 7281.35A \
      -18.8686  
\end{verbatim}
In this input, the {\tt no photoionization} option is added to {\tt case b}
to suppress photoionization from excited levels. Then, the pumping of the levels
by the resulting photon field is removed by turning off the destruction of
Lyman lines with the {\tt no Pdest} option. The density is set to
$10^4\, \text{cm}^{-3}$ using the {\tt hden 4} command, and the temperature is
kept constant with {\tt constant temperature 4}. The commands:

\begin{verbatim}
stop zone 1
set dr 0
\end{verbatim}
define a layer of gas of 1~cm thickness. The commands

\begin{verbatim}
no scattering escape
no induced processes
\end{verbatim}
prevent losses due to scattering, so that all Lyman lines are degraded into Balmer lines. 
The {\tt monitor line luminosity \ldots}
command compares the computed luminosities (\erg\s$^{-1}$)
of selected \hei{} lines against the reference values given by the PS method.
Luminosities can be directly translated into emissivities as the thickness of the gas is 1~cm.

Figure \ref{fig:helike} has been computed extending to the $n=200$ shell using the
{\tt database he-like resolved/collapsed levels} commands described in section
\ref{sec:modelsize}. Most of the higher
$n$-shells are collapsed (C13, Figure 1) assuming a statistical population for the $l$-subshells,
while the low-$n$
levels are resolved. The number of resolved levels needed is determined by the critical density for
$l$-mixing, where collisional transition rates exceed radiative decay rates, as shown in figure 4 of
\citet{1964MNRAS.127..165P}. 
The number of resolved $n$-shells used to predict the lines in Figure \ref{fig:helike} has been
varied from $n=60$ at the lowest density to $n=20$ at the highest density.

Commands are provided to select the preferred $l$-changing theory in the input file for \Cloudy{}. 
The command to use PS-M is:
\begin{verbatim}
database he-like collisions l-mixing S62 \
      no degeneracy Pengelly
\end{verbatim}
\noindent
The keyword \verb+S62+ in this command tells \Cloudy{} to use the \citet{Seaton1962}
electron impact cross sections for the $l$-changing collisions of the highly
non-degenerate $l<3$ subshells \citep[see][]{Guzman.II.2017}.
The keyword \verb+no degeneracy+ uses an energy criterion \citep{1964MNRAS.127..165P} to account for
the non-degeneracy of the $l$-subshells of  \hei{} Rydberg levels (see above).
Calculations using the original formulas provided by \citet{1964MNRAS.127..165P}
can be used by adding the keyword \verb+Classic+.

VOS12-QM rate coefficients can be used with the command:

\begin{verbatim}
database he-like collisions l-mixing S62 \
      no degeneracy thermal VOS12 Quantal
\end{verbatim}

\noindent
where the keyword \verb+thermal+ tells \Cloudy{} to perform a Maxwell average
for the cross sections. By default, the effective coefficients will not
be Maxwell averaged, and energies of the collision particles will be taken to 
be $kT$.
The evaluation at a single energy 
$kT$ is significantly faster.

The VOS12-QM theory needs a larger number of operations than the analytic PS-M approach. 
It also needs a numerical integration of the collision probability to obtain the cross sections. 
These may be integrated once more to obtain the Maxwell averaged coefficients, 
making this method computationally slow. 
Simulations using VOS12-QM calculations are $\sim60$ times slower than those using PS-M. 
The computational cost of VOS12-QM calculations makes PS-M method the preferred one.
The VOS12-QM method is recommended when very high-precision results are required.

Finally VOS12-SC and VOS12-SSC can be obtained using the commands:

\begin{spverbatim}
database he-like collisions l-mixing S62 \
      thermal Vrinceanu 
database he-like collisions l-mixing S62 \
      VOS12 semiclassical
\end{spverbatim}

\noindent respectively. While VOS12-SSC cross-sections can be obtained with an 
analytical formula, VOS12-SC need a double integration making them as computationally slow as VOS12-QM.

It is not now possible to experimentally determine which of the theories mentioned above 
is more correct, although we prefer the PS-M approach.
\citet{Guzman.2017.TwoPhoton} outline an astronomical observation that,
while difficult, could conclusively determine which $l$-changing theory holds.

\subsubsection{Adjusting the size of the model}
\label{sec:modelsize}
Our models of the H- and He- like iso-sequences have a mixture of resolved and collapsed levels, 
as shown in Figure 1 of C13.
Resolved levels are relatively expensive to compute due to the need to
evaluate  the $l$-changing collision rates described above.
Collapsed levels assume that the $l$-levels are populated according to their statistical weight
within the $n$ shell, so this expense is avoided.

The number of resolved and collapsed levels are controlled by
a family of commands similar to

{\footnotesize
\begin{verbatim}
database H-like hydrogen levels resolved 10
database H-like hydrogen levels collapsed 30
database H-like helium levels resolved 10
database H-like helium levels collapsed 30
database He-like helium levels resolved 10
database He-like helium levels collapsed 30
\end{verbatim}
}
This example resolves $n\leq 10$ for \hi, \hei, and \heii{}
and adds 30 collapsed levels to make each atom extend through $n=40$.
These commands work for H-like and He-like ions of all elements up through zinc (Z=30).

The command
\begin{verbatim}
database print
\end{verbatim}
generates a report summarizing all databases in use during the current calculation.
This includes the number of resolved and collapsed levels for the iso-sequences.
By default we resolve $n\leq 10$ with an additional 15 collapsed
levels for \hi{} and \heii, and $n\leq 6$ as resolved with an additional 20 collapsed
levels for \hei. 

In C13 and earlier versions the collapsed levels were intended to ``top off'' the model and their treatment
did not have spectroscopic accuracy.  
\citet{2005ApJ...628..541B} discuss top off, the need to use a finite number of 
levels to approximate an infinite-level system, and its effects on predictions.
We did not report  lines from collapsed levels.
A great deal of effort has gone into improving the physics of these levels.
Their
emission is now highly accurate if the implicit assumption that the $l$ levels within the
$n$ shell are populated according to their statistical weight is valid.
The densities required for this ``well $l$-mixed'' assumption can be derived from
Figure 4 of \citet{1964MNRAS.127..165P}.
We now report emission from collapsed levels up to $n < n_{\rm highest}-4$.
This limit was chosen to avoid the ``edge'' effects discussed in the next subsection.

\subsubsection{Comparisons in the Case B limit}

It is possible to judge the accuracy of the predicted lines by comparing with the textbook ``Case B'' spectrum
\citep{Baker1938}.
Case~B is a well defined limit that is a fair approximation to nebulae 
(AGN3 Section 4.2) and can serve as an important benchmark.
\citet{1987MNRAS.224..801H} and \citet{1995MNRAS.272...41S} compute Case~B emission
 and the second paper includes a series of machine-readable tables.
We interpolate on these tables to include Case~B predictions for  H-like ions in the \Cloudy{} output.
This makes comparison of our predictions with Case~B simple.

While Case~B is an idealized limit, it gives a fairly good representation of emission
for low to moderate density photoionized nebulae (AGN3).  The assumptions break
down if Balmer or higher lines become optically thick,  when collisional excitation from the ground state
becomes important, or if Lyman lines can escape.  Optical depth effects become important in high radiation-density
environments such as the inner regions of active galactic nuclei.
Collisional contributions become important when suprathermal electrons are present
as in X-ray ionized neutral gas \citep{1968ApJ...152..971S}, or for hot regions such 
as very low metallicity nebulae.
The Lyman lines may not be optically thick in low column density clouds.
These processes are treated self-consistently in any complete \Cloudy{} calculation.

Larger models, with more levels, make the spectrum more accurate, at the
cost of longer execution times and higher memory requirements.
The default \hi{} model is a compromise between performance and accuracy.

\begin{figure}
\centering
\includegraphics[width=\linewidth]{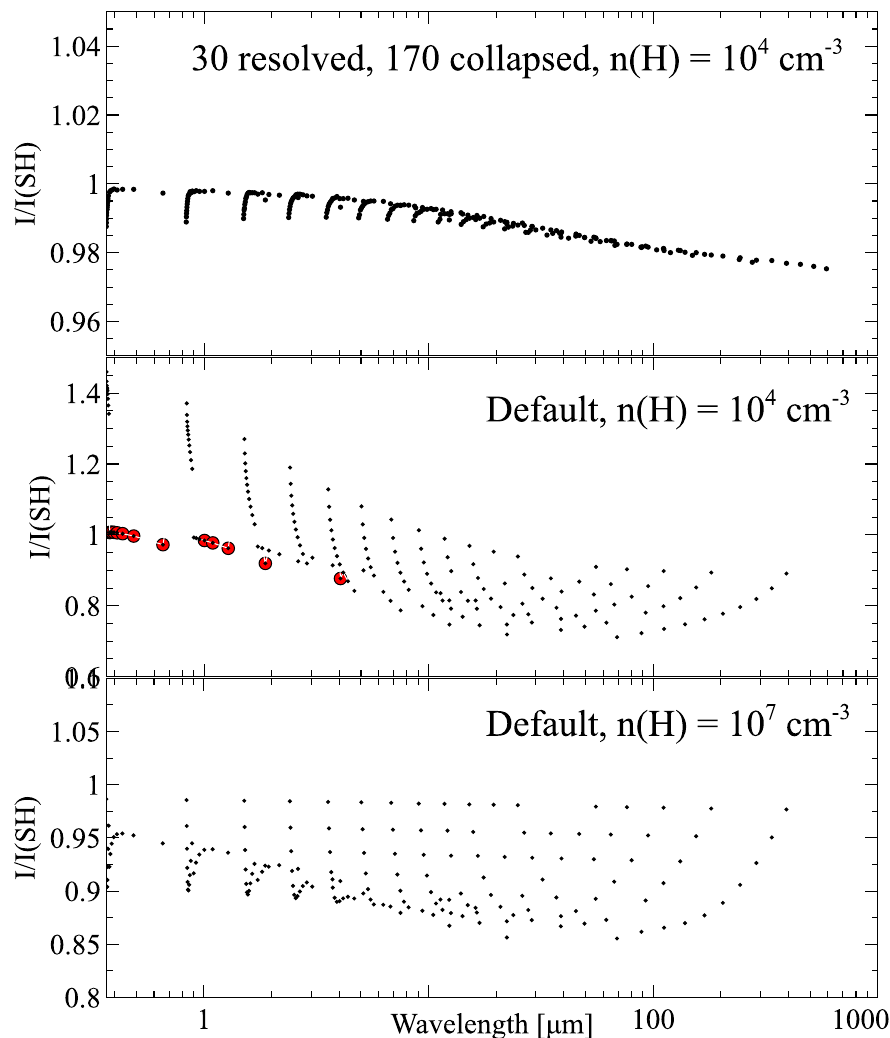}
\caption[l-changing lines]{
  \label{fig:casebAccuracy}
This shows ratios of our predicted \hi{} emission to the \citet{1995MNRAS.272...41S} Case~B tables.
Calculations are for $T_e = 10^4$~K and the indicated densities.
The upper panel, our test case {\tt limit\_caseb\_h\_den4\_temp4.in.}, 
shows that we reproduce their results to high accuracy when a large
model is used.  The default model, chosen as a compromise between speed and accuracy,
is shown in the lower two panels.
The default model is designed to give higher accuracy for the brighter optical and near-IR lines,
plotted as the larger filled circles.
Note that each panel has a different vertical scale.
}
\end{figure}

Figure \ref{fig:casebAccuracy} compares our predictions with  \citet{1995MNRAS.272...41S}
for a typical ``nebular'' temperature, $T_e = 10^4$~K, and two densities,
$n($H$) = 10^4 \pcc$ and $10^7 \pcc$.
The lower two panels show predictions of our default
model at the two densities while the upper panel shows predictions at the lower
density with a greatly increased  number of
resolved and collapsed levels.

In a normal calculation, \Cloudy{} determines the line optical depths self-consistently, assuming
the computed column densities, level populations, and line broadening.
A {\tt Case~B} command exists to create these conditions and
make these comparisons possible while computing a single ``zone'',
a thin layer of gas.  
The command sets the Lyman line
optical depth to a large number and suppresses collisional excitation out of $n=2$,
to be consistent with the \citet{1995MNRAS.272...41S} implementation of Case~B.
This command was included in the input script used to to create Figure \ref{fig:casebAccuracy}.
In previous versions of \Cloudy{} we also recommended using the  {\tt Case~B}  command in certain
simple PDR (photodissociation region, or photon-dominated region) 
calculations to block Lyman line fluorescence. 
As described in Section \ref{sec:LymanPDR} below, we now 
recommend using a different command in the PDR case, reserving the  {\tt Case~B} command
for this purpose only.

The large model reproduces the \citet{1995MNRAS.272...41S} results to high precision.
There are differences at the $\sim 2$\% level, which we believe to be caused by recent improvements
in the collision and recombination data.  
This will be the subject of a future paper (Guzm{\'a}n et al. in preparation).

The middle and lower panels show the results of applying our default \hi{} model 
($n \leq 10$ resolved with an additional 15 collapsed) to the same density, and to a higher density case.
The default model was chosen to reproduce the intensities of the brighter \hi{}
lines to good precision.
The large red filled circles in the middle panel indicate lines with intensities 
greater than 5\% that of H$\beta$.
These have deviations of $\lesssim 10\%$.

The eye picks up a trend for the error ratio to move away from unity along a spectroscopic
(that is, Balmer, Paschen, etc)  series,
as the upper level $n$ increases.
This is produced by two effects.
The first is an ``edge'' effect as the upper level approaches the upper limit of the model.
(We do not report lines arising from $n \geq n_{\rm highest}-4$ for this reason.)
The level populations for the very highest levels are inaccurate because of their proximity
to the large ``gap'' that exists between the highest level and the continuum above.
The errors at lower $n$ are due to the  fact that the lower collapsed levels,
$10 < n \leq 15$, are not actually $l$-mixed at the lower density of $10^4 \pcc$.
The density of $10^7 \pcc$ is high enough to $l$-mix $n=11$ so this model is better behaved.
These tests show that the implementation gives reliable results when the number of
levels is made large enough.

\begin{figure*}[t!]
\centering
\includegraphics[width=\linewidth]{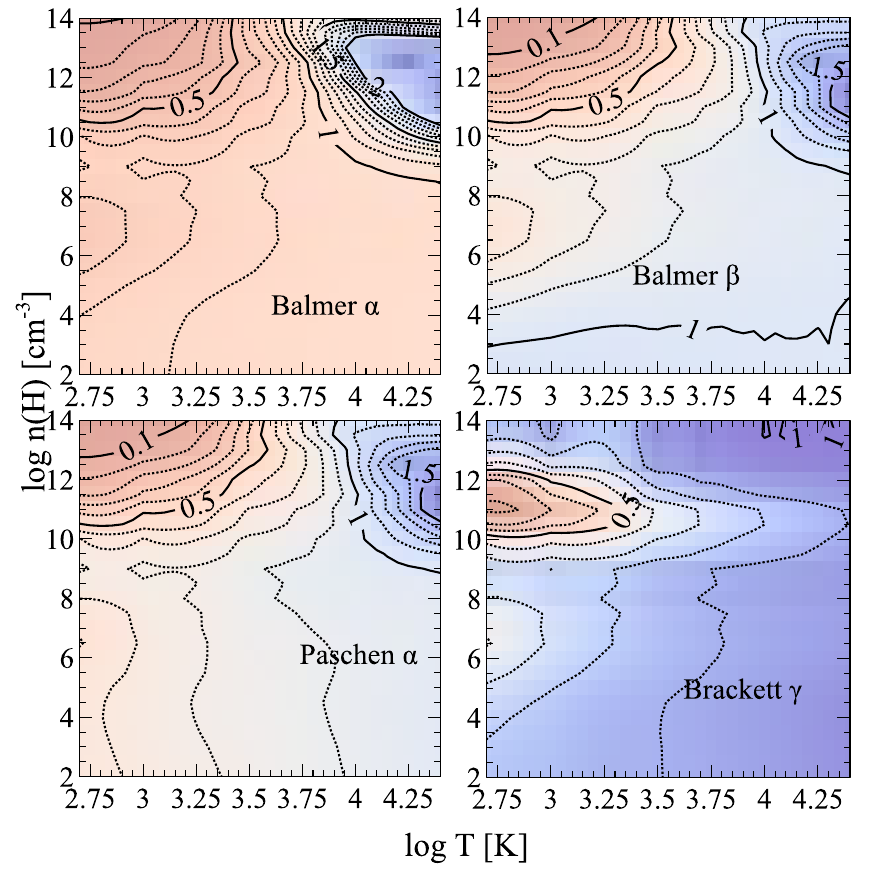}
\caption[l-changing lines]{
  \label{fig:nt}
This shows ratios of our predicted \hi{} emission to the \citet{1995MNRAS.272...41S} Case~B tables for our default \hi{} model.
Calculations are for the full temperature and density range they provide.
Major contours are at $I/I_{Case B} =0.5$,1, and 1.5 while minor contours, shown as the dotted lines,
are at 10\% incremental values.
The differences at the higher densities are due to the use of more recent collision rates in
our calculations.
}
\end{figure*}

The default model was designed to reproduce the intensities of the brighter \hi{} lines
while being computationally expeditious.
As a test we computed the intensities of four of the most commonly observed \hi{} lines
over the density and temperature given in the  \citet{1995MNRAS.272...41S} tables including
the {\tt Case B} command.
The ratio of our predictions to their Case~B values is given in Figure \ref{fig:nt}.
The largest differences are at the higher densities where Case~B would not be 
expected to apply.  These differences are due to recent improvements in the \hO{} collision rates. 
At the lower densities where Case~B might apply the agreement is good; 
the default model is generally within 10\% of \citet{1995MNRAS.272...41S}.

This calculation used our {\tt grid} command to compute the required range of density and temperature
and the {\tt save linelist ratio} command to save predictions into a file.  The input script is

{\footnotesize
\begin{verbatim}
set save prefix "nt"
case B
hden 4 vary
grid 2 14 .5 log
constant temperature 4 vary
grid 2.7 4.4 0.05 log
#
stop zone 1
set dr 0
laser 2
ionization parameter 0
init "honly.ini"
save grid ".grd"
save linelist  ratio ".rat" "nt.lines" last no hash
\end{verbatim} }

The {\tt save linelist ratio} command reads the list of lines in the file {\tt nt.lines} and
saves them into the file {\tt nt.rat}.  The list of lines in {\tt nt.lines} are ordered pairs and
the ratio of intensities of the first to second is reported.
(The ``\#'' lines are comments added to aid the user and are ignored.)
The predictions in the {\tt nt.rat} file were combined with
the grid model parameters saved in the file {\tt nt.grd} to create the plot.
The  {\tt nt.lines} file contained the following set of line ratios:

{\footnotesize
\begin{verbatim}
H  1     4340.49A  
Ca B     4340.49A  
#
H  1     4861.36A 
Ca B     4861.36A 
#
H  1     6562.85A
Ca B     6562.85A
#
H  1     1.87511m 
Ca B     1.87511m 
#
H  1     2.16553m  
Ca B     2.16553m  
\end{verbatim} }
\noindent
Similar tests can be made for other lines of interest.

The number of resolved levels must be increased when higher precision is needed at low densities
(Figure \ref{fig:nt})
or for faint IR / FIR lines (Figure \ref{fig:casebAccuracy}).
Figure 4 of \citet{1964MNRAS.127..165P} can be used as a guide in deciding how many
resolved levels are needed.  Their vertical axis is, in effect, the negative log of the hydrogen density.
The lines indicate the critical density, defined in eqn 3.30 of AGN3, where $l$-changing rates are equal to the radiative decay rate.
The $l$ levels within the indicated $n$ will be well mixed when the density is significantly higher than this critical density.
For instance, at a density of $10^4 \pcc$, the figure shows that the $n=15$ shell 
is at its critical density.  Our default model uses collapsed levels for
$11 \leq n \leq 15$, causing
the residuals in the center panel of Figure \ref{fig:casebAccuracy}.
Our lowest collapsed level, $n=11$, has a critical density of $\sim 10^{5.5} \pcc$.
This is  why the model is so much more accurate, but still not excellent, at the higher density
in the lower panel, $10^7 \pcc$.

The user could  adjust the model to suit requirements at a particular density 
using Case~B predictions as a guide.
We recommend creating a one-zone Case~B simulation with a density and temperature
set to the conditions under study.
Then, adjust the size of the model to achieve the desired accuracy
by comparing the lines of interest with the \citet{1995MNRAS.272...41S} Case~B predictions
that are included in the output.

Making the model large does come at some cost.  Tests show that the test suite
example {\tt pn\_paris}, one of the original Meudon meeting test cases \citep{1986mone.work..363P}, 
takes $\sim 27$s using the default \hone{} model on a modern Xeon, while the large model
in Figure \ref{fig:casebAccuracy} takes 44m 25s.

\subsubsection{Convergence of lines onto radiative recombination continua}

\begin{figure}
\centering
\includegraphics[width=\linewidth]{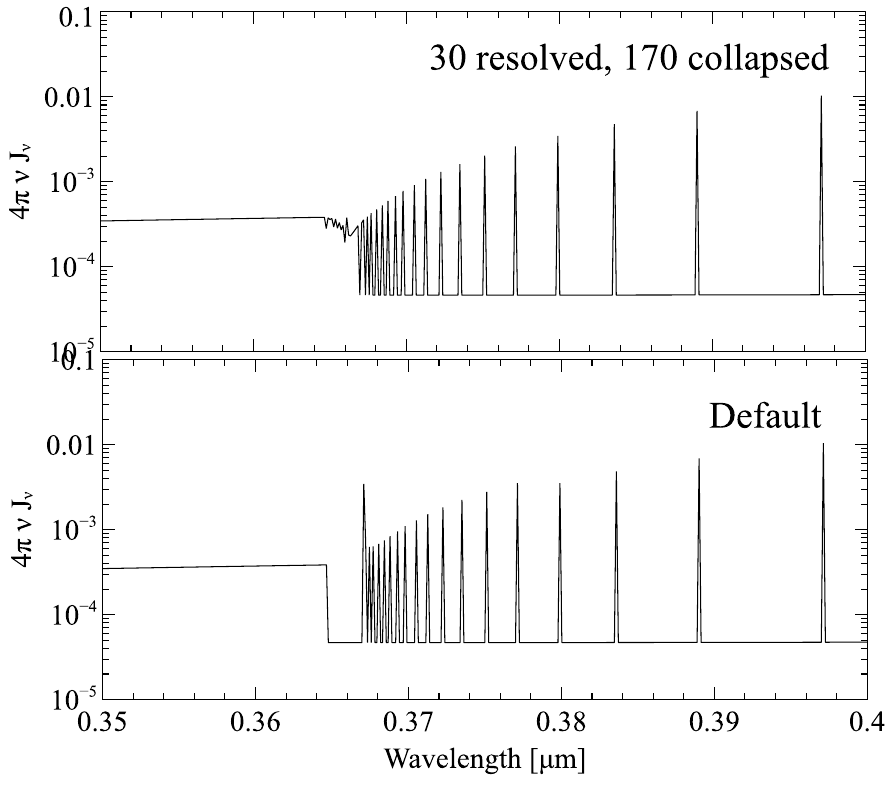}
\caption[l-changing lines]{
  \label{fig:BalmerJump}
This compares emission around the Balmer jump for the Case~B models
used in Figure \ref{fig:casebAccuracy}.
The continuum resolution is increased by a factor of ten above our default
with the command
{\tt set continuum resolution 0.1}.
The upper and lower panels compare the emission predicted by the large and default models.
The large model reproduces the correct smooth merging of the lines and continuum,
while the ``gap'' introduced by the finite size of our default \hone{} model is
obvious in the lower panel.
}
\end{figure}

As mentioned above, the finite
size of the \hone{} model is one source of
 deviations from Case~B line predictions in Figure \ref{fig:casebAccuracy}.
Any finite model will have a ``gap'' between the highest
level and the continuum above.
This gap is also present in the predicted spectrum, as shown in Figure \ref{fig:BalmerJump}.
This shows the converging high-$n$ Balmer lines and the Balmer jump
corresponding to radiative recombination captures to $n=2$.
The upper panel shows the large model used in Figure  \ref{fig:casebAccuracy} while the
lower is the default model.
There is no ``gap'' in the large model, or in nature, but rather the Balmer lines merge onto the Balmer jump.
This is correct and due to the fact that the oscillator strength is continuous
between high-$n$ Balmer lines and the Balmer continuum
\citep{1998MNRAS.297.1073H}.
The ``gap'' in the default model is obvious at this high resolution.  We note that
\citet{2016PASP..128k4001S} presents similar figures.

\begin{figure}
\centering
\includegraphics[width=\linewidth]{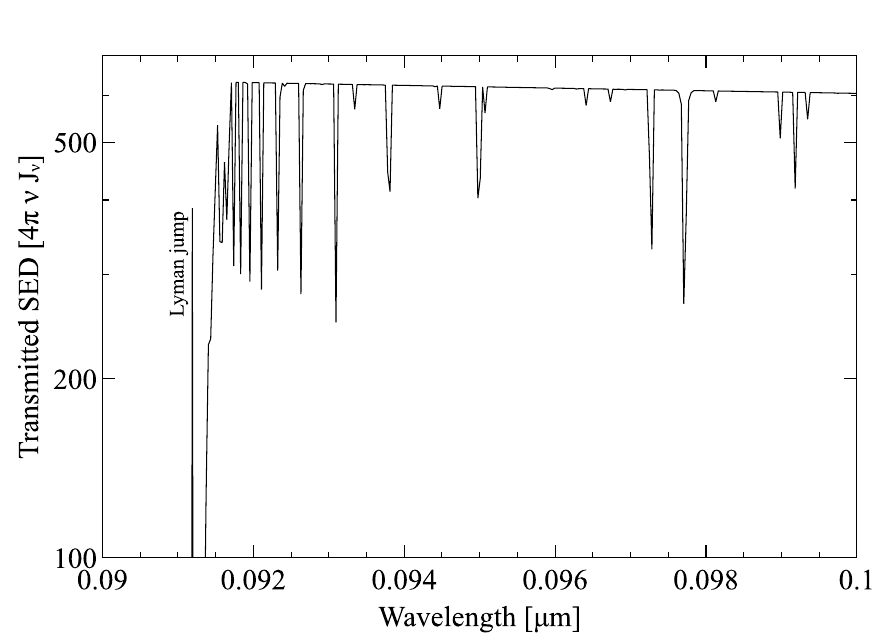}
\caption[l-changing lines]{
  \label{fig:LymanLimit}
The transmitted continuum in the region around the Lyman jump.
This is for our default model which includes``extra'' Lyman lines extending to $n=100$.
The  high-$n$ Lyman lines merge into the Lyman continuum.
The vertical line indicates the wavelength of the Lyman jump.
The continuum resolution has been increased by a factor of ten above our default.
}
\end{figure}

Lyman absorption lines do not suffer from the complexities of \hi{} emission lines since
absorption lines depend on the population of the lower level, the ground state in this case.
It is then simple to include an arbitrary number of ``extra'' Lyman lines, lying above the explicit model,
so that ``gaps'' do not appear.
An example is shown in Figure \ref{fig:LymanLimit}, a high-resolution blow-up of the spectral
region around the Lyman jump.
The calculation used our default \hone{} model.
It is a solar abundance cloud illuminated by our generic AGN SED,
with an ionization parameter of $\log U = -2$,  a column density of
$N(\hone)=10^{18}~\pscm$, and does not assume Case~B.
The smooth blending into the Lyman jump is the correct behavior.
All iso-sequence models are topped off with these extra Lyman lines.

\subsubsection{The H- and He-like ions in the X-ray}

The entire H-like and He-like series of ions between H I and Zn XXIX
are treated with a common code base and have the same commands
to change their behavior.
\citet{2007ApJ...664..586P} discuss the treatment with an emphasis on 
the changes in the He-like X-ray emission due to UV photoexcitation
of the metastable $2\,^3$S level.
\citet{2016A&A...596A..65M} compare the X-ray spectral predictions of
C13 with the Kaastra (SPEX) and Kallman (XSTAR) codes and find reasonable agreement.

\begin{figure}
\centering
\includegraphics[width=\linewidth]{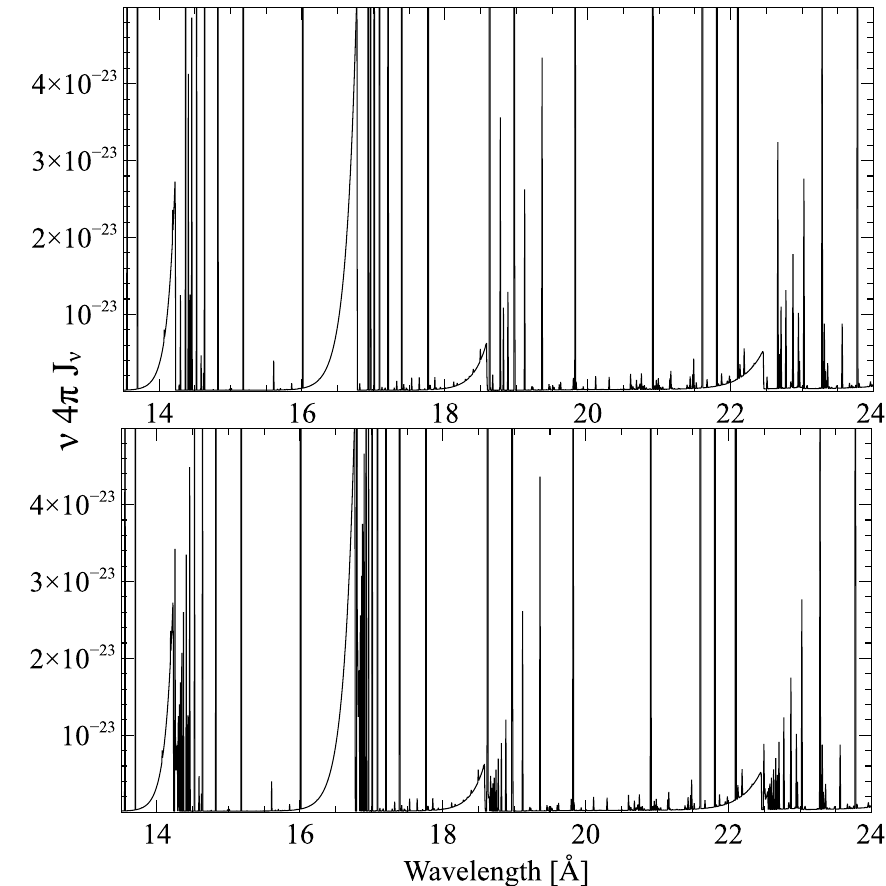}
\caption[l-changing lines]{
  \label{fig:xray}
The emitted spectrum of an optically thin photoionized cloud with solar abundances.
The upper panel uses our default models while the lower panel is an
enlarged model.
The continuum resolution is increased by a factor of ten above our default.
}
\end{figure}

The default number of levels for the H-like and He-like iso-electronic sequences
are summarized in Tables \ref{tab:HLevels} and  \ref{tab:HeLevels}.
The issues discussed above carry over into the X-ray.
Figure \ref{fig:xray} shows a small portion of the emission spectrum
of a solar abundance cloud photoionized by our generic AGN continuum.
The ionization parameter was adjusted to $\log U = 10^{0.75}$ to insure
that C, N, and O were present as H-like and He-like ions.

The ``gap'' between the recombination edges and the converging Lyman lines
is evident in the upper panel.
It is larger than in the \hi{} case because we use relatively few levels for
these high-ionization species
(see Tables \ref{tab:HLevels} and \ref{tab:HeLevels}).
The number of collapsed levels was increased to predict the spectrum in the lower panel.
This larger model was computed with the input stream

{\footnotesize
\begin{verbatim}
set save prefix "large"
set continuum resolution 0.1
c only include H, C, N, and O, with an increased 
c number of collapsed levels
database H-like levels collapsed 20
database He-like levels collapsed 20
table agn
ionization parameter 0.75
hden 0
stop zone 1
set dr 0
iterate 
print last iteration
save emitted continuum ".con" last units Angstroms
\end{verbatim}
}
\noindent
The extra levels produce enough lines to fill in the ``gap'' at this resolution.
Regions of significantly increased emission produced by the additional levels 
are also evident in the larger model.

\begin{table}[t!]
\centering
\small
\caption{Default number of levels for the H-like iso-electronic sequence
}
\label{tab:HLevels} 
\tablecols{4}
\setlength\tabnotewidth{\linewidth}
\setlength\tabcolsep{2\tabcolsep}
\begin{tabular}{cccc}
\toprule
Element & $n$(res) & $nls$(res) & $n$(coll)
\\ \midrule
H & 10 & 55 & 15 \\
He &10 & 55 & 15 \\ 
Li & 5 & 15 & 2  \\
Be & 5 & 15 & 2  \\
B &  5 & 15 & 2  \\
C &  5 & 15 & 5  \\
N &  5 & 15 & 5  \\
O &  5 & 15 & 5  \\
F &  5 & 15 & 2  \\
Ne & 5 & 15 & 5  \\
Na & 5 & 15 & 2  \\
Mg & 5 & 15 & 5  \\
Al  &  5 & 15 & 2  \\
Si & 5 & 15 & 5 \\
P & 5 & 15 & 2 \\
S & 5 & 15 & 5 \\
Cl & 5 & 15 & 2 \\
Ar & 5 & 15 & 2 \\
K & 5 & 15 & 2 \\
Ca & 5 & 15 & 2 \\
Sc & 5 & 15 & 2 \\
Ti & 5 & 15 & 2 \\
V & 5 & 15 & 2 \\
Cr & 5 & 15 & 2 \\
Mn & 5 & 15 & 2 \\
Fe & 5 & 15 & 5 \\
Co & 5 & 15 & 2 \\
Ni & 5 & 15 & 2 \\
Cu & 5 & 15 & 2 \\
Zn & 5 & 15 & 5 \\
\bottomrule \\
\end{tabular}
\end{table}

\begin{table}[t!]
\centering
\small
\caption{Default number of levels for He-like iso-electronic sequence
}
\label{tab:HeLevels} 
\tablecols{5}
\setlength\tabnotewidth{\linewidth}
\setlength\tabcolsep{2\tabcolsep}
\begin{tabular}{ccccc}
\toprule
Element & $n$(res) & $nls$(res) & $n$(coll)
\\ \midrule
He & 6 & 43 & 20 \\
Li & 3 & 13 & 2 \\
Be & 3 & 13 & 2 \\
B & 3 & 13 & 2 \\
C & 5 & 31 & 5 \\
N & 5 & 31 & 5 \\
O & 5 & 31 & 5 \\
F & 3 & 13 & 2 \\
Ne & 5 & 31 & 5 \\
Na & 3 & 13 & 2 \\
Mg & 5 & 31 & 5 \\
Al & 3 & 13 & 2 \\
Si & 5 & 31 & 5 \\
P & 3 & 13 & 2 \\
S & 5 & 31 & 5 \\
Cl & 3 & 13 & 2 \\
Ar & 3 & 13 & 2 \\
K & 3 & 13 & 2 \\
Ca & 3 & 13 & 2 \\
Sc & 3 & 13 & 2 \\
Ti & 3 & 13 & 2 \\
V & 3 & 13 & 2 \\
Cr & 3 & 13 & 2 \\
Mn & 3 & 13 & 2 \\
Fe & 5 & 31 & 5 \\
Co & 3 & 13 & 2 \\
Ni & 3 & 13 & 2 \\
Cu & 3 & 13 & 2 \\
Zn & 5 & 31 & 5 \\ 
\bottomrule \\
\end{tabular}
\end{table}

\subsection{A modified two-level approximation for other ions}
\label{sec:TwoLevelApproximation}

\subsubsection{The two-level approximation}

Textbooks on the interstellar medium  (ISM), e.g.\@
\citet{Spitzer1978},
\citet{Tielens2005},
\citet{AGN3},
and
\citet{Draine.B11Physics-of-the-Interstellar-and-Intergalactic-Medium},
write the ionization balance of an ion as the equivalent two-level system:
\begin{equation}
\frac{n(i+1)}{n(i)} =
\frac{\Gamma(i)}{\alpha(i+1) n_e }
\label{eq:TwoLevel}
\end{equation}
where $n(i+1)$ and $n(i)$ are the densities of two adjacent ionization stages, 
$\alpha(i+1)$ is the total recombination rate coefficient of the ion
(cm$^3$ s$^{-1}$) and 
$\Gamma(i)$ is the ionization  rate (s$^{-1}$).
In photoionization equilibrium 
\begin{equation}
\Gamma(i) = \int \phi_{\nu} \sigma_{\nu}\ d\nu \, ,
\end{equation}
where $\phi_{\nu}$ is the flux of ionizing photons [photons s$^{-1}$ cm$^{-2}$ Hz$^{-1}$],
$\sigma_{\nu}$ is the photoionization cross section [cm$^{-2}$] and the integral
is over ionizing energies.  
On the other hand, in collisional ionization equilibrium 
\begin{equation}
  \Gamma(i)=q(i) n_e\,,
  \label{eq:collion}
\end{equation}
where
$q(i)$ is the collisional ionization rate coefficient.

In its simplest form, the two-level approximation
assumes that recombinations to all excited states will eventually decay to the ground state, 
and that all ionizations occur out of the ground state.
Only the  ionization rate from the ground state and
 the sum of recombination coefficients
to all excited states need be considered, a great savings in data needs.

This two-level approach extends over to the chemistry.  Most codes
use databases similar to the UMIST Database for Astrochemistry
\citep{2013A&A...550A..36M}.
Reactions between complex molecules are treated as a single channel without 
detailed treatment of internal structure.
In this approximation,
rate coefficients do not have a strong density dependence and 
do not depend on the internal level populations of the molecule.
Most of the chemical data needed to implement a more complete model simply do not currently exist.

In some fields, the two-level approximation is called the ``coronal'' approximation when
collisional ionization is dominant.
The solar corona has low density and is collisionally ionized.  The low densities
insure that most population is in the ground state and that recombinations to excited states
decay to ground.
Emission from the
solar photosphere is too soft to affect the ionization.
\Cloudy{} has long included a {\tt coronal} command which sets a gas kinetic
temperature, informs the code that it is acceptable for no incident radiation field to be specified,
and calculates the ionization and emission including thermal collisions
and any light or cosmic rays that may also be specified.

\subsubsection{The independent ionization / emission approximation}

Together with the two-level approximation,
we can further assume that emission from 
low-lying levels are not affected by the ionization / recombination process, 
so that they can be treated as separate problems.
As examples,
the  C IV $\lambda$1549, Mg II $\lambda$2978, and [O III] $\lambda\lambda$ 5007, 4959
multiplets are produced by the lowest excited levels of their ions in Figure \ref{fig:EnergyLevels}. 
These levels are much closer to ground than to the continuum,
so they should be most directly coupled to the ground state.
The fact that, at low particle and photon densities, nearly all of the population of a species is in the ground
state, further justifies this assumption.  

This ``independent ionization / emission'' approximation (IIEA)  is also suggested 
from consideration of the relevant timescales.  
Consider the simple model of the Orion Nebula described by \citet{Ferland16.kappa}.
The recombination time of a typical ion is $\sim 7$~yr at a density of $n\sim 10^{4}$~cm$^{-3}$.  
Electrons tend to be captured into
highly excited states, which have lifetimes of about
10$^{-5}$~s to 10$^{-8}$ s, so the electron quickly falls
down to the ground state. The electron remains in
the ground state for about five hours before another
ionizing photon is absorbed and the process starts
again.

Line-emission timescales are much faster, with collisional excitation timescales of $\sim 10^5$~s 
at a density of $10^4 \pcc$
and photon emission occurring within $\tau \sim 10^{-7}$~s for a typical permitted transition.  
Collisional / emission processes within the low-lying levels occur on timescales that are
$\tau \geq 4$~dex faster than ionization - recombination.
As a result, most codes first solve for the ionization distribution of an element, 
then for the line emission from each ion.
They are treated as separate problems.

This is equivalent to the use of photo-emission coefficients (PEC), 
introduced by \citet{2006PPCF...48..263S} and commonly used in fusion plasmas. 
The line emissivity for a transition between excited levels $i$ and $j$ can
then be expressed as 
$\text{PEC}_{\sigma,i\to j}^{\text{(exc)}}n_\sigma$, where the excited levels $i$
and $j$ are  assumed to be in equilibrium with the ground or metastable state $\sigma$. 
Here, $\text{PEC}_{\sigma,i\to j}^{\text{(exc)}}=A_{i\to j}\mathcal{F}_{i\sigma}^{\text{(exc)}}$
is the excitation photo-emission coefficient for the $i\to j$ transition,
where $\mathcal{F}_{i\sigma}^{\text{(exc)}}$ accounts for the collisional-radiative 
excitation to level $i$ from $\sigma$.

Systems with an especially complex structure, such as Fe~II,
are a major exception to the discussion so far. 
Fe~II has levels extending, nearly uniformly, between the ground state and the continuum,
as shown in the right of Figure \ref{fig:EnergyLevels}.  
The atomic physics of Fe~II is especially complex due to the fact that it has a half-filled
$\it d$ shell, combined with the near energy degeneracy of the $3d$ and $4s$ electrons.
Unfortunately Fe~II emission is strong in a number of astrophysically important 
classes of objects, including quasars and shocked regions.
This is a worst case, with our treatment discussed by \citet{Verner1999}.

The remainder of this section discusses our implementation of a modified two-level
approximation for many-electron systems.  The following section discusses our treatment
of the bound levels and their emission.

\subsubsection{Ionization / recombination rates}

Our sources for ionization and recombination data for many-electron systems
are summarized in C13. 
Ground and inner-shell photoionization cross sections are given 
by \citet{1996ApJ...465..487V},  and summed recombination rate coefficients
are computed as in \citet{BadnellEtAl03,Badnell06}
 and  listed on Badnell's web site\footnote{see http://amdpp.phys.strath.ac.uk/tamoc/RR/, 
http://amdpp.phys.strath.ac.uk/tamoc/DR/}.

We have long used collisional ionization rate coefficients presented by
\citet{Voronov1997}.
Two recent studies,  \citet{Dere07.CollIon} and \citet{KwonSavin14.CollIon},
have presented new rate coefficients for collisional ionization of some ions.
These studies are in very good agreement with \citet{Voronov1997}
for temperatures around those of the peak abundance of the ion.  
Unfortunately, the fitting equations used by   \citet{Dere07.CollIon} and \citet{KwonSavin14.CollIon}
only return positive values for temperatures around the peak abundance of the ion in
collisional ionization equilibrium.  The 
\citet{Voronov1997} fits are well behaved over the full temperature range we cover,
2.7~K to 10$^{10}$~K.

As described by \citet{Lykins.M12Radiative-cooling-in-collisionally-and-photo},
we improved the \citet{Voronov1997} fits by rescaling by the ratio of
the new to Voronov values at temperatures near the peak abundance
of the ion.  This correction was usually small, well less than 20\%, 
so has only modest changes in results.

\subsubsection{Collisional suppression of dielectronic recombination (DR)}

Dielectronic recombination (DR), a process where a free electron is 
captured by exciting a bound electron, forming an autoionizing state 
that can decay into bound levels, is the dominant recombination process for most many-electron ions.  
We mainly use data from Badnell\footnote{http://amdpp.phys.strath.ac.uk/tamoc/DR}, which is also the largest collection available for this process.

Energetically, the DR process occurs via levels with energies within ~kT of the ionization limit, 
so, as shown in the center of Fig \ref{fig:EnergyLevels}, 
it will mainly populate levels that are close to the ionization limit. 
Of the ions shown in the Figure, only \oiii{} has low-lying autoionizing levels,
the levels above the ionization limit.
For many ions, experimental data for such low-lying autoionizing levels are either completely missing
or incomplete.
The summed DR rates listed on the Badnell web site 
assume that all these populations eventually decay to the ground state,
an approximation that must fail at high densities, as quantified below.

It has long been known that DR is suppressed by collisional ionization 
at moderate to high densities 
\citep{1968ApL.....1...69S, 1969ApJ...157.1007B, 1975ApJ...195..285D}.  
\citet{2013ApJ...768...82N}  extended previous work to estimate the extent of this
suppression for various iso-electronic sequences.
Their figure 7 shows that the ionization of low stages of iron 
can change by nearly 1 dex at densities of $\approx 10^{10}$~cm$^{-3}$ when suppression is included. 
As stressed in that paper, these results are highly approximate, 
with the uncertainty in the suppression being of order the correction itself. 
A full solution of the populations of the excited states must be done to get the right answer
\citep{2006PPCF...48..263S}.
Such corrections do allow the two-level approximation to be applied at higher densities, although with
this uncertainty.

Nikoli{\'c} et al.{} (2017, in preparation) revisited the problem and estimated new suppression factors that
offer an improvement over the 2013 values.  We use those suppression factors in 
this release of \Cloudy{}.

\subsubsection{Limits of the two-level approximation for the ionization}

Over what density range is the two-level approximation valid?
How does it fail outside this range?
Despite the wide application of this approximation, we do not know of a discussion of 
these questions.

The two-level approximation will fail when highly excited states do not decay to the ground,
most often due to collisional ionization at high densities.
Highly-excited levels are long lived due to smaller spontaneous decay rates;
for hydrogenic systems the lifetime scales as $\tau \propto n^5$
\citep{1990JPhB...23..661H}. At the same time, the cross section for collisions
increases for higher levels as $n^4$. 
So, compared to electrons in low levels, those in highly-excited levels decay to lower levels slowly, 
and a large probability for collisional ionization is the result.
Further, at high density and temperature the highly-excited levels of all ions will have significant
populations, related to the well-known divergence of the partition function 
in stellar atmospheres \citep{HubenyMihalas14.StarAtmos}.
Collisional ionization from excited states becomes important, and the use of summed 
recombination coefficients becomes highly approximate, 
so both the two-level approximation and ``independent ionization / emission''
assumptions break down.

\begin{figure}
\centering
\includegraphics[width=\linewidth]{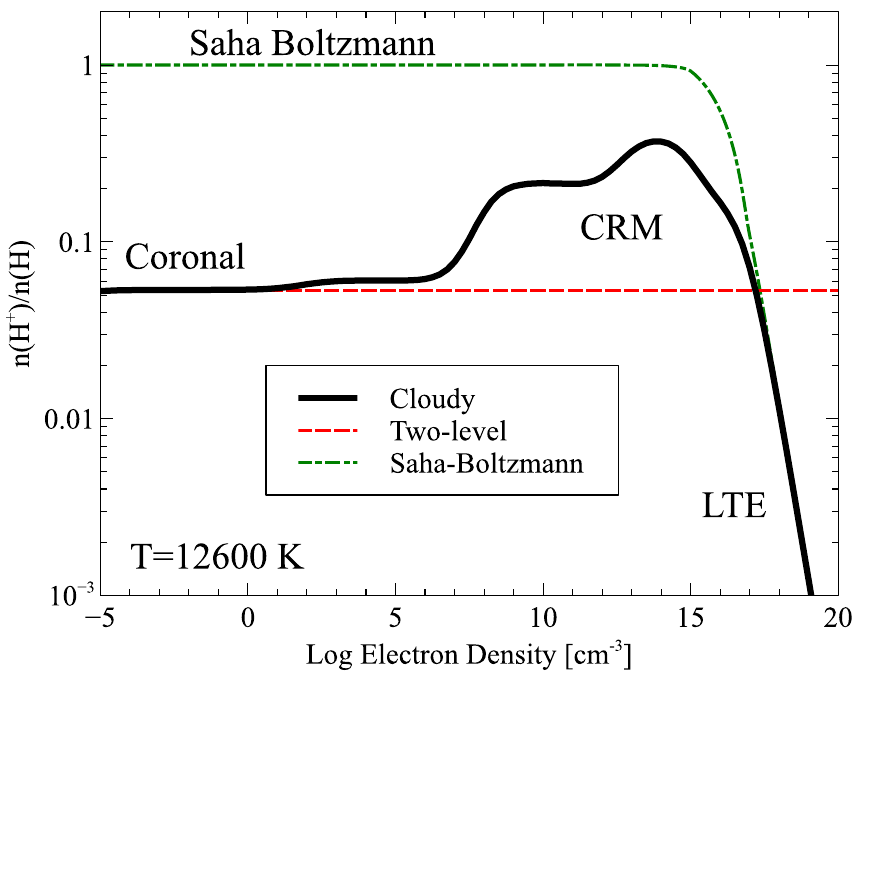}
\caption[Effects of density on hydrogen ionization]{
\label{fig:HIonVsDensity}
Ionization of hydrogen as a function of density.
The low-density 
coronal approximation limit occurs on the left,
while the thermodynamic statistical equilibrium limit applies
at high densities.
The solid black line is the full numerical CRM solution, 
the red dashed line is the ionization predicted by the two-level approximation,
and the dashed-dotted
line is the prediction of the Saha-Boltzmann equation.
}
\end{figure}

We can use our complete CRM solutions for the H- and He-like isoelectronic-sequences 
to show where the  two-level approximation fails.
In Figure \ref{fig:HIonVsDensity} we consider
the ionization of hydrogen in a collisionally-ionized gas at the indicated temperature,
shown over a very wide range of density.  
The Figure, based on one shown in \citet{2014MNRAS.440.3100W},
shows how collisional-radiative effects, mainly involving highly-excited levels,
cause the hydrogen ionization to go from  
coronal equilibrium at low-densities to
Local Thermodynamic Equilibrium (LTE) at high densities. 
Figure 5 of \citet{2006PPCF...48..263S} shows a somewhat similar effect.
The kinetic temperature was chosen so that the gas would be partially ionized at
the lowest densities,
so that a wide range in ionization can result.

In the low-density limit, 
the solution to the coronal equilibrium two-level system given by Equation \ref{eq:TwoLevel}, is
\begin{equation}
\frac{n(i+1)}{n(i)} =
\frac{n_e q(i)}{ n_e \alpha(i+1)} =
\frac{q(i)}{\alpha(i+1) } 
\label{eq:CollTwoLevel}
\end{equation}
where $n(i+1)$ and $n(i)$ are the densities of the ion and atom, 
and $n_e$ is the electron density, which cancels out.
The two-level solution is given as the red-dashed line in Figure \ref{fig:HIonVsDensity}.  
It does not depend on density since the electron density cancels out 
when collisional ionization and recombination are in balance.
It does have an exponential dependence on temperature, as does
the Saha-Boltzmann equation described next, because of the  exponential
Boltzmann factor that enters in the collisional ionization rate coefficient, $q(i)$.

In the high-density limit, as might be found in lower parts of some stellar atmospheres, 
accretion disks near black holes, or certain  laboratory plasmas,
the gas comes into LTE and
the ionization balance is given by the Saha-Boltzmann equation:

\begin{equation}
\frac{n(i+1)}{n(i)} =
\frac{g_e}{n_e}
\frac{(2\pi mkT)^{3/2}}{h^3}
\frac{u(i+1)}{u(i)} 
\exp (-\chi_i/kT)
\label{eq:SahaBoltzmann}
\end{equation}
where $g_e$ is the electron 
statistical weight, 
the $u$'s are partition functions, and $\chi_i$ is the ionization potential of the atom 
\citep{1960ratr.book.....C, HubenyMihalas14.StarAtmos}.  
In this limit, shown as the green dashed-dotted line in Figure \ref{fig:HIonVsDensity}, 
the ionization depends exponentially on the temperature and inversely linearly on the electron density.  At the microphysical level this can be understood as a balance between 
collisional ionization, $n(i) + e \rightarrow n(i+1) +2e$, 
and three-body recombination, 
$n(i+1)+2e \rightarrow n(i) + e$, the inverse process.  

The black line in Figure \ref{fig:HIonVsDensity} shows the \Cloudy{} collisional-radiative solution. 
It  goes between 
the coronal approximation, 
Equation \ref{eq:CollTwoLevel}, valid at $n \lesssim 10^{7}$~cm$^{-3}$, 
and the Saha-Boltzmann limit at high densities, 
Equation \ref{eq:SahaBoltzmann}, valid at $n \gtsim 10^{16}$~cm$^{-3}$.  
The density ranges where the coronal, CRM, and LTE limits apply are indicated by the labels.

Many sources, such as the Orion Nebula or planetary nebulae, are safely in the
limit where the two-level approximation holds, and lower regions of the 
solar atmosphere or accretion disks are dense enough to be
in LTE.  However, many regions, including the emission-line clouds in quasars, or the upper
layers of accretion disks, have intermediate densities, where neither approximation holds.
This is the density range over which full collisional radiative models must be applied.
The density is high enough for collisional ionization from excited levels to be important.
Indeed, the modest increase in ionization at densities of $\sim 10^3 \pcc$ is caused
by collisional ionization from the metastable $2s$ level, which is highly overpopulated
due to its small radiative decay rate.
At high densities, the ionization increases as the density increases and collisional ionization from
more highly-excited levels becomes more important.  Eventually, when collisional 
ionization and three-body recombination comes into balance, the atom comes into LTE
\citep{Seaton64.Hpops}.

For more highly-charged ions, 
the scaling introduced by charge dependencies means their populations also
will behave much like Figure \ref{fig:HIonVsDensity}, but at considerably higher densities.
For hydrogenic systems, the lifetime of a level goes as
$\tau \propto Z^{-4}$ \citep{2006PPCF...48..263S},
and the energy $\chi$, to ionize a level $n$, varies as $Z^2$.
Because of this, we expect the two-level approximation to apply at higher densities
for more highly-charged ions.

\begin{figure}
\centering
\includegraphics[width=\linewidth]{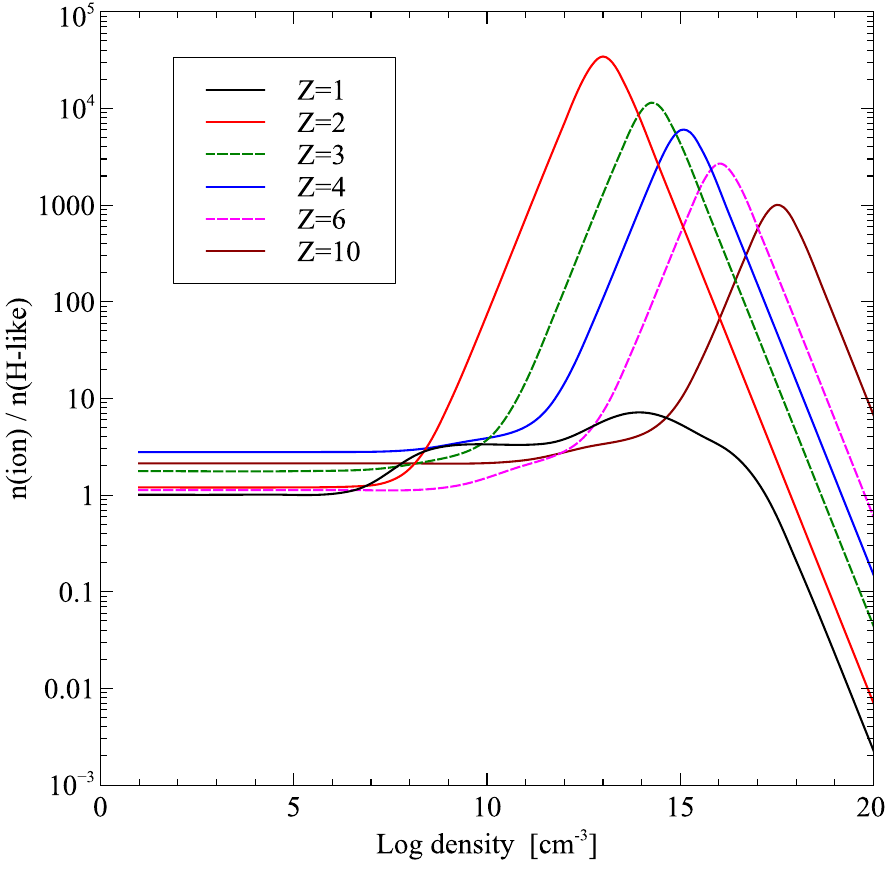}
\caption[Effects of charge on coronal and LTE limits]{
\label{fig:2levelColl}
Density ratio of fully-ionized to single-electron species
as a function of electron density.
The plot is for different elements with nuclear charge indicated.
The effects of increasing charge on the range of validity of the coronal and LTE
limits are dramatic.
}
\end{figure}

We quantify this in Figure~\ref{fig:2levelColl}.
This shows a plot of the ratio of densities of the fully-ionized to single-electron
ions that is very similar to Figure~\ref{fig:HIonVsDensity},
but for different elements, indicated by the nuclear charge $Z$.
Here, the $Z^7$ scaling of density effects can be seen on the shift in peak
densities on going from $Z=2$ to 10.
For each element we have chosen a temperature so that the ratio of ion densities 
at the low-density limit is $\sim$1--5.  

The \Cloudy{} test suite includes a series of example programs which drive \Cloudy{}
in various applications.
One such test, {\tt collion}, creates tables giving the ionization balance
in the low-density collisional-ionization limit.
These are formatted to be very similar to the tables in Carole Jordan's
classic paper \citep{1969MNRAS.142..501J}.
An example, for helium, is given as Table \ref{tab:HeIoniz}.
The first column gives the log of the gas kinetic temperature, while the
remaining columns give the logs of the fractional abundances of
He$^0$, He$^+$, and He$^{2+}$. 
Only portions of the full temperature range are shown.
Such tables, covering the lightest 30 elements and our full temperature range,
are present in the {\tt cloudy/tsuite/programs/collion} folder.
These tables were consulted to choose the kinetic temperature that resulted
in the appropriate low-density limit ionization.

\begin{table}
\centering
\caption{Collisional equilibrium ionization fractions}
\label{tab:HeIoniz} 
\small
\renewcommand\arraystretch{0.65}
\begin{tabular}{ c c c c }
\hline
Log Te & He & He+1 & He+2 \\
$[\rm{K}]$ \\
\hline
3.00 & 0.00 & -- & -- \\ 
3.10 & 0.00 & -- & -- \\ 
\vdots & \vdots & \vdots & \vdots \\
3.90 & 0.00 & -- & -- \\ 
4.00 & 0.00 & -- & -- \\ 
4.10 & -0.00 & -6.45 & -- \\ 
4.20 & -0.00 & -3.93 & -- \\ 
4.30 & -0.00 & -2.16 & -- \\ 
4.40 & -0.07 & -0.81 & -8.73 \\ 
4.50 & -0.55 & -0.15 & -5.71 \\ 
4.60 & -1.35 & -0.02 & -3.70 \\ 
4.70 & -2.07 & -0.01 & -2.16 \\ 
4.80 & -2.67 & -0.05 & -0.97 \\ 
4.90 & -3.33 & -0.34 & -0.26 \\ 
5.00 & -4.15 & -0.95 & -0.05 \\ 
\vdots & \vdots & \vdots & \vdots \\
7.00 & -- & -5.89 & -0.00 \\ 
7.10 & -- & -6.00 & -0.00 \\ 
7.20 & -- & -6.11 & -0.00 \\ 
\vdots & \vdots & \vdots & \vdots \\
8.70 & -- & -7.78 & 0.00 \\ 
8.80 & -- & -7.90 & 0.00 \\ 
8.90 & -- & -8.02 & 0.00 \\ 
9.00 & -- & -8.14 & 0.00 \\ 
 \hline
\end{tabular}
\end{table}

A typical input script, for $Z=10$, is

{\footnotesize
\begin{verbatim}
set save prefix "z10"
init "honly.ini"
element neon on
database H-like levels collapsed 50
database He-like levels collapsed 50
database H-like continuum lowering off
database He-like continuum lowering off
hden -5
element neon  abundance 5 
coronal, temperature 6.8
eden 1 vary
grid 1 20 .25
stop zone 1
set dr 0 
save element neon ".ion" last no hash
save grid ".grd" last no hash
\end{verbatim} }

\noindent
This script first ``turns off'' all elements except hydrogen, to save time, 
then turns neon, $Z=10$, back on.
Next, the number of collapsed levels is increased to improve the precision of the solution.
The {\tt database H-like neon} and similar commands give the iso-electronic sequence and element
to change the treatment of particular species.
Neon is given a very large abundance, 10$^5$ times that of hydrogen,
to prevent H - Ne charge exchange from affecting
the ionization balance.
This model is not in charge equilibrium since the electron density is set with the {\tt eden} command.
This is used to show one way of driving \Cloudy{}.
The {\tt grid} command was used to vary the electron density over a wide range.
We set the temperature to $\log T = 6.8$  with the {\tt coronal} command, to produce
the desired ionization at low densities.
Table \ref{tab:CoronalLTEdensity}, which summarizes results, lists the chosen temperature
for each charge.
The model is in collisional ionization equilibrium so the temperature is, in the simplest case,
the only important parameter.
The two {\tt save} commands produce files that were later used to create the Figures.
Continuum lowering  is
 treated  following \citet{Bautista00} and is
disabled with the command
{\tt database H-like continuum lowering off}.

\begin{table}[t!]
\centering
\caption{Densities for coronal and LTE limits}
\label{tab:CoronalLTEdensity} 
\small
\tablecols{4}
\setlength\tabnotewidth{\linewidth}
\setlength\tabcolsep{2\tabcolsep}
\begin{tabular}{@{} cc@{\hspace*{1em}} cc @{\quad}}
\toprule
Charge
&
$T_{kinetic}$ 
& 
$n(\mathrm{Coronal})$
& 
$n(\mathrm{LTE})$ \\
&
[K] &
[$\pcc$] &
[$\pcc$]
\\ \midrule
1 & 1.2\e{4} & $\lesssim10^7 $    & $\gtsim 10^{16}$ \\
2 & 7.9\e{4} & $\lesssim10^8 $    & $\gtsim 10^{14}$ \\
3 & 2.2\e{5} & $\lesssim10^{10} $ & $\gtsim 10^{15}$ \\
4 & 5.2\e{5} & $\lesssim10^{11} $ & $\gtsim 10^{16}$ \\
6 & 1.3\e{6} & $\lesssim10^{12} $ & $\gtsim 10^{17}$ \\
10 & 6.3\e{6} & $\lesssim10^{14}$ & $\gtsim 10^{18}$ \\
\bottomrule \\
\end{tabular}
\end{table}

The shift of the CRM peak to higher densities for higher $Z$ is dramatic, 
and is a result
of the changing lifetimes, level energies, and collisional rate coefficients, as described above.
Table~\ref{tab:CoronalLTEdensity}  indicates  approximate density limits
over which the coronal and LTE limits apply.  
We choose the lower limit where the ratio was higher by
$\sim$30\% than the value at the lowest density.
The LTE limit was chosen as the density when the solution goes over to the asymptotic
Saha-Boltzmann  power-law.
The actual results do depend on the chosen temperature.  The values in the Table
are highly approximate and only intended to serve as a guide.

The photoionization case behaves quite differently from the collisional case.
In collisionally-ionized plasmas the peak abundance of an ion with
ionization potential IP occurs at temperatures roughly equal to the IP \citep{1969MNRAS.142..501J}.  
At these temperatures, collisions can fully populate all levels from the
ground state (see Figure~\ref{fig:EnergyLevels})  so the
populations of excited levels are far larger than for a photoionized plasma.
Collisional ionization out of these levels produces the CRM peak seen in
Figures \ref{fig:HIonVsDensity} and \ref{fig:2levelColl}.

\begin{figure}
\centering
\includegraphics[width=\linewidth]{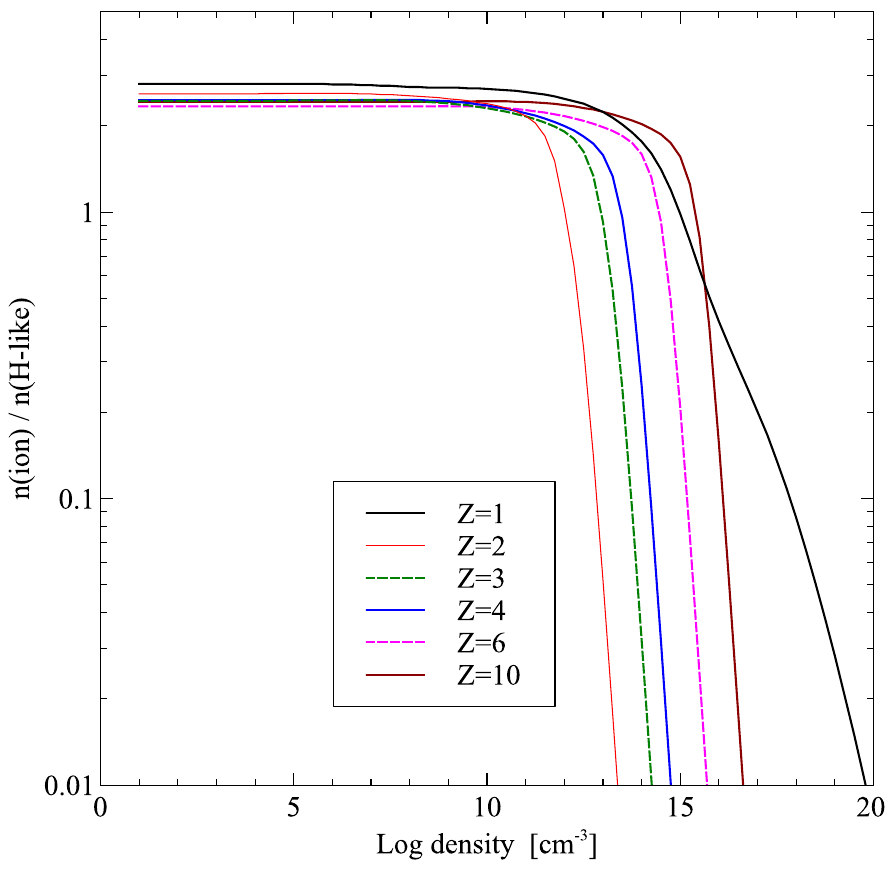}
\caption[Effects of density in photoionization equilibrium]{
\label{fig:2levelPhoto}
As in Figure~\ref{fig:2levelColl}, but for the photoionization case.
The CRM peak, prominent in Figure~\ref{fig:2levelColl} and produced by
 collisional ionization from highly-excited levels, is not present because the
Rydberg levels that enhance the ionization have a smaller population in photoionization equilibrium
}
\end{figure}

We computed a series of photoionization models for the same elements,
density range, and ionization ratio, with results
shown in Figure~\ref{fig:2levelPhoto}.
The gas is illuminated by a blackbody with a temperature 
$T_{BB} = 2\e{4} \times Z^2$~K.  
The ionization parameter (AGN3, eqn 14.7) was adjusted to establish the desired
ionization  ratio in the low-density limit.
The gas kinetic temperature, which plays a minor role in establishing the ionization,
was set to values typical of the ionization parameter.

The input scripts that produced this figure were similar to

{\footnotesize
\begin{verbatim}
set save prefix "z10"
init "honly.ini"
element hydrogen ionization -9 0
element neon on
element neon abundance -2
hden -5 vary
grid 1 20 .25
constant temperature 2e5 K
blackbody 2e6
ionization parameter 0.3
database H-like neon levels collapsed 50
database He-like neon levels collapsed 50
database H-like continuum lowering off
database He-like continuum lowering off
stop zone 1
set dr 0 
save element neon ".ion" last no hash
save grid ".grd" last no hash
\end{verbatim} }

\noindent
Many of the commands are similar to the collisional ionization case.  
This example shows a second way to vary the electron density over the desired range.
Only H and Ne are included in the model, and Ne has a modest abundance.
The hydrogen ionization fraction is set so that the \hone{} density is $10^{-9}$
smaller than the total hydrogen density.

The CRM peak is not present because excited levels are not overpopulated at these
lower temperatures.  
The decreasing ionization at high densities is produced by three-body recombination
bringing higher levels into LTE.
Table~\ref{tab:2levelPhoto} gives the density limit for the two-level approximation to
be valid.  This is taken as the density where the ionization ratio has fallen by 30\%.
As also happened in Table \ref{tab:CoronalLTEdensity}, hydrogen behaves differently
because electron - atom collision rate coefficients are so much smaller than electron-ion
rate coefficients.

\begin{table}[t!]
\centering
\small
\caption{Densities for photoionization coronal limit}
\label{tab:2levelPhoto} 
\tablecols{2}
\setlength\tabnotewidth{\linewidth}
\setlength\tabcolsep{2\tabcolsep}
\begin{tabular}{@{} cr@{\hspace*{2em}} cr @{\quad\quad}}
\toprule
Charge
&
Coronal \\
&
[$\pcc$]
\\ \midrule
1 &  $\lesssim10^{10}$ &  \\
2 &  $\lesssim10^{10}$ &  \\
3 &  $\lesssim10^{12}$  \\
4 &  $\lesssim10^{14}$  \\
6 &  $\lesssim10^{16}$  \\
10&  $\lesssim10^{16}$  \\
\bottomrule \\
\end{tabular}
\end{table}

Tables \ref{tab:CoronalLTEdensity}  and \ref{tab:2levelPhoto} are meant as an indication
of where our methods of solving for the ionization of many-electron systems will fail.
The ranges of validity 
are highly approximate for several reasons.
The collisional ionization results are sensitive to the gas kinetic temperature while the 
photoionization predictions depend on the SED shape.
Further, we only have compete CRM models, extending to high Rydberg levels, for one
and two-electron systems.
The energy level structure, as shown in Figure~\ref{fig:EnergyLevels}, does not
resemble  the structure of many-electron systems, which tend to have a higher density
of levels with low energies.

We are working to incorporate the CRM methods developed by the laboratory plasma
community to have a better representation
of the highly excited levels.
A number of physics codes have been developed, mainly in support of experimental facilities,
to do just that.
Many are summarized in the NLTE7 and NLTE9 code-comparison workshop
summaries of \citet{ChungNLTE7} and \citet{PironNLTE9}.
Few of these codes are openly available.
The ADAS data collection, part of the ADAS code described by  \citet{2006PPCF...48..263S}, is available
through the OPEN-ADAS\footnote{\url{http://open.adas.ac.uk/}} portal, and the FLYCHK code of
\citet{ChungFLYCHK}
is hosted at NIST\footnote{\url{http://nlte.nist.gov/FLY/}}.

\section{Atoms and ions}

The greatest changes to our treatment of atoms and ions involved the move to large databases, 
as described in  Section \ref{sec:databases},
and the development of our Stout database \citep{LykinsStout15}.
This section describes other advances, mainly involving Gaunt factors.

Gaunt factors, named after John Arthur Gaunt, are corrections 
to classical electron-ion interaction theory to incorporate quantum results
\citep{GarstangGaunt}.  Three
Gaunt factors are  encountered \citep{MenzelPekeris35, AllerAtmos63}.
Following the \citet{MenzelPekeris35} numbering and Chandrasekhar's notation \citep{ChandrasekharStructure57},
$g_I$ corrects bound-bound transitions,
$g_{II}$ describes free-bound processes and is
discussed in Section \ref{sec:GauntFreeBound} below,
while $g_{III}$ adjusts free-free or bremsstrahlung emission 
and is discussed in the next Section.
\citet{BurgessSummers87} summarize numerical methods of calculating 
radiative Gaunt factors for complex ions.

\subsection{Free-free Gaunt factor $g_{III}$}
\label{sec:GauntFreeFree}

Free-free or bremsstrahlung emission is the dominant coolant at high temperatures, 
is usually the main contributor to radio emission in nebulae,
and can produce emission across 
the full electromagnetic spectrum if the gas is hot enough.  
As figure 5.2 of \citet{Rybicki1979} shows,
different ranges of temperatures and photon energies require
different asymptotic expressions for $g_{III}$.
These approximations do not
join continuously with one another.
For this reason,
it is not possible to use such expansions across the full temperature and spectral
range we cover with \Cloudy{}.

\citet{2014MNRAS.444..420V} derived averaged and wavelength-specific free-free Gaunt factors $g_{III}$, 
limited to  non-relativistic energies.
This work was extended by \citet{2015MNRAS.449.2112V} to include parameters 
where relativistic effects were important.  
Results from both studies were presented as tables that can be downloaded, 
and these are now used by \Cloudy{} to predict free-free emission.

\begin{figure*}[t!]
	\centering
	\includegraphics[width=\linewidth]{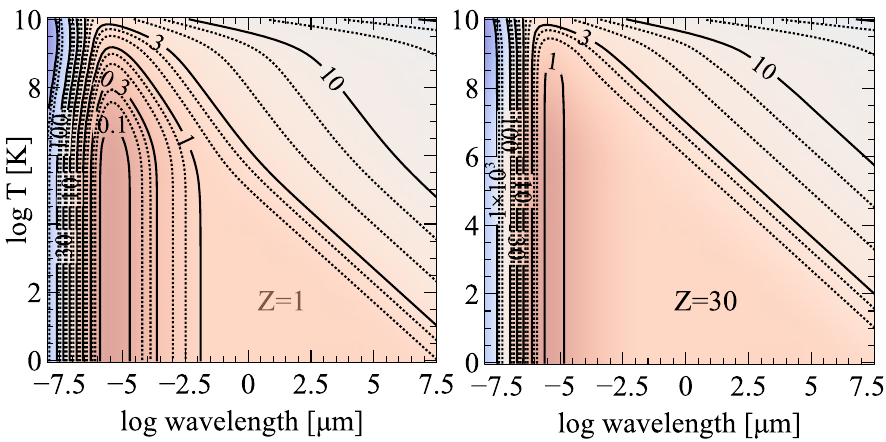}
	\caption
	{
		Values of the $g_{III}$ Gaunt factor over the full wavelength, temperature, and
		nuclear charge range \Cloudy{} considers.
		\label{fig:gaunt}
	}
\end{figure*}

Values of $g_{III}$ covering the full range of kinetic temperature, photon energy, 
and nuclear charge considered by \Cloudy{} are shown in Figure \ref{fig:gaunt}.
As is well known, $g_{III} \sim 1$ across ``nebular'' temperatures and 
UV, optical, and NIR wavelengths.
The closeness to unity shows that emission  is roughly classical over these parameters.
We see that $g_{III}$ initially drops below 1 towards shorter wavelengths. But then,
at yet shorter wavelengths, it starts rising towards infinity when the relativistic effects
become important. We also see that $g_{III} > 1$ at long wavelengths, which is the well-known
non-relativistic result for radio emission.
The values depend on nuclear charge, 
as shown by comparing the left panel, for hydrogen ($Z=1$), and
the right panel, for zinc ($Z=30$). This is because the relativistic effects
break the simple $Z$-scaling that exists in non-relativistic calculations.
As explained in the original papers, we believe that these values are highly accurate.

\subsection{Free-bound Gaunt factors $g_{II}$}
\label{sec:GauntFreeBound}

Rate coefficients for radiative recombination are derived from the photoionization
cross section, using detailed balance, often called  the Milne relation 
(see, for instance, Appendix 2 of AGN3).  
The $g_{II}$ factor allows classical expressions for the photoionization cross section
to be used to obtain accurate recombination rate coefficients.

\citet{Seaton1959a}, hereafter S59, derived hydrogenic radiative recombination rate coefficients over a broad range of temperature using the $g_{II}$ approach. 
\citet{SutherlandDopita93}, hereafter SD93, claimed that this work was numerically flawed 
and offered corrections to Seaton's theory. 
One of us has revisited the theory of hydrogenic radiative recombination twice 
\citep{Ferland1980c, FerlandPeterson1992}, referred to as F80 and F92,
and found only excellent agreement with S59,
so this claim is puzzling.  

We use a finite model to approximate the internal structure of \h0{}, so we need to 
``top off'' the model by accounting for the difference between the 
infinite-level total radiative recombination
and the summed recombination to the modeled levels.
This requires accurate Case B recombination coefficients over a very broad temperature range.
This brings up the SD93 correction to the S59 theory.
SD93 is {\it very} highly cited, and its results are widely used, 
so it is important to understand whether there is a problem in S59.

S59 developed analytical expansions, in the small- and high-temperature limits, of quantities that enter into $g_{II}$.
No analytical expansion was possible at intermediate energies and direct numerical integration was performed instead. 
S59 presented the expansions as equations and the numerical results as tables.
S59 estimated the errors due to the high-temperature expansion to not exceed 
2\% for $T \leq 10^4$~K, and that it increased for higher temperatures.

\begin{figure}
	\centering
	\includegraphics[width=\linewidth]{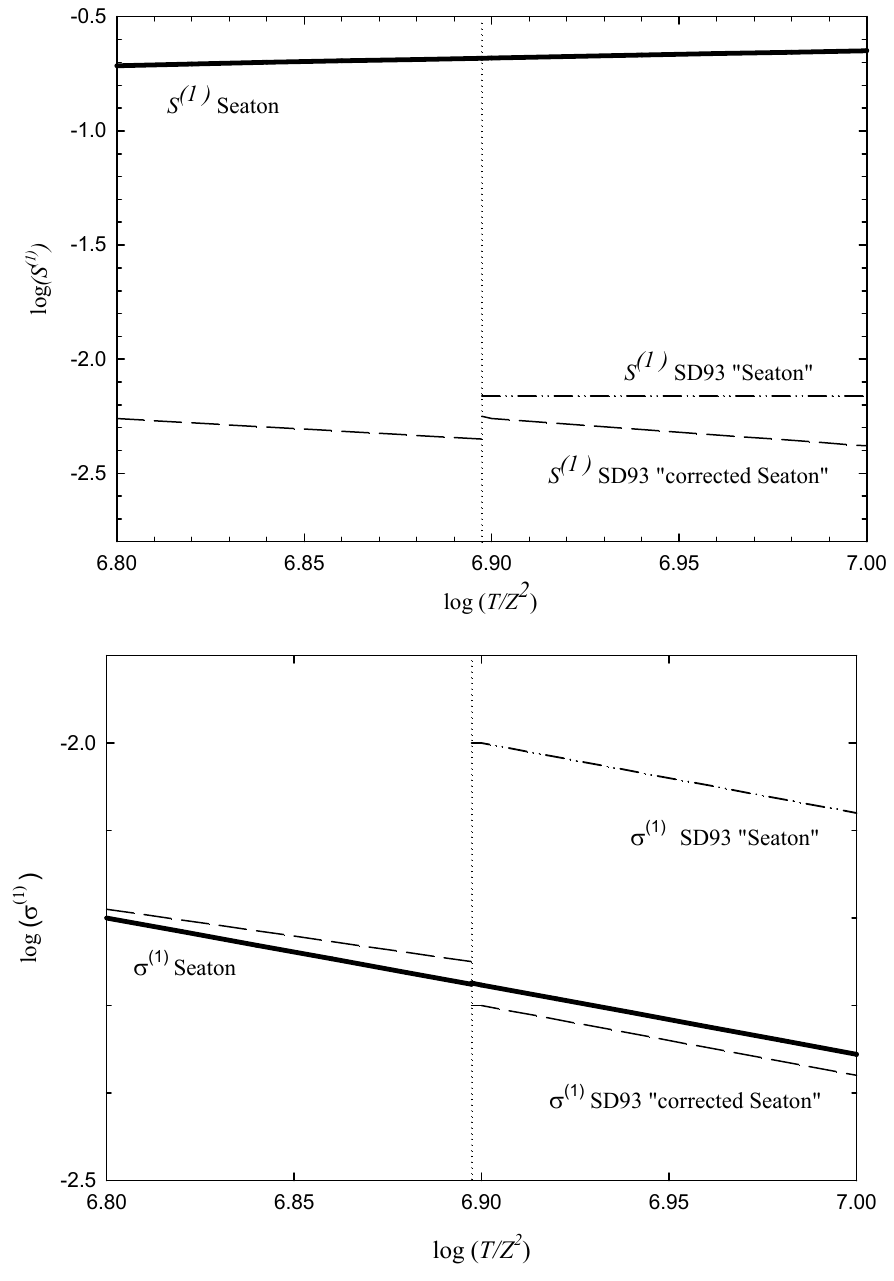}
	\caption
	{
		This shows two of the quantities that enter into the S59 calculation
		of $g_{II}$.  The vertical dotted 
		line marks the point where SD93 claim that Seaton's theory is discontinuous. 
		The dashed-dotted lines give the quantities plotted in the
		lower panel of figure 1 of SD93. 
		The dashed curves give  SD93's ``corrected Seaton'' theory.
		The heavy solid lines give our recalculation of the terms 
		$S$ and $\sigma$  using Seaton's original expressions and tables. 
		\label{fig:SeatonSutherlandDopita}
	}
\end{figure}

\begin{figure}
	\centering
	\includegraphics[width=\linewidth]{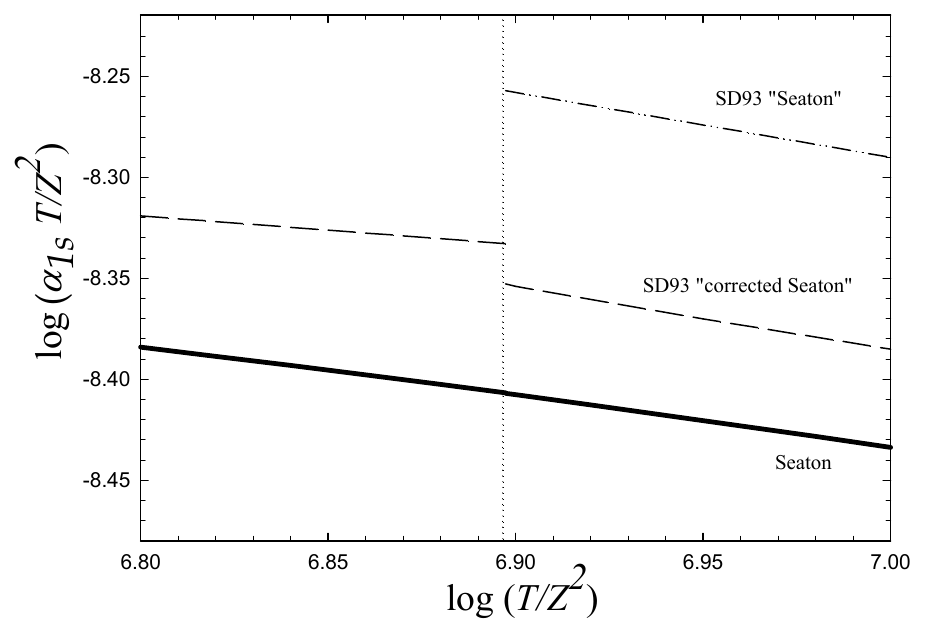}
	\caption
	{
		Hydrogenic recombination rate coefficients are shown vs. reduced
		temperature, $T/Z^2$, for the ground state, $n=1$. 
		The vertical dotted line marks the SD93 
		discontinuity, where the
		theory changes from the numerical to analytical forms. The dash-dotted line gives
		SD93's evaluation of the Seaton rate, 
		taken from the upper panel of their figure 1. The dashed line
		is SD93's ``corrected Seaton'' theory. 
		The heavy solid line gives
		our recalculation of the rates using Seaton's original expressions and tables. 
		All the rates are multiplied by the reduced temperature.
		\label{fig:SeatonSutherlandDopita-total}
	}
\end{figure}

SD93, section 3.4.1 and Figure 1, report that the S59 intermediate-temperature 
numerical, and high-temperature analytical,
results do not smoothly join one another, 
and offer an alternative expression for the high-temperature part. 
We show results from their tables and figure as our Figures  \ref{fig:SeatonSutherlandDopita}
and  \ref{fig:SeatonSutherlandDopita-total}.
Figure  \ref{fig:SeatonSutherlandDopita} shows one of the S59 intermediate results and taken from
their tables.
The line marked ``SD93 Seaton'' is their version of S59 while ``corrected Seaton''
is their altered version.  
F80 closely followed S59, so we have some experience in working with this paper.  
The heavy line is our evaluation of S59.
It is continuous and somewhat displaced from SD93. 

Figure \ref{fig:SeatonSutherlandDopita-total} compares total
recombination rate coefficients over the narrow temperature
range where SD93 report problems.
The discontinuities are obvious.
The heavy line is our evaluation of S59.
It is continuous and displaced from SD93.

\begin{figure}
	\centering
	\includegraphics[width=\linewidth]{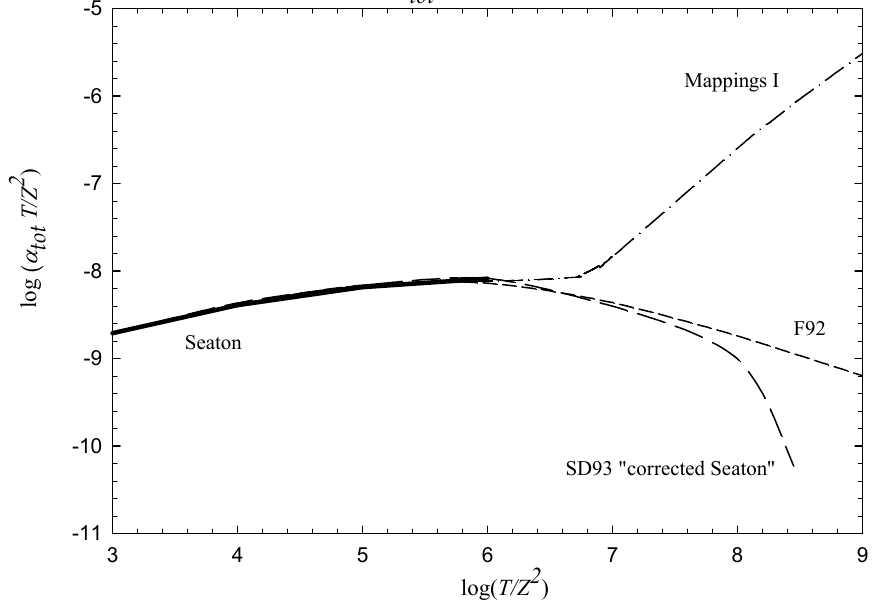}
	\caption
	{
		Comparison of various hydrogenic total Case B recombination rate coefficients. 
		Our reevaluation of Seaton's theory
		is shown as the heavy solid line drawn over the temperature range where he said it was
		applicable. The dot-dashed line gives the rate used in Mappings I, equation (7) of SD93.
		The SD93 ``corrected Seaton'' rate is shown as the long dashed line. 
		The short dashed line gives the F92 rates derived by direct numerical 
		integration of the Milne equation over hydrogenic photoionization cross sections. 
		The F92 curve lies behind the Seaton curve for $T < 10^6$~K, 
		the temperature range he indicated the theory was accurate. 
		All rate terms are multiplied by a factor of $T/Z^2$.
		\label{fig:SeatonSutherlandDopita-total-highT}
	}
\end{figure}

Figure \ref{fig:SeatonSutherlandDopita-total-highT} shows total recombination
rate coefficients over a broad range of temperature.
The heavy line gives
our reevaluation of the Seaton results.
S59 notes that his expressions are not valid above 10$^6$~K
and the heavy line ends at that temperature.
The lines marked ``Mappings I'' and ``corrected Seaton'' are taken from SD93.
The line marked ``Mappings I'' is an extrapolation of the S59
expression as used by \citet{Arnaud1985} and the Mappings I code.
The line marked ``corrected Seaton'' is the expression proposed by SD93.

Today the $g_{II}$ approach is seldom used because large repositories
of photoionization cross-sections, such as the Opacity Project \citep{Seaton1987},
are readily available, so the Milne relation can be applied directly to photoionization cross sections.
F92 took this approach.
The F92 results are believed to be of high accuracy for all temperatures presented,
as was recently confirmed by \citet{2017A&A...599A..10M}.
S59 reported that his results should be
``in error by less than 0.5 per cent for $(T/Z^2) \le 10^5$~K, by 3 per cent for $(T/Z^2) = 10^6$~K
and by 31 per cent for $(T/Z^2 ) = 5\times 10^6$~K.''
F92 confirms these error estimates.

Our models of the H- and He-like isoelectronic sequences require level-resolved 
recombination rate coefficients and photoionization cross sections.
We now use hydrogenic expressions, for H-like, or results from the Opacity Project
as described by \citet{2012MNRAS.425L..28P}, to obtain the needed data.
But we use F92 to provide the correction needed to ``top off'' the hydrogenic systems.
We believe that this is the correct approach and do not use the SD93 theory.

\section{The grain and chemical models}
\label{sec:GrainMolec}

Our treatment of the chemistry and molecular emission is described in C13,
\citet{2005ApJS..161...65A},  \citet{2005ApJ...624..794S}, and \citet{Abel2008}.  Grain physics,
which has a great impact on the chemistry, is described by \citet{2004MNRAS.350.1330V},
while  \citet{2005ApJS..161...65A} outlines how we do ionization / recombination of ions on grain surfaces.

\subsection{\htwo{} formation on grain surfaces}
The treatment of \htwo{} formation via catalysis on grain surfaces due to chemisorption and physisorption, 
the Eley - Rideal mechanism \citep{2012A&A...541A..76L},
has been modified since C13.  Our revised treatment incorporates the rates from 
\citet{Cazaux2002, Cazaux.S04H2-Formation-on-Grain-Surfaces},
including the corrections of \citet{2010ApJ...715..698C}.
 Our implementation of the Cazaux \& Tielens rates  utilizes equations 
 C1, C2, and C3 given in the appendix of \citet{Rollig13}.  
 However, while comparing the latter paper to the Cazaux \& Tielens work, 
 two typesetting errors in equation C2, which describes chemisorption, were found,
 as confirmed by private communications with the authors.  First, the entire right side of 
 equation C2, not just the 2F term, should be raised to the -1 power.  
 The second error in C2 involves the exponential dependence of the \htwo{} formation rate 
 on the desorption energy of chemisorbed hydrogen, which should have a 
 negative sign.  
 
 The new rates were first  used in the chemical modeling of 
 Orion's Veil \citep{2016ApJ...819..136A}, a low \htwo{} abundance environment where the revised rates 
 were expected to have the greatest impact through increased \htwo{} formation \citep{2012A&A...541A..76L}.  
 However, since most of the increased rate  occurs on PAHs 
\citep[polycyclic aromatic hydrocarbons,][]{Rollig13} and since small grains are 
 largely absent in the Veil, the effects on the \htwo{} abundances were found to be less than 10\%.  
 
 \subsection{\htwo{} collisions and \htwo{}, HD cooling}

Our complete \htwo{} model is described by  \citet{2005ApJ...624..794S}.
We include  three datasets for \htwo\ -- \hO\ collisions, as controlled by
options on the command {\tt database H2 collisions}.
The default is to use the rates given by \citet{Wrathmall2007}.
The rates given by \citet{LeBourlot1999} are also available.   
The recent dataset, \citet{2015MNRAS.453..810L}, which includes ortho-para-changing reactive collisions,
is available in this release although  \citet{Wrathmall2007} remains the default.
The \htwo{} cooling is the explicit sum of the difference between the collisional excitation and
deexcitation rates given in Equation \ref{eq:SpeciesCooling} below.

We fall back on simple and approximate models of \htwo{} formation and cooling when
our complete \htwo{} model is not used.
Our model of the dissociation physics is inspired by the \citet{Tielens1985a} \htwo{} model
and is described by \citet{Elwert2006}.
\htwo{} cooling is approximated with the function given by \citet{Glover2008}.
HD cooling is given by \citet{FlowerHD00}.
There are cases where these are significantly different from the approximations used in C13.

\subsection{How does the PAH abundance vary across the \hplus{}, \hone{}, and \htwo{} regions of a nebula?}
\Cloudy{} can consider the adjacent \hplus{}, \hone{}, and \htwo{} regions of a nebula,
the so-called \hii{} region, PDR, and molecular cloud.
This brings in questions concerning how the PAH abundance changes across these regions.
Observations of the Orion Bar originally showed \citep{Sellgren1990} that PAHs emit in a narrow region
that might most closely be associated with the \hone{} region (AGN3 Section 8.5).
The obvious interpretation is that PAHs are destroyed in the H$^+$ region of the nebula.
The  lack of PAH emission at deeper, more molecular, regions of the PDR could be due to 
a lack of PAHs in \htwo{} regions, or simply an illumination effect, where PAHs do not
fluoresce because UV light does not penetrate into the well-shielded high \Av{} regions
where \htwo{} is found.  This is an area of active research

We provide three options to describe how the PAH abundance varies with physical conditions.
These are controlled with the commands 

\begin{verbatim}
set PAH constant
set PAH "H"
set PAH "H,H2"
\end{verbatim}

\noindent
The {\tt constant} option will keep the PAH abundance constant across the   \hplus{}, \hone{}, and \htwo{} regions.
This allows PAHs to exist in the \hplus{} region.
We do this for completeness, even though observations suggest that
PAHs are not present there.
The {\tt "H"} option will scale the PAH abundance as the atomic hydrogen fraction,
$n({\rm PAH}) \propto n({\rm H})/n_{\rm total}$, where $n_{\rm total}$ is the total density
of hydrogen in all forms.  This is the simplest interpretation of the Orion Bar observations.
The {\tt "H,H2"} option will scale the PAH abundance as the fraction of hydrogen that is
either atomic or molecular,
$n({\rm PAH}) \propto [n({\rm H}) + 2n({\rm H}_2))]/n_{\rm total}$.
This is consistent with the Orion Bar observations if UV extinction prevents starlight
fluorescence of PAHs in deeper molecular regions.
By default, we assume the {\tt "H"} case.

\begin{figure}
\centering
\includegraphics[width=\linewidth]{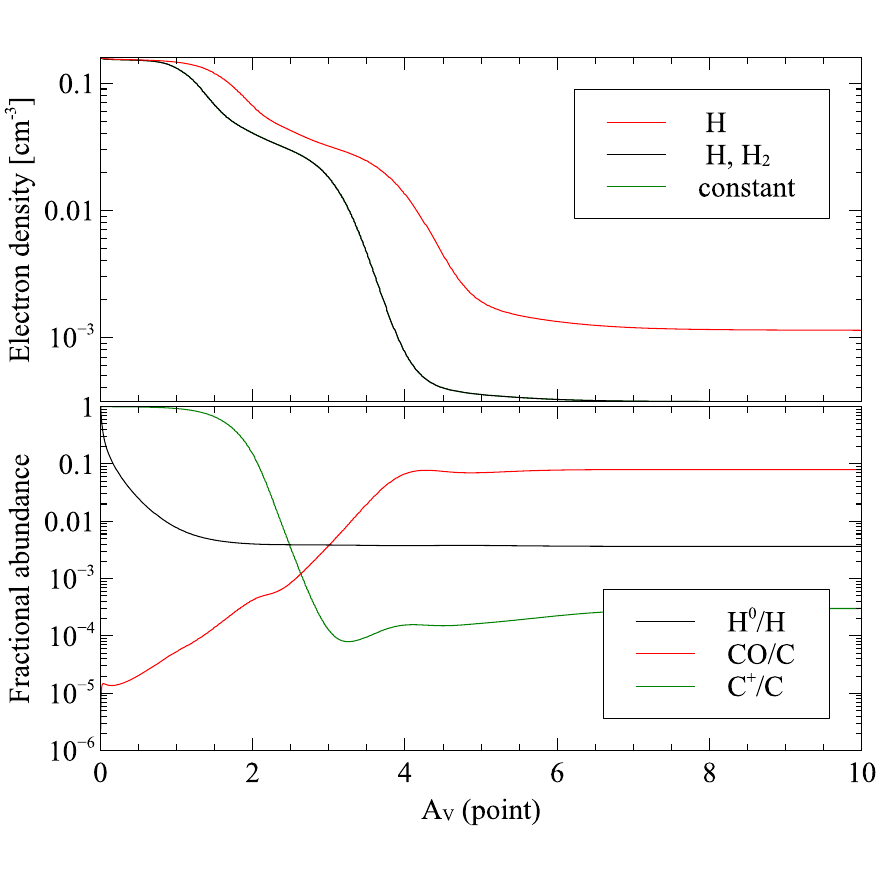}
\caption[Number of lines, default and maximum databases]{
\label{fig:PDR-PAH}
This compares  the effects of our different PAH abundance laws 
in the Leiden F1 PDR model.
The upper panel shows the electron density for the three cases,
see text for details.
The  {\tt constant} and {\tt "H,H2"} cases overlap because there is very little \hplus{} in this model.
The lower panel shows fractional abundances, for the {\tt "H"} case, for some of the important
species in a PDR.
}
\end{figure}

Figure \ref{fig:PDR-PAH} explores the effects that the PAH abundance has on the electron density
in the F1 PDR model  of the Lorentz Center workshop on PDRs \citep{Roellig2007}.
The model is discussed extensively in that paper, so only the relevant details are given here.
We added PAHs with three times our default abundance, to be consistent with current estimates,
and computed the three PAH abundance cases.
The panels show the abundances of some very important species 
as a function of extinction \Av{} in the lower panel.  
Extinction depends on whether the light source is a point source or extended (AGN3 Section 7.6)
and we use the point-source definition here.

The  main regions of a PDR are evident in the lower panel of Figure  \ref{fig:PDR-PAH},
which is  for the {\tt "H"} case.
Hydrogen is atomic at very shallow regions but quickly becomes molecular.  Carbon
transitions from C$^+$ to CO in a narrow range around \Av{}$ \approx 2 - 4$ mag.
\citet{Roellig2007} give further details.
The SED used in this model is only defined over a narrow spectral range, 
see Figure \ref{fig:ISM_background}, and includes no H-ionizing photons,
so the only \hplus{} present is created by cosmic ray secondary ionization.
The next subsection describes some effects this simple SED has our full \hi{} model.

The upper panel shows the predicted electron density.  Cosmic rays are the dominant
source of hydrogen ionization in this cloud while carbon is a dominant electron donor
in some regions.
Our treatment of electron capture on grain surfaces is described by 
\citet{2004MNRAS.350.1330V} and \citet{2005ApJS..161...65A}.
Although we show all three options, there is little \hplus{} in this model, so the
{\tt constant} and {\tt "H,H2"} cases are identical.  The PAH abundance does not
vary across Figure  \ref{fig:PDR-PAH} in that case.
In the {\tt "H"} case, PAHs are present near the illuminated face but have very low abundance
at \Av{}$ > 1$ mag, where H is mostly \htwo.

Table \ref{tab:PDRPAH} shows the logs of some predicted column densities for the {\tt "H,H2"} 
and {\tt "H"} cases.  
The differences, which can be large, are caused by the changed electron density.
These changes suggest a method to determine observationally how the PAH abundance
varies with the hydrogen molecular fraction.

\begin{table}[t!]
\centering
\caption{Effects of PAH abundance laws
}
\label{tab:PDRPAH} 
\normalsize
\tablecols{3}
\setlength\tabnotewidth{\linewidth}
\setlength\tabcolsep{2\tabcolsep}
\begin{tabular}{cccc}
\toprule
Species & {\tt "H"} & {\tt "H,H2"} \\
\midrule
H$^+$ & 15.43 & 14.99 \\
H$_2$ & 21.90 & 21.89 \\
H$_3^+$ & 14.63 & 14.91 \\
C$^+$ & 17.41 & 17.28 \\
CH & 16.02 & 16.31 \\
CH$_4$ & 15.06 & 15.78 \\
C$_3$ & 15.28 & 15.86 \\
OH & 13.84 & 13.68 \\
H$_2$O & 13.94 & 13.73 \\
CO & 16.90 & 16.52 \\
SiO & 14.93 & 14.54 \\ 
\bottomrule \\
\end{tabular}
\end{table}

\subsection{Lyman line pumping in PDR models}
\label{sec:LymanPDR}

\begin{figure}
\centering
\includegraphics[width=\linewidth]{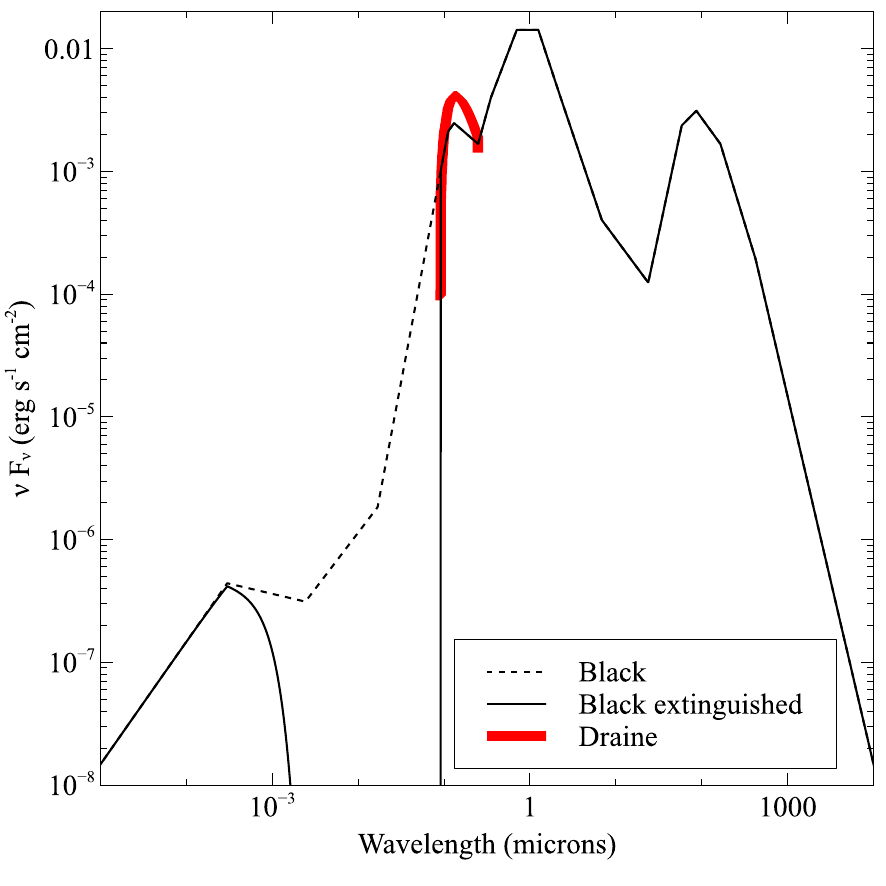}
\caption[ISM radiation field]
{\label{fig:ISM_background}The SED produced by the {\tt table ISM} command is the lighter
line.
The infrared cirrus is the peak at $\lambda \sim  100\, \micron $ and starlight
dominates at shorter wavelengths.
The points just shortward of the Lyman
limit ($0.0912 \micron$) are interpolated---actually it is thought that interstellar
extinction removes most of this continuum.
The dashed line shows the
interpolated SED and the solid line shows the effects of absorption
introduced by adding the {\tt extinguish} command.
The heavy red line is SED produced the {\tt table Draine} continuum
and used in many PDR calculations. 
}
\end{figure}

We participated in the 2004 Leiden PDR meeting where we compared our predictions 
with  codes specifically designed to model PDRs \citep{Roellig2007}.  
The comparison models employed an SED that was only defined over a very narrow range of wavelengths, as shown by the heavy red line in Figure \ref{fig:ISM_background}.
This uses our {\tt table Draine} command to enter the galactic background radiation field given by
equation 23 of \citet{Draine1996}.

Early in the comparisons we noticed that our calculations predicted a thin moderately-ionized layer of warm gas on the illuminated face of the PDR that was not seen by the other codes.  Some investigation revealed that it was due to hydrogen continuum fluorescence through the Lyman lines, followed by photoionization of the metastable $2s$ level.  The process, and our 
method of removing it to allow comparison with conventional PDR codes, is described next.

Our preferred method of generating the local interstellar radiation field is to use the command
{\tt table ISM}.
This uses Figure 2 of \citet{Black1987} to represent the
\emph{unextinguished}
local interstellar radiation field (see Figure \ref{fig:ISM_background}). The continuum
generated by \Cloudy\ is exactly that given by Black, except that the radiation
field between 1 and 4 Ryd is interpolated from the observed or inferred
values.  Actually it is thought that this part of the radiation field is
heavily absorbed by gas in the ISM so that little radiation
exists where the dotted line appears, at least in the galactic plane.  
Such absorption can be introduced
with our {\tt extinguish} command.
This SED does not include the cosmic microwave background so that it can be used at any redshift.
The {\tt CMB} command is used to add this component.

\begin{figure}
\centering
\includegraphics[width=\linewidth]{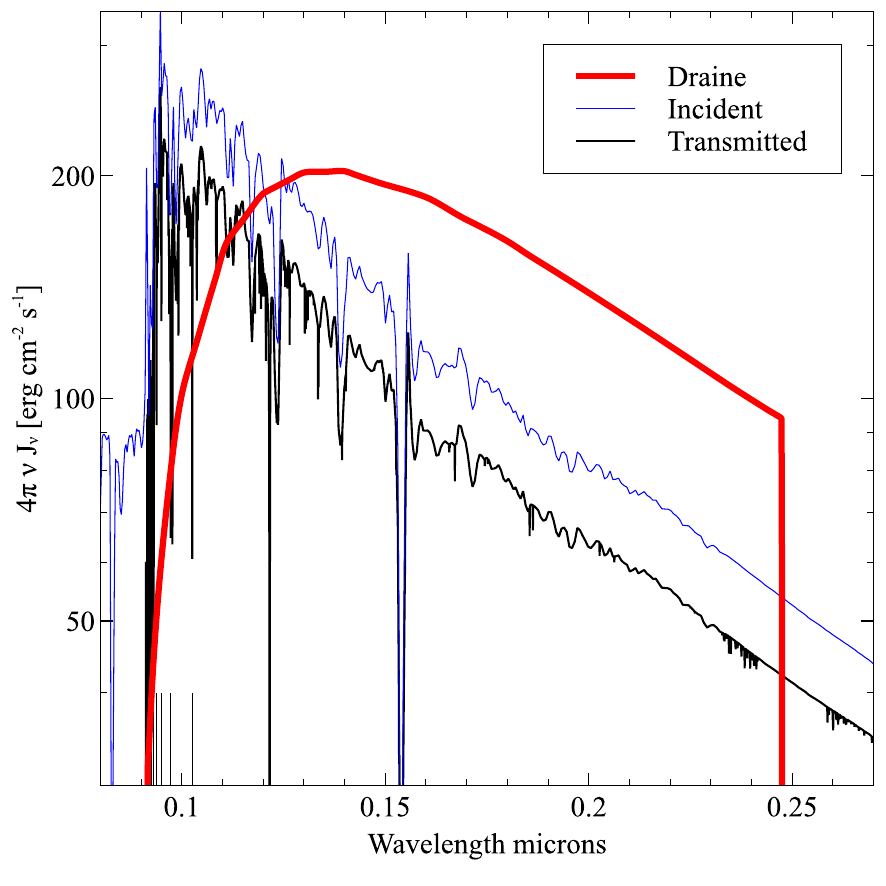}
\caption[SED incident on PDR]
{\label{fig:PDRcontinua}The thick red line shows the limited SED produced by the 
{\tt table Draine} command and used in many PDR calculations.
The SED of the Orion Trapezium cluster is shown as the higher blue line while
the SED transmitted through the \hplus{} region and incident on the PDR is
the lower black line.
The continuum resolution is ten times higher than our default to increase the
contrast in the absorption lines.
The Lyman line optical depth through the \hplus{} region is large and the Lyman lines
are quite optically thick.
The vertical lines at the lower left indicate positions of Lyman lines.
 The incident SEDs were normalized to the same radiation parameter
$G_0$.
}
\end{figure}

Figure \ref{fig:PDRcontinua} shows a closeup of the UV/ FUV portion of the SED with
line styles similar to Figure  \ref{fig:ISM_background}.
Conventional PDR calculations of, for instance, the Orion PDR, scale up the Draine 
isotropic radiation field
to mimic the SED of the illuminating stars, in this case the Trapezium cluster.

Our preferred method of treating a PDR near a region of active star formation is to
include models of the full SED of the central star cluster to model successive \hplus{}, \hone{}, and \htwo{}  layers
\citep{2005ApJS..161...65A}.
The higher blue line in Figure  \ref{fig:PDRcontinua} shows a portion of the SED of the 
Trapezium cluster, as implemented with our new {\tt table SED "Trapezium.sed"} command.
There is a strong Lyman jump and significant Lyman continuum radiation.
The lower black line shows the SED transmitted through the \hplus{} region and entering the PDR.
The stellar continuum is harder than the Draine field, especially at the short wavelengths
which cause electron photoejection from grains.

This SED was produced by creating the following model of the Orion Nebula in the style of \citet{BFM}:

{\footnotesize
\begin{verbatim}
set continuum resolution 0.1
set save prefix "full"
phi(H) 13.0721 range 0.25 1
table SED "Trapezium.sed"
hden 4
abundances Orion
grains PAH 3 function
stop  efrac -2
stop  temperature off
constant pressure
save overview ".ovr"
save continuum units microns last ".con"
\end{verbatim}}

\noindent
This script increases our default continuum resolution by a factor of ten,
to make the Lyman absorption line contrast, as discussed next, more prominent.
Orion grains and our default PAHs, using the {\tt function} keyword, are included.
Note that it was necessary to recompile the grain opacities since the continuum mesh
was changed by increasing the resolution.  This was done with the {\tt compile grains} command.
The incident SED is the Trapezium cluster with the flux of ionizing photons, set with the
{\tt phi(H)} command, deduced by \citet{BFM}.
Their density was used, and the equation of state assumes constant total pressure.
The {\tt stop temperature off} command prevents the code from stopping at the point
where the kinetic temperature falls below 4000 K, one of our default stopping criteria.
Instead, the calculation stops when the electron fraction, $n_{\rm e}/n_{total}$, falls
below $10^{-2}$, a point near the illuminated face of the PDR.

Orion grains, as used here, are fairly grey, apparently due to a deficiency of small particles.
As a result, the UV portion of the transmitted SED is fainter than the incident SED,
but the shape is not otherwise greatly changed. 
Nearly all radiation shortward of the Lyman limit is extinguished.
The increased continuum resolution used in Figure  \ref{fig:PDRcontinua} allows Lyman
absorption lines formed in the \hplus{} layer to stand out.
The vertical bars at the lower left of the Figure mark the positions of the higher Lyman lines,
which are strongly in absorption.
Because of this Lyman line self-shielding, very little light reaches the PDR in the cores of the
Lyman lines.

\begin{figure}
\centering
\includegraphics[width=\linewidth]{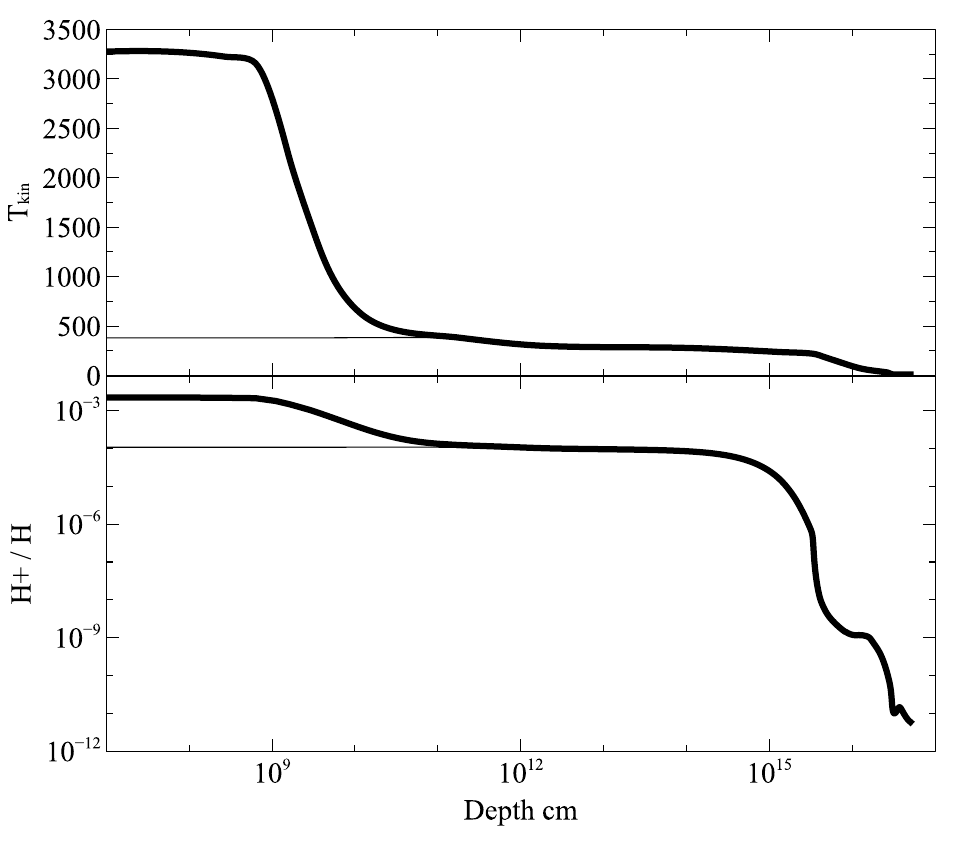}
\caption[The PDR ionization process]
{\label{fig:PDR_TeHp}
The gas kinetic temperature (upper panel) and ionized hydrogen fraction (lower panel)
for one of the Leiden meeting test cases.
The thick line shows predictions for the simple PDR SED striking an unshielded PDR.
A thin layer of warm ionized gas, produced by the process sketched in Figure \ref{fig:PDR_Hpump},
is produced.
The thin line is the same calculation but with the {\tt database H-like Lyman pumping off}
command included to block Lyman line fluorescence, as is done in the {\tt pdr\_leiden\_v4.in}
test case in our test suite.
 }
\end{figure}

Conventional PDR calculations use the smooth \citet{Draine1996} SED shown in
Figure   \ref{fig:PDRcontinua}.
We do participate in PDR meetings and include several PDR models in our test suite,
so this SED must be used. 
This SED, when coupled with our full model of \hi, produces several unexpected results.
Figure \ref{fig:PDR_TeHp}, 
computed using our version of the standard Leiden \citep{Roellig2007} V4 test case,
shows the problem.
The heavy line shows our predicted temperature and hydrogen ionization as a function
of depth from the illuminated face of the PDR.
We predict a thin, $\delta r \sim 10^{10}$~cm, layer of warm ionized gas.

\begin{figure}
\centering
\includegraphics[width=\linewidth]{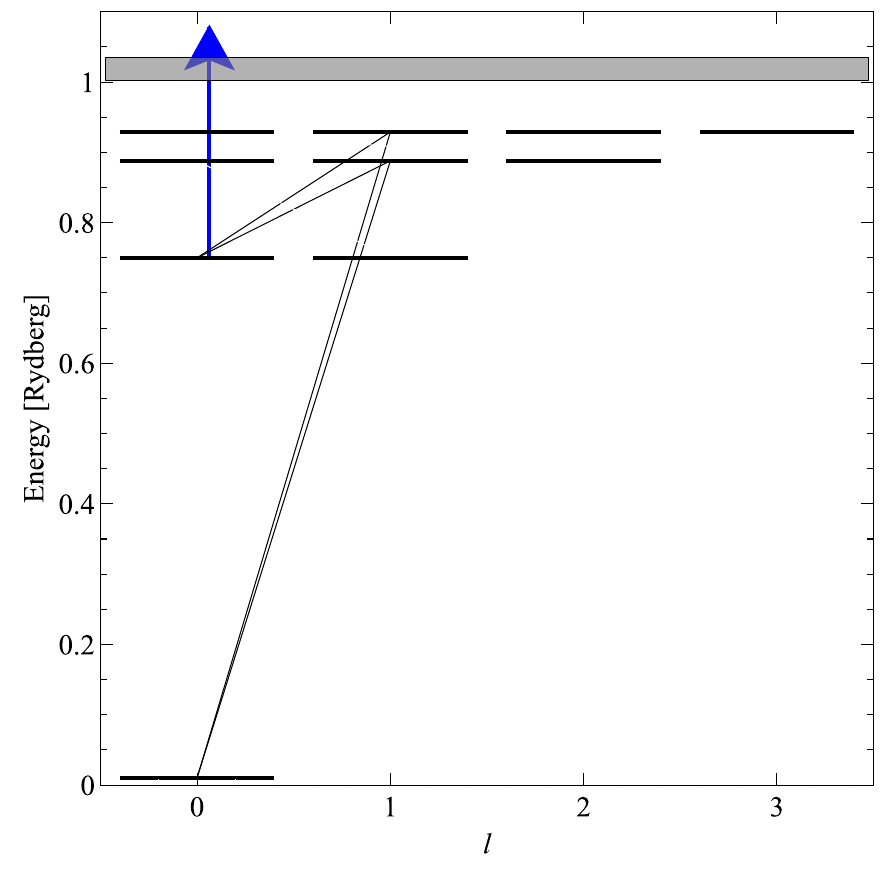}
\caption[The PDR ionization process]
{\label{fig:PDR_Hpump}
This shows the process which produces the thin warm layer in 
Figure \ref{fig:PDR_TeHp}. Radiation in the Lyman lines
photoexcites the $np$ levels via high-$n$ Lyman lines.
Many decay to the metastable $2s$ level, which is then photoionized
by the Balmer continuum.
The process is an artifact of the smooth SED used in PDR simulations.
In nature, there is sufficient \hi{} column density in the \hplus{} layer to make
the Lyman lines self shielded.}
\end{figure}

Examination revealed that this layer is produced by photoionization of hydrogen by
the Balmer continuum.  The process is outlined in Figure \ref{fig:PDR_Hpump}.
Radiation in the Lyman lines pumps \hi{} to excite the $np$ levels.
Some decay to the metastable $2s$ level while others
decay back to $1s$ producing Lyman lines which then scatter within the \hone{} gas.
These eventually decay to produce a Balmer line and populate either $2s$ or $2p$.
The metastable $2s$ level is long lived and can have a significant population.
The Balmer continuum  photoionizes $2s$, producing the warm, more ionized, layer.
The thickness of the layer is set by the column density needed for the Lyman lines to become self shielded.

In previous versions of \Cloudy{}, we recommended adding the {\tt case B} command to
artificially make the Lyman lines optically thick.
That command has other effects and we introduce a new command,
{\tt database H-like Lyman pumping off},
to make all Lyman lines optically thick.
The actual effect is to prevent the pumping rather than change the radiative transfer of the Lyman lines.
When this command is included in the input deck we obtain the results shown as the light line in
Figure \ref{fig:PDRcontinua}.
This not only makes our predictions more closely match those of conventional PDR codes,
but also results in a significant time savings since resolving the thin warm layer is fairly expensive.

\subsection{The LAMDA database}
\label{sec:LAMDA}

Since C13, we have updated our version of the LAMDA 
\citep{Schoier.F05An-atomic-and-molecular-database-for-analysis}
database of molecular structure and emission.  
LAMDA does not use version numbers.  
Our version of LAMDA was downloaded on 2015 June 30.

In addition to updating collisional rates for several species, 
which have been modified by LAMDA since the last integration of \Cloudy{} 
with the LAMDA database in C10, we added the ability to predict the intensity of 
five of their new species: OH$^+$, H$_2$S, H$_2$CS, C$_2$H, and NH$_2$D.  

\Cloudy{} uses the stored level energies to predict line wavelengths,
photon energies, and collision rates \citep{LykinsStout15}.
Some of the energy levels present in LAMDA are not precise enough to 
compute wavelengths with sufficient accuracy.  
LAMDA does give the frequency of all transitions to higher precision than the energy levels.  
Therefore, for the species where more precision is needed, we  modified the
LAMDA data to include higher precision for the energy levels.  
This was especially important for hyperfine transitions, such as in OH, 
where computing the wavelengths using the energy level spacing present in 
LAMDA leads to multiple transitions having the same wavelength.  
For C17, these issues have been resolved and 
all hyperfine transitions for OH are now specified with unique wavelengths.  

Overall, the improved rates from the LAMDA update did change some 
molecular intensities in our test suite of simulations.
The largest and most notable were H$_2$O, CS, and 
some of the high rotational transitions of CO.

\subsection{The grain data}

In order to model grains in photoionized environments and XDRs, \Cloudy{}
needs grain refractive index data covering the entire wavelength range from
the X-ray to the far-IR and sub-mm regime. The reason for this is that
\Cloudy{} creates a self-consistent model of the physical state of the grains.
The X-ray data are needed because X-ray photons can be important for heating
and charging of the grains in environments with a hard radiation field, while
the IR and sub-mm data are obviously needed to correctly predict the thermal
emission from the grains. Refractive index data covering such a vast
wavelength region are very hard to find, which is the reason why \Cloudy{}
only offers data for a limited set of grain materials. In C17 we added
refractive index data for $\alpha$-SiC using the results from
\citet{Laor1993}. The enthalpy data needed to model stochastic heating of
these grains were taken from \citet{Chekhovskoy1971}. We created a custom fit
to the data from Table 4 for $T < 300$~K, while we retained Equation~3 for
higher temperatures, resulting in the following formulas for the heat capacity
$C_p$ of $\alpha$-SiC in cal\,mol$^{-1}$\,K$^{-1}$:
\[ C_p = 13.25 \times [ 0.398 f_3(T/\Theta_1) + 0.602 f_3(T/\Theta_2) ] \]
\[ \hspace*{54mm} {\rm for}~T < 300~{\rm K}, \]
\[ C_p = 13.25 - 2035/T + 288\times10^5/T^2 \exp(-5680/T) \]
\[ \hspace*{54mm} {\rm for}~T \geq 300~{\rm K}. \]
Here $T$ is the temperature of the grain in kelvin, $\Theta_i$ are the Debye
temperatures of the material given by $\Theta_1 = 747$~K and $\Theta_2 =
1647$~K, and the function $f_n$ is given by:
\[ f_n(x) = \frac{1}{n} \int_0^1 \frac{y^n{\rm d}y}{\exp(y/x) - 1}. \]
To convert the heat capacity to erg\,mol$^{-1}$\,K$^{-1}$, the formulas above
need to be multiplied by $4.184\times10^7$~erg\,cal$^{-1}$.

\Cloudy{} includes data for AC and BE amorphous carbon taken from
\citet{Rouleau1991}. These data used an erroneous $\beta=2$ extrapolation of
the laboratory data towards long wavelengths. This extrapolation has been
removed in C17, so that the data now have the expected $\beta \approx 1$
behavior in the sub-mm regime.

The numerical integration scheme for grain opacities that is part of the Mie
code has been rewritten, resulting in higher accuracy results. This was needed
for the new SiC refractive index data, but will also result in more accurate
data for other grain materials. We also modified the code to allow it to
converge opacity data for larger grains. All the grain opacity files in C17
have been updated using the new code.

\section{The Cooling Function}

The gas kinetic temperature is the only temperature used in a non-equilibrium plasma.
A temperature could be defined for individual level populations or  the degree of ionization,
but  each would have a different value if the gas is not in LTE.
The electron velocity distribution in a fully ionized gas is assumed to be Maxwellian 
because of frequent electron-electron elastic collisions
\citep{Spitzer1962, Ferland16.kappa}.
This is often not true in a partially-ionized gas, especially one exposed to energetic radiation
or cosmic rays  \citep{1968ApJ...152..971S}.
In this case, a population of suprathermal electrons can be present, and is treated as
described in C13 and AGN3.
The kinetic temperature is set by the energy exchange, the rate at which inelastic
collisions between particles convert kinetic energy into light, which then escapes.
The cooling function, described here, represents the total cooling due to all species.

\subsection{Species cooling}

Nearly all species are now treated with multi-level models of the internal structure
of the atom or molecule.
In earlier versions of the code, with much simpler model atoms, it was easy to 
determine the cooling due to a particular line.
For a two-level atom the net energy exchange, the rate at which kinetic energy
is converted into light, is
\begin{equation}
\Lambda_1 = (n_l q_{l,u} - n_u q_{u,l}) h \nu_{u,l}
\end{equation}
where the $q$'s are collision rates, $\ps$, $u$, $l$ indicate the upper and lower
levels, and $h\nu_{u,l}$ represents the energy difference, 
which corresponds to the photon energy for a two-level system.

With a large multi-level model this single-line concept loses meaning. 
The total cooling due to a species with $N$ levels is
\begin{equation}
\Lambda_{species} = \sum_{u=2}^{N} \sum_{l=1}^{l<u} (n_l q_{l,u} - n_u q_{u,l}) h \nu_{u,l}
\label{eq:SpeciesCooling}
\end{equation}
We now report the total cooling due to a species
with the species label followed by a wavelength of ``0'' to 
indicate that it represents the entire species.
Examples include the following entries which were
taken from the output of the {\tt save line labels} command
(Section \ref{sec:SaveLineLabels} above):

{\footnotesize
\begin{verbatim}
 30      H  1c   0  net cooling due to iso-seq species
 31      H  1h   0  heating due to iso-seq species
 32      He 2c   0  net cooling due to iso-seq species
 33      He 2h   0  heating due to iso-seq species
450      Sc16c   0 		net cooling due to database species
451      Sc16h   0 		heating due to database species
452      Sc17c   0 		net cooling due to database species
453      Sc17h   0 		heating due to database species
\end{verbatim} }

\noindent
Each entry has either ``c'' or ``h'' after the line label.
The ``c'' indicates that the sum in Equation \ref{eq:SpeciesCooling} is
positive and that the species contributes net cooling.
If the level populations are inverted, corresponding to a negative temperature,
the species will heat rather than cool the gas, the sum will be negative,
and will appear with the ``h'' character.
This often occurs when levels are predominantly photoexcited by the attenuated 
incident radiation field.

\subsection{Time-steady non-equilibrium cooling function}

Nearly all of the calculations presented in this paper are non-equilibrium
in the sense that the ionization and level populations are not described by
the gas kinetic temperature.  The Boltzmann level population and Saha-Boltzmann
ionization equations do not apply.  
Equilibrium only occurs at the highest densities in Figures  
\ref{fig:HIonVsDensity}, \ref{fig:2levelColl}, and \ref{fig:2levelPhoto}.

By default, \Cloudy{} solves for the non-equilibrium conditions
in a gas cloud in steady state, i.e., assuming that its
properties do not evolve over times that are more rapid than those needed
for atomic processes to come into equilibrium.
This is accomplished by taking the difference between
the recombination and ionization rates in the ionization
balance equation for each species to be zero.
We provide an {\tt age} command that will check whether this is a valid assumption.

\begin{figure}
\centering
\includegraphics[width=\linewidth]{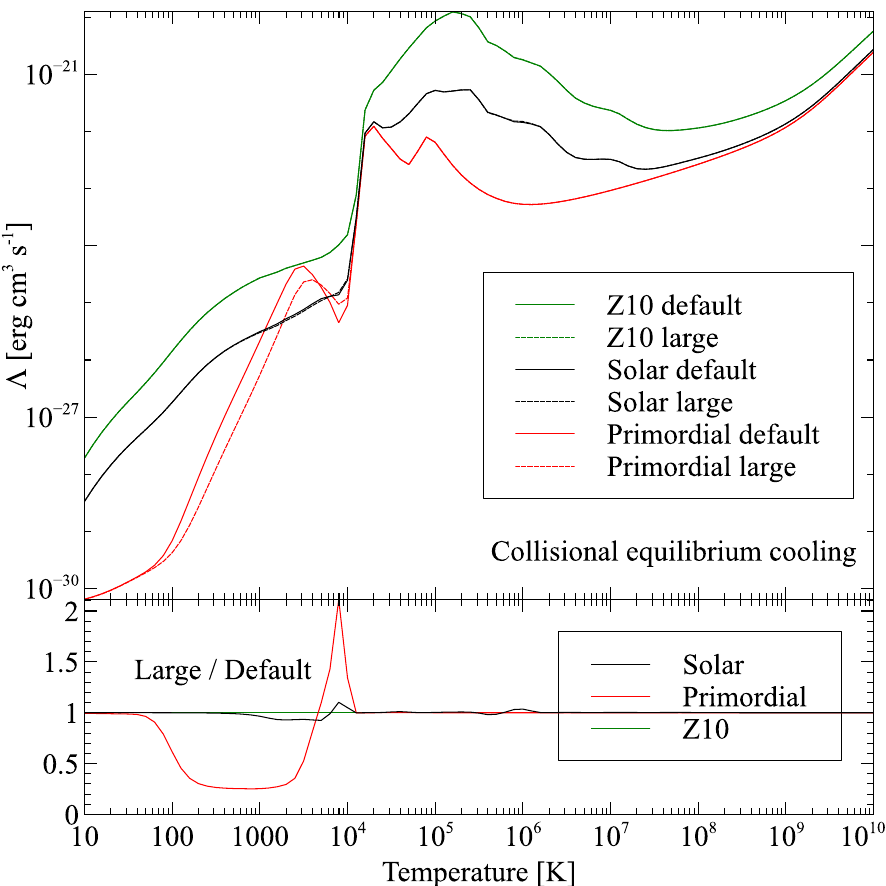}
\caption[Solar time steady cooling function]{
\label{fig:cooling}
Time-steady non-equilibrium cooling for three sets of chemical abundances.
The gas is collisionally ionized by thermal collisions although background cosmic rays are
included to allow the chemistry to converge at very low temperatures.
Calculations are done for both our default and maximum atomic models.
The upper panel shows the total cooling for each set of abundances while the
lower panel shows the ratio of cooling computed with the large to default models.
}
\end{figure}

\citet{GnatFerland12},
\citet{Lykins.M12Radiative-cooling-in-collisionally-and-photo}, and
\citet{2014MNRAS.440.3100W}
use \Cloudy{} to compute time-steady cooling.
Figure \ref{fig:cooling} shows cooling functions for three very different sets
of chemical abundances.
The cooling in the primordial case is dominated by H, He, and their molecules, while
the solar case has been discussed in many papers.  The original
review by \citet{Dalgarno1972} is still among the best.
The cooling in the high-metallicity case, marked ``Z10'', is dominated by the heavy elements.

The input script used to create the large solar case shown in Figure \ref{fig:cooling} is given below.
This set of calculations uses our {\tt grid} command to vary the kinetic temperature
between 10~K and $10^{10}$~K.
The {\tt grid} and {\tt optimize} command now use the {\tt *nix fork}
system call to compute a series of independent models, one on each available thread.
This method of computing in parallel is discussed further in Section \ref{sec:OtherTechnicalChanges} below.

{\footnotesize
\begin{verbatim}
set save prefix "all"
database H2 
database CHIANTI levels maximum
database stout levels maximum
database lamda levels maximum
coronal t=3 vary
grid 1 10 .1 log
hden 0
stop zone 1
set dr 0
cosmic rays background
save grid ".grd"
save cooling ".col"
\end{verbatim} }

This example may have problems on some computers.
A single calculation uses nearly 4GB of RAM since all the models are set to the largest number of levels.
However, this is a grid run which launches many models simultaneously.
If the total memory requirements of these models exceed the physical
memory in the system,
the OS may start using swap space (which is very slow)
to compute the model, or (if that fails) abort the sim.
This is an inevitable consequence of computing many large models
on a multi-core system.
In such a case, the solution is to run the model sequentially
by adding the keyword
{\tt sequential} to the {\tt grid}  command.
Then, only a single thread will be started requiring 4 GiB of memory in total.
This is another example of the necessary compromise between
model fidelity and resource utilization
which permeates this review.

The {\tt coronal} command sets the kinetic temperature and 
tells the code to bypass its normal check that the incident SED
is correctly specified, since a purely collisional model may be intended.
The {\tt hden} command sets the hydrogen density to $1 \pcc$ and the {\tt set dr} and
{\tt stop zone 1} commands result in a
single zone with a 1 cm thickness.
We include the chemistry network originally based on 
\citet{2005ApJS..161...65A} and
 \citet{Roellig2007}, and originally based on UMIST 
 \citep{Le-Teuff.Y00The-UMIST-database-for-astrochemistry-1999}.
 That chemistry will not converge at low temperatures 
 without a source of ionization to drive the ion-molecule reactions.
 The background cosmic rays provide that ionization.
 
 Figure  \ref{fig:cooling} compares the cooling with both our default and largest data sets,
and shows their ratio in the lower panel.
 The largest differences are due to including the complete \htwo{} model in the large models.
 The complete model includes detailed treatment of formation and dissociation processes,
 which results in somewhat different predictions for the chemistry, compared with the default model.
 The explicit calculation of thousands of \htwo{} lines changes the cooling per \htwo{}.
 Both produce the large differences around 100~K - 1000~K in the primordial case.
 
 There are only small differences at moderate to high temperatures.  
 The default model is surprisingly accurate for these temperatures due to the 
 level selection criteria described by \citet{Lykins.M12Radiative-cooling-in-collisionally-and-photo}.
 
\begin{figure}
\centering
\includegraphics[width=\linewidth]{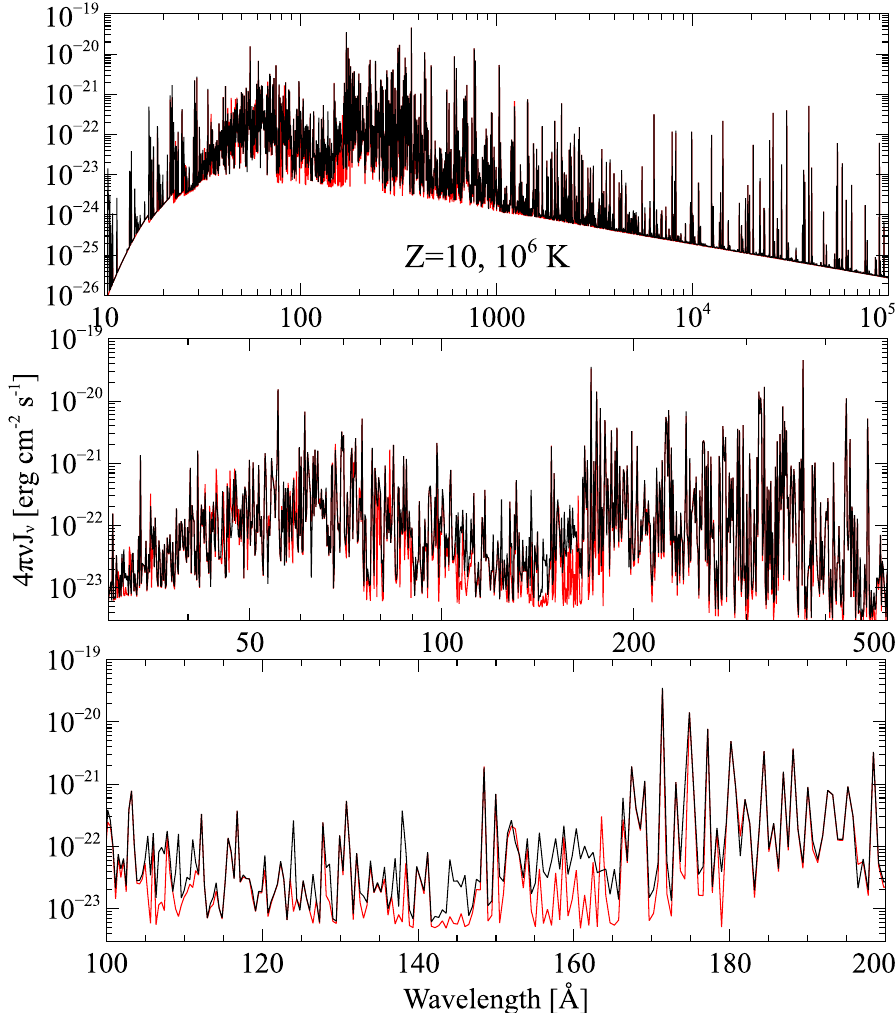}
\caption[Solar time steady cooling function]{
\label{fig:Z10Emission}
The spectrum of a $10Z_\odot$ plasma using our default (red) and large database (black).
The top panel shows the full emission and each panel below it is a zoom into 
smaller regions of spectrum.}
\end{figure}

 Figure \ref{fig:Z10Emission} compares the total emission for the ``Z10'' case and
 a temperature of $10^6$~K for our default (red) and large (black) models.
 Over much of the spectrum they are similar and the default model includes the strongest lines.
 However, there are significant ``gaps'' in some portions of the spectrum as shown in the
 lowest panel.
 This shows that our default setup is good enough for most purposes, but that there will
 be circumstances when a larger model is needed.

\subsection{Time-dependent non-equilibrium cooling}

There are applications where the time-steady assumption
is not valid \citep[e.g.,][]{GnatSternberg07, Gnat17}.
\Cloudy{} has long included the option to model time-steady non-equilibrium dynamical flows,
\citep{2005ApJ...621..328H, 2007ApJ...671L.137H},
in which advected material enters gas at a particular location, altering
the balance equations. 

\citet{Chatzikos2015} has extended this to treat the time-dependent non-equilibrium case.
This extended
the steady-state treatment to permit solving the class
of time-dependent problems related to the cooling of a
parcel of gas from a definite initial state.
In the time-dependent scheme, ionization and
recombination are not in balance, which introduces
a net rate of change for a species density.
This permits advancing the ionization for each species
over time-steps $\Delta t$ with a first-order implicit
integration scheme, as explained in mode detail in
\citet{Chatzikos2015}.
\par

Example calculations, available as of C17,\footnote{Test suite
models {\tt time\_cool\_cd.in} and {\tt time\_cool\_cp.in}.}
evolve a unit volume of gas from a preset temperature (30 million K)
to a definite floor (10,000 K).
The commands
\begin{verbatim}
coronal 3e7 init time
stop time when temperature falls below 1e4
\end{verbatim}
are used to set the initial and final temperatures.
\par

\begin{figure}
	\centering
	\includegraphics[scale=0.46]{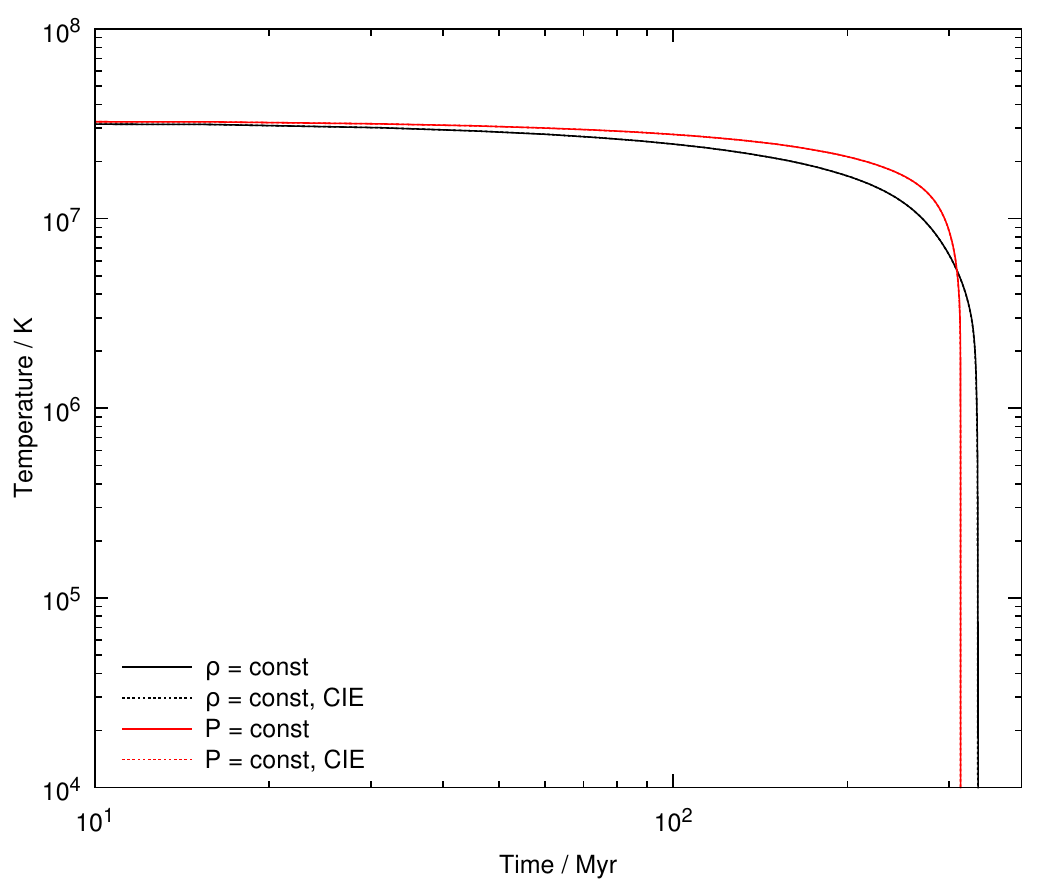}
	\includegraphics[scale=0.46]{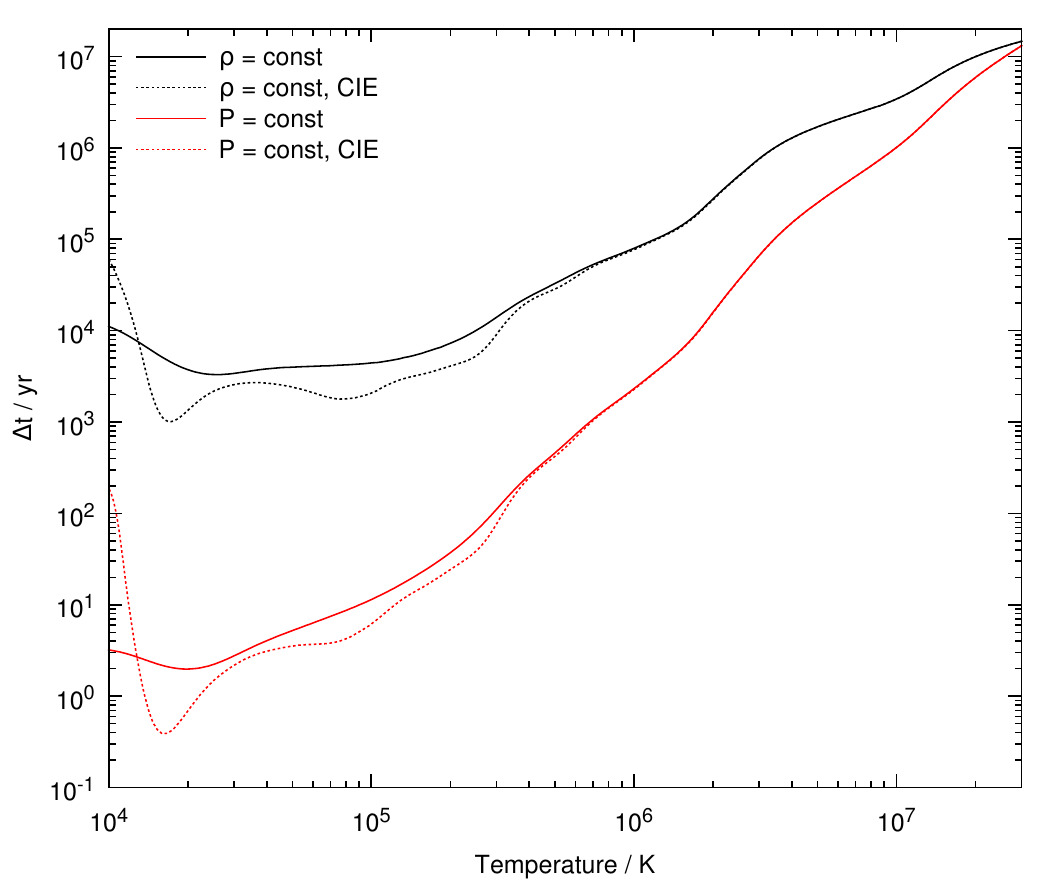}
	\caption
	{
		Time-dependent cooling of a unit volume of gas.
		Evolution of temperature with time {\em (top panel)} and
		time-step advance with temperature {\em (bottom panel)}
		under isobaric and isochoric conditions.
		The dashed curves show the evolution assuming collisional
		ionization equilibrium (CIE).
		\label{fig:cooling-time-vs-temp}
	}
\end{figure}

Figure~\ref{fig:cooling-time-vs-temp} presents the evolution of
temperature with time in the top panel, and of the time-step advance with
temperature in the right.
The solver adapts the time-step to track the changing conditions in the gas.
At high temperatures, the system evolves slowly, and large time-steps
are taken, while at lower temperatures, the time-step size is reduced
and a larger number of steps is required for fixed fractional
change in temperature.
In both isochoric and isobaric cases, cooling down to 10 million K
overwhelms the total cooling time of the gas, which subsequently cools
down to 10,000 K about 4--15 times faster.
This is related to the peak of the cooling function around 300,000~K,
see Figure~\ref{fig:cooling}.
For isobaric conditions, the density rises with decreasing temperature,
which enhances the cooling rate of the gas, and leads to the shorter
cooling time-steps seen in the bottom panel of Figure~\ref{fig:cooling-time-vs-temp}.
Compared to the collisional equilibrium ionization case,
the system generally takes longer time-steps, because the
cooling function is suppressed, as explained in \citet{GnatSternberg07}.
\par

\begin{figure}
	\centering
	\includegraphics[scale=0.40]{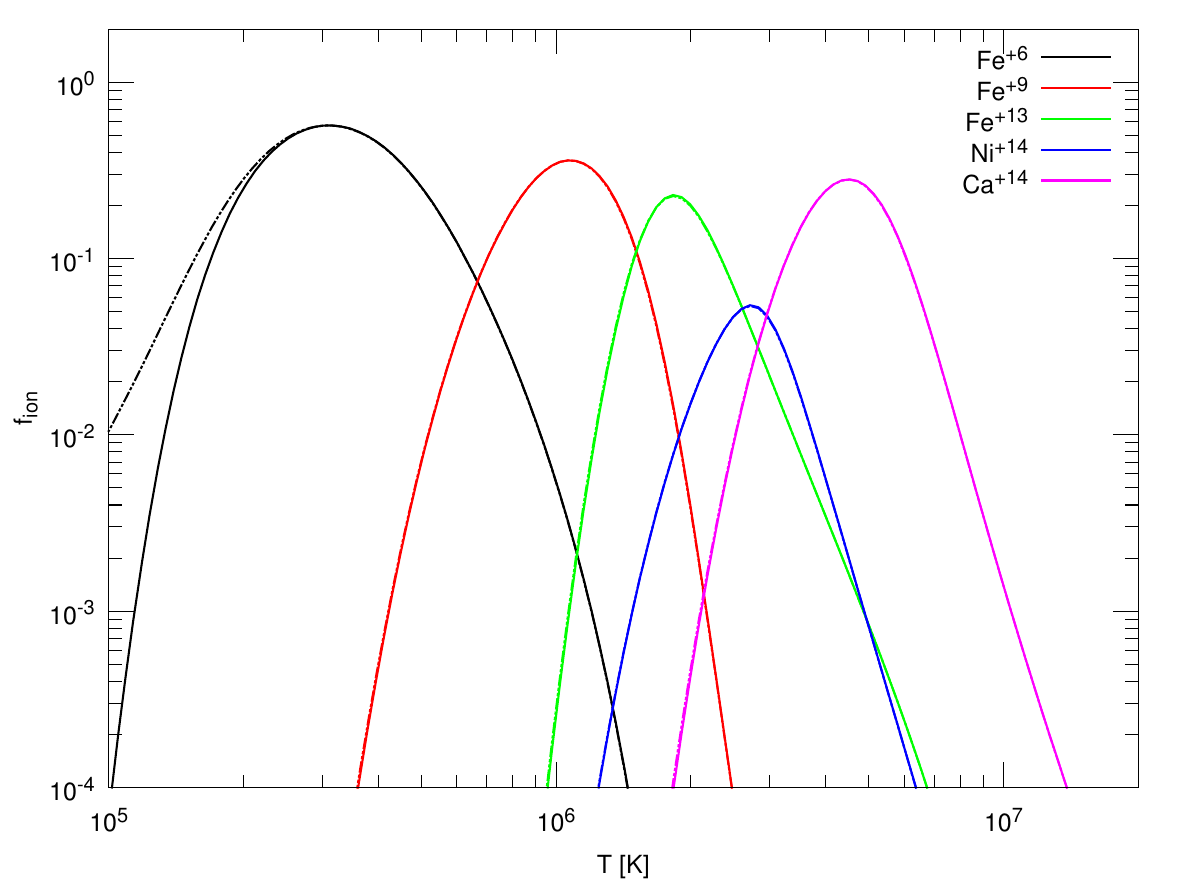}
	\caption
	{
		Comparison of ionization fractions in collisional equilibrium
		(solid lines) and non-equilibrium ionization (dot-dot-dashed lines).
		The effects of the latter, become important below 300,000 K.
		\label{fig:ion-frac-cie-vs-nei}
	}
\end{figure}

As Figure~\ref{fig:ion-frac-cie-vs-nei} shows,
the time-dependent scheme is able to reproduce
the known behavior that deviations from equilibrium
at fixed temperature become important only below the
peak of the cooling curve at $\sim$300,000~K
\citep{Chatzikos2015}.
We illustrate this effect by forcing equilibrium conditions
with the command\footnote{See test
suite model {\tt time\_cool\_cp\_eq.in}.}
\begin{verbatim}
set dynamics populations equilibrium
\end{verbatim}
In that Figure, the equilibrium ionization fraction of Fe$^{6+}$
drops quickly with decreasing temperature.
This comparison highlights the non-equilibrium origin
for the tail to moderate ionization fraction values
at lower temperatures.
\par

In time-steady calculations, \Cloudy{} reports the instantaneous
intrinsic and emergent emission produced by the cloud, where the
latter has been corrected for extinction occurring outside the
line-forming region.
These are reported by default in the main output of the program.
\par

In time-dependent calculations, integrations over time of the
instantaneous intrinsic and emergent emission are performed,
referred to as ``cumulative'' emission.
The general form of the integration is
\begin{equation}
	F = \int \, dt \, j(t) \, w(t)	\,	,
	\label{eqn:time-integration}
\end{equation}
where $w(t)$ is a weighting function.
Its application to line emission is presented in
\citet[equation~(3)]{Chatzikos2015}.
The integration may be performed in two fashions,
or not done at all, through the commands
\begin{verbatim}
set cumulative mass
set cumulative flux
set cumulative off 
\end{verbatim}
By default, weighting by \verb+mass+ is performed,
which uses a weighting function $w(t) = 1 / \rho(t)$,
where $\rho(t)$ is the mass density at time $t$.
Notice that the units of the time-integral of the emissivity
are erg g$^{-1}$.
Obviously then, this choice closely traces the predictions
of the cooling flow model \citep{Fabian1984}.
Specifically for lines, the integral reduces
to the usual line $\Gamma$ factors
\citep{Chatzikos2015,Graney:1990aa}.
Luminosities may be obtained from these estimates by
multiplication by the mass cooling rate, $\dot{M}$.
Notice, however, that these considerations apply
only to unit volume calculations; for extended clouds
the meaning of the computed emission is unclear.
\par

On the other hand, when weighting by \verb+flux+, the weighting function
is taken to be unity, $w(t) = 1$, which suggests that the units of the
integrated emission are erg cm$^{-3}$.
This is equivalent to the power emitted per unit volume over
the course of the integration.
For extended models, the units become erg cm$^{-2}$,
and have the meaning of the accumulated flux over
the duration of the simulation.
Such weighting is more appropriate for calculations on fixed domains,
as will be discussed in a future paper.
\par

For a cooling flow, we follow a parcel of gas as it cools down,
in the manner shown in Figure \ref{fig:cooling-time-vs-temp}.
We integrate over time the intrinsic and emergent emission
to capture the emission from all the regions that contribute
to the observed spectrum.
By default, \Cloudy{} reports both the instantaneous emission
from the gas at each time, as well as the cumulative emission
up to that point in the simulation, when time-dependent
simulations are enabled.
\par

\begin{figure}
	\centering
	\includegraphics[scale=0.45]{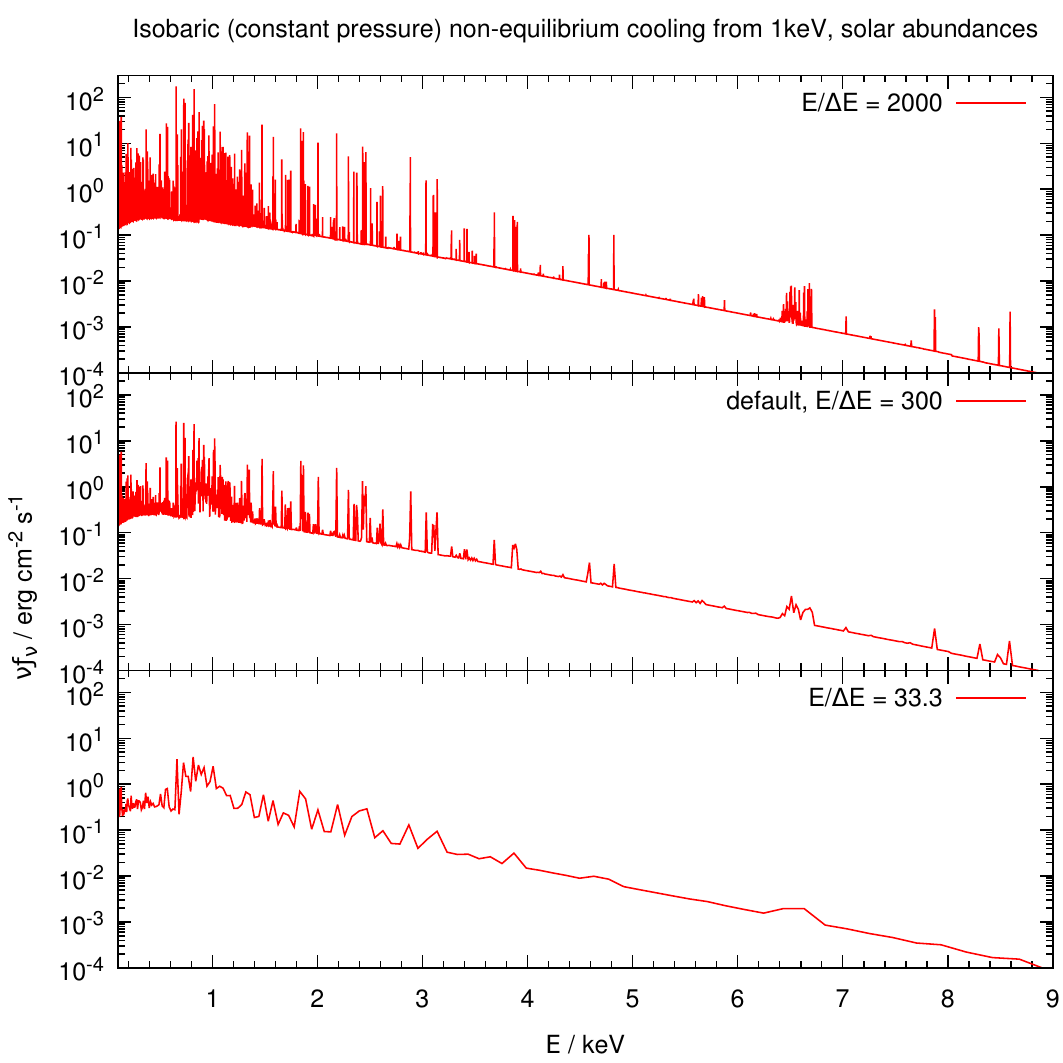}
	\caption
	{
		Cooling flow spectra in the energy range 0.1--9 keV,
		at different energy resolutions.
		The top panel corresponds to the proposed resolution of {\em Athena},
		the middle panel corresponds to the current state-of-the-art
		resolution obtained with dispersed X-ray spectrometers, such
		as {\em RGS} onboard \xmm{}, while the bottom panel corresponds
		to the typical energy resolution in X-ray imaging observations.
		\label{fig:cooling-flow}
	}
\end{figure}

Figure~\ref{fig:cooling-flow} presents the predicted
soft X-ray spectrum from a cooling flow, as it would
be observed by X-ray telescopes at three different
energy resolutions.
The bottom panel corresponds to the current state
of affairs for X-ray imaging spectroscopy with
the detectors onboard \chandra{} or \xmm{}.
The current state of the art, obtained with X-ray dispersed
spectroscopy observations, is shown in the middle panel.
Finally, the top panel shows the spectrum that can
be obtained with microcalorimeters such as those
onboard the unfortunate {\em Hitomi} mission, and
the proposed {\em Athena} mission.
\par

For the purposes of that Figure, the cumulative emission
(line and continuum) was obtained by using the keyword
\verb+cumulative+ with the family of \verb+save continuum+
commands.
Note that the instantaneous cloud emission can still
be obtained by invoking the \verb+save+ commands without
the \verb+cumulative+ keyword.
For reference, a minimum set of commands to reproduce
the high resolving power spectrum in the top panel of
the Figure is the following
\begin{verbatim}
coronal 1.1e7 K init time 
hden 5.88e-2 linear
constant pressure reset
set dr 0 
set nend 1
stop zone 1
iterate 400
stop time when temperature falls below 6e5 K
cosmic rays background
set save resolving power 2000
save cumulative continuum units Angstroms \
      last "cooling.concum"
\end{verbatim}
We point out that the \verb+set save resolving power+ command increases the
line-to-continuum contrast without adjusting the line width and therefore
violates energy conservation in the saved spectrum.
\par


\section{Other physics changes}
\label{sec:OtherPhsicsChanges}

\subsection{Corrections for isotropic radiation}

\begin{figure}
	\centering
	\includegraphics[scale=0.5]{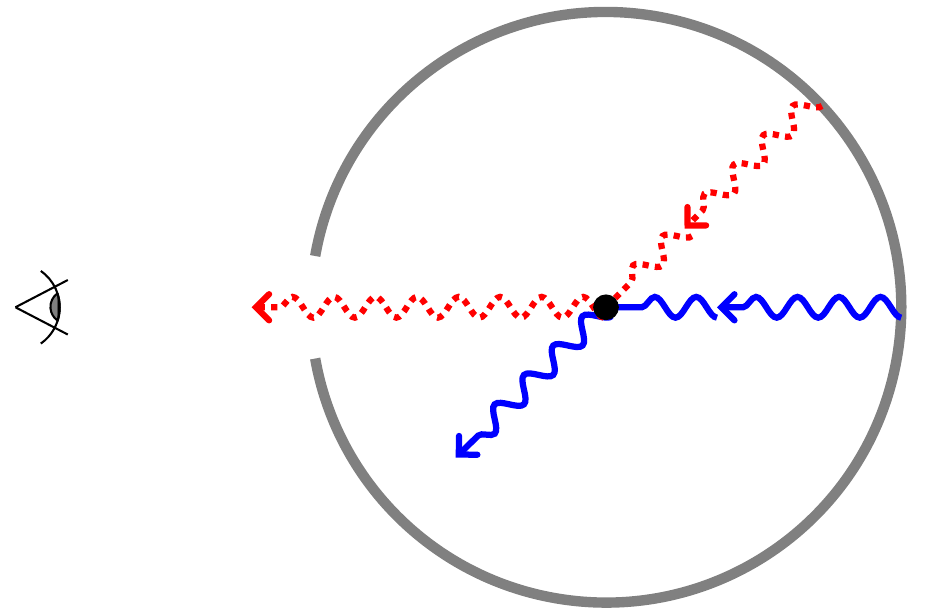}
	\caption
	{
		Photons of an isotropic continuum are absorbed and reemitted in
		random directions by an idealized isolated absorber (atom or
		molecule).
		This process does not change the isotropic character of the
		radiation field, but it does populate the upper level of the
		transition for an ensemble of absorbers, and therefore moderates
		the line emissivity.
		\label{fig:cmb-pumping}
	}
\end{figure}

\citet{Chatzikos2013} expanded our treatment of isotropic radiation fields
to produce new reporting options in which their effects are removed,
to better approximate what is measured at the telescope.
Figure~\ref{fig:cmb-pumping} presents an idealized situation,
in which an isolated absorber (atom or molecule) is exposed to
isotropic radiation, and undergoes two successive excitations,
each followed by a radiative deexcitation.
An example might be the CMB photo-exciting the CO rotation ladder.
Because of the isotropic character of the field, the number of
photons scattered out of the line of sight is equal to the number
scattered into the line-of-sight, so the radiation field remains
isotropic despite the presence of the absorber.
No net emission or absorption feature is produced.

A population of absorbers behaves similarly.
Isotropic photons at the frequency of an atomic
(or molecular) transition are absorbed and reemitted in
random directions, leaving the field isotropic.
Although pumping by an isotropic continuum does not produce a spectral
line in the two-level system, it does affect the level populations,
which then affects the line emissivity.
This has been traditionally parameterized as a diminution
factor on the emergent line intensity \citep[e.g.,][]{DCruz1998}.

It can be shown that this diminution factor is connected 
to the escape probability formalism employed for the
transfer of radiation.
The essence of the escape probability theorem lies in
the statement that the net emission from a parcel of
gas at some location in a cloud is proportional to the
probability of escape from the cloud starting from that
location in a {\em single} flight.
Photons that require several scatterings before
escape are assumed to be absorbed {\em in situ},
that is, \Cloudy{} does not track their propagation
through the gas cloud.
The first proof of the theorem was offered by
\citet{Irons.F78On-the-equality-of-the-mean-escape-probability},
who showed that it had the character of energy conservation for
an emitting volume.
The theorem was subsequently extended by
\citet{Rybicki.G84Escape-probability-methods-pp-21-64},
who showed that the theorem was far more detailed,
as it applied to individual rays.
In \citet{Chatzikos2013}, Rybicki's formalism was
refined for the case of isotropic radiation, and it was
shown that the net emission
is proportional to the usual escape probability times
a factor that depends on the field intensity
\citep[see equations (39) and (40) in][]{Chatzikos2013}.
It was also shown that this form reduces to the traditional
diminution factor.

Because isotropic radiation may affect line emissivities
across the spectrum, as of C17, line intensities are corrected
for isotropic radiation by default.
This resembles the usual practice in far infrared,
millimeter, and radio observations to automatically
correct for the CMB by some method (position, beam,
or frequency switching), so that the observer is
provided with net line or continuum fluxes.
This behavior may be disabled with the aid of the
command
\begin{verbatim}
no lines isotropic continuum subtraction
\end{verbatim}
which causes line intensities uncorrected
for isotropic radiation to be reported.
\par

On the other hand, the transmission of the
continuum through the cloud involves only
the attenuation of radiation at some frequency
due to optical depth effects.
By default, \Cloudy{} reports the total
continuum radiation, including any attenuated
isotropic radiation, unless the command
\begin{verbatim}
no isotropic continuua report
\end{verbatim}
is issued, or the keyword \verb+no isotropic+
is used with any of the \verb+save continuum+
commands.
Either of these options will cause the continuum to
be corrected for any (attenuated) isotropic radiation.
\par

\begin{figure}
	\centering
	\includegraphics[scale=0.45]{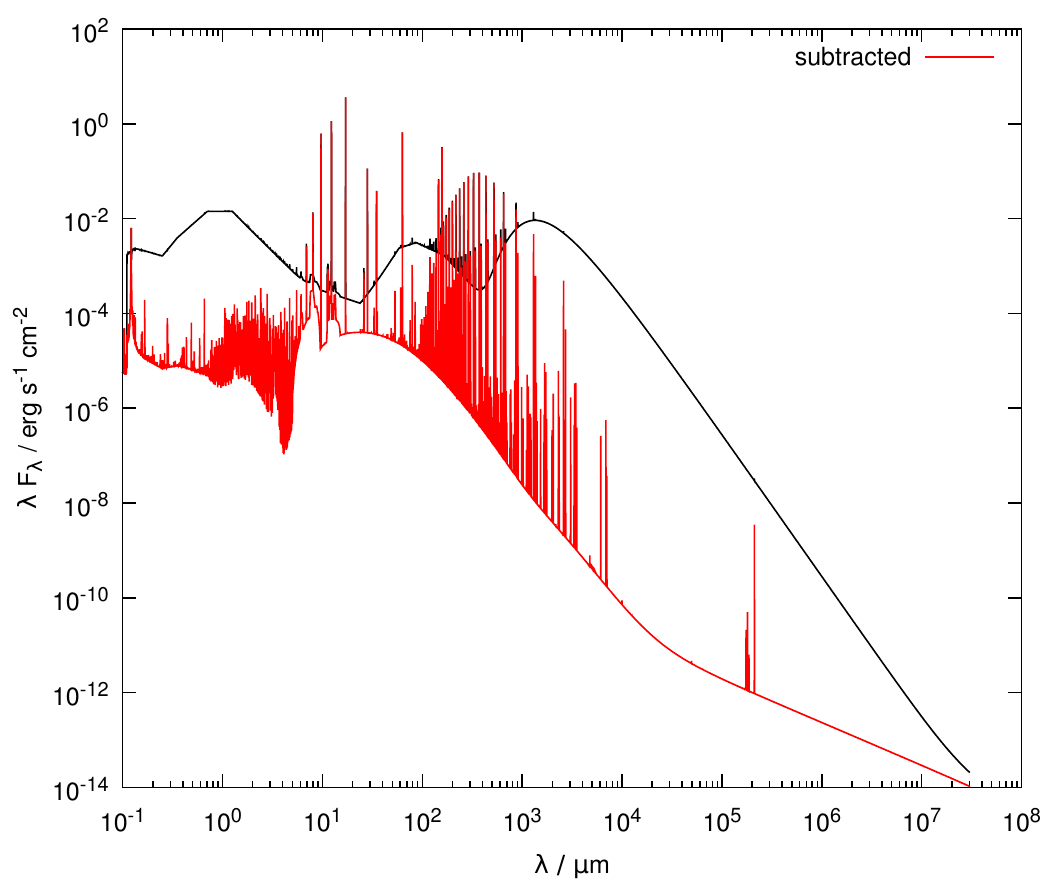}
	\caption
	{
		Example of the correction for isotropic continua.
		An interstellar cloud is exposed to a radiation
		field that is the superposition of two isotropic continua:
		the local interstellar radiation field, and the CMB.
		The black curve shows the spectrum emergent from the cloud.
		This is the light distribution incident on a telescope.
		In the radio and the far infrared, observations are corrected
		in real time for isotropic continua (e.g., by position, beam, or
		frequency switching).
		The red curve shows the spectrum that such observations would
		yield.
		Notice that at the position of the \hi{} 21cm line,
		the correction is about 4 dex.
		\label{fig:iso-cont-sub}
	}
\end{figure}

As an illustration of these capabilities,
Figure~\ref{fig:iso-cont-sub} presents the emergent
spectrum of an interstellar cloud irradiated by two
isotropic continua: the cosmic microwave background
and the local interstellar continuum.
The black and red curves show the total emergent
spectrum, and that corrected for isotropic
radiation, respectively.
The Figure highlights the magnitude of isotropic
radiation  in the mm, where the CMB overwhelms the
diffuse fields by 6--8 dex.
The calculation may be reproduced with the following
commands:

{\footnotesize
\begin{verbatim}
cmb
table ism
extinguish 21
cosmic ray background
database H2
grains PAH 3
abundances ISM
hden 3
stop thickness 1 linear parsec
constant temperature 500 K
stop temperature off
save continuum units microns "iso.con"
save continuum units microns no isotropic \
     "iso.con-noiso"
\end{verbatim}}
\par


\subsection{Chemical composition}

The built-in abundance sets have been exported to external files located in the {\tt cloudy/data/abundances} directory.
These are much easier to add or change since updates no longer require editing the C++ source code.
There is no change in the {\tt abundances}-type commands which use these files.

The default composition
is given in the file {\tt cloudy/data/abundances/default.abn}.
It can be updated by merely changing this file.
A new {\tt abundance "filename"} command has been introduced.
The chemical composition will be read from the specified file,
which may be located in the current directory or in {\tt cloudy/data/abundances}.
The {\tt abundance "filename"} command has a {\tt print} option to report the gas abundances and grain types used.

The command {\tt abundances isotopes} may be used to specify isotope fractions for the species of astrophysical interest.
By default, the isotope abundances of  \citet{Asplund2009} are used, with a couple of modifications
(see {\tt cloudy/data/abundances/default-iso.abn}).

The command {\tt element [name] isotopes} may be issued to modify the default isotopic abundances to user-specified values.
For instance, the commands to specify isotope fractions for hydrogen and carbon are now
\begin{verbatim}
element hydrogen isotopes (1, 1) (2, 2e-5)
element carbon isotopes (12, 29) (13, 1)
\end{verbatim}
For each element, isotopes are specified as pairs of the atomic weight
and its fractional abundance by number.
The sum of isotopic abundances is renormalized to unity.

\subsection{The table SED command and stellar grids}

A number of built-in SEDs are available by a series of {\tt table ...} commands.  
These have been exported into data files
in the {\tt cloudy/data/SED} directory.
This makes the external SEDs simple to maintain and easy to build upon.
The {\tt table ...} commands present in previous versions of the code work as they did before.

The {\tt table HM12} command has been added to implement the 
\citet{HaardtMadau2012} grid of background continua. 
The command works the same way as the {\tt table HM05} command, 
except that the keyword quasar is not supported 
(since data files for the quasar-only case do not exist). 
This command uses the stellar grid infrastructure described in C13 to do the interpolation. 
The {\tt table HM05} command has also been moved over to use this infrastructure, 
but this should be transparent to the user.

To implement the new Haardt \& Madau grids, a start has been made to convert
the stellar grid code so that it can work directly on ascii files without the
need to convert them into binary files first. In C17 the following commands
are already supported.
\begin{verbatim}
table star "somegrid.ascii" <par1> ...
table star list "somegrid.ascii"
\end{verbatim}
This concept will be developed further in future releases with the aim to make
recompilations of binary files unnecessary when the frequency mesh changes.

\subsection{Optical depth solution}

The total optical depth across the structure is required in order to
evaluate escape probabilities from the rear face of the cloud.  The
algorithm used to update these estimates has been improved, as has the
method used to provide fall-back estimates when the optical depth
scale has been overrun, that is, the current optical depth is greater than
the optical depth in the previous iteration.  

One side-effect of the new methods is that localized spikes or troughs
can appear in the population densities of species strongly affected by
high optical depth transitions, at the position where the optical
depth scale is over-run.  This may occur if the solution has not been
allowed to iterate fully to convergence, and are symptom of a problem
which requires further iterations to obtain accurate solutions.  The
methods used previously would have led to similar inaccuracies in
the populations, but with a less obvious pathology.

Overall, these changes have greatly reduced the number of
iterations required to converge models with significant line optical
depths.

\section{Other technical changes}
\label{sec:OtherTechnicalChanges}

\subsection{The frequency mesh}

The method for generating the frequency mesh has been rewritten from scratch. 
This should solve the long-standing problem where generating a frequency mesh with 
non-standard parameters would lead to obscure crashes while running the code. 
Since this change invalidates all existing compiled stellar atmosphere and grain opacity 
files on its own, we decided to combine it with a reduction of the lower frequency limit to 10 MHz 
(roughly the lowest frequency LOFAR can observe). 
We also increased the standard resolving power to 300 and extended the range over which this 
resolving power is used to $Z^2 = 30^2 = 900$~Ryd, 
chosen to include all lines of the first thirty
elements. 
As a result, the number of frequency cells has risen from 5277 to 8228 in the standard setup.

\subsection{Grid runs now fork multicore}

In a previous release of \Cloudy{} we added support for parallel execution of model grids using MPI.
In this release we added support for parallel execution based on the {\tt fork} system call.
The big advantage of this approach is that no external packages or support libraries are needed.
The disadvantage is that this method can only work on a single shared-memory machine.
We offer this possibility enabled by default on all UNIX, Mac and Cygwin systems.
On UNIX and Cygwin machines, grid runs will run parallel using all available threads unless the user alters this behavior.
On a Mac the number is limited to the available physical cores.
The number of cores can be adjusted using the {\tt grid sequential} and {\tt grid ncpus} options.
We will continue to support MPI-based grid runs.
This will be the preferred method for very large grids requiring more threads than a single node can offer.

At the end of a grid run, the code will gather the output for each grid point
into a single file. For very large grids this can take a substantial amount of
time, especially for the main output as these files are already quite large
individually. Several options now exist to alleviate this problem. First of
all, gathering the main output has been parallelized using MPI I/O. Second,
the user now has the option to suppress printing either the intrinsic or
emergent line stack (or both) using the
\begin{verbatim}
print lines [ intrinsic | emergent ] off
\end{verbatim}
command. This will reduce the size of the main output
substantially Last but not least, we have added a keyword {\tt separate} to
the {\tt grid} command that tells the code to skip gathering of the main
output files altogether and leave them as separate files.

\subsection{Data layout optimizations}

Some of the code's internal data structures have been reorganized to
improve locality of data access.  Storing numerical data as
structures-of-arrays, rather than arrays-of-structures, can
dramatically increase the performance of codes on modern CPU
architectures, if this results in improvements to the utilization of
processor caches.  This also offers better opportunities for
vectorization.
We have also added vectorization primitives for reduction loops
and vectorized versions of commonly used math library functions.
These changes have led to substantial improvements for
test suite problems which emphasize these areas, particularly
collisional excitation processes, and worthwhile improvements for more
typical cases.

\subsection{Other changes}

A large number of small changes have occurred.
These are summarized at \url{http://trac.nublado.org/wiki/NewC17}.
Some of the more significant changes are:

\begin{itemize}

 \item
 Commas are no longer allowed to be embedded in numbers, they are treated as separators along with most other symbols. Standard exponential format 3.14159e16 can be used instead. Using commas embedded in numbers was deprecated since C10.
 
 \item 
 The {\tt tlaw} command now has a {\tt table} option, 
 similar in function to the existing {\tt dlaw table} command.

\item
We are converting the {\tt save} command output to report linear quantities. 
The output had reported a mix of log and linear quantities. 
A {\tt log} option has been added to report quantities in the old style, for backwards compatibility. 

\item
The fundamental constants have been updated to the CODATA 2014 values.

\item
  The {\tt run\_parallel.pl} script can now run only a subset of the scripts
  in the test suite. We have added a new script {\tt rerun\_parallel.pl} that
  works the same way as {\tt run\_parallel.pl}, except that it doesn't delete
  the output from a previous run.

\item
  We have added support for platforms where {\tt g++} is missing, but the
  {\tt clang++} compiler is present. This includes systems with a FreeBSD
  or Mac OS X operating system.
  
\end{itemize}

\subsection{Online access}

The primary access to all versions of \Cloudy{} remains the \url{http://nublado.org} web site.
The general organization has not changed since the C13 review.
A full set of changes to \Cloudy{} since C13, many too technical to be included here,
are on the page \url{http://trac.nublado.org/wiki/NewC17}.

\section*{Acknowledgments}

Many suggestions for improving \Cloudy{} came from the discussion forum hosted at 
\url{https://groups.yahoo.com/neo/groups/cloudy_simulations/info}.
Gabriel Altay provided valuable input for implementing the {\tt table HM12} command.
Testing for the beta versions of this release was provided by
Christophe Morriset,
Eric Pellegrini,
\& Mitchell Revalski.
We are grateful for their help.

GJF acknowledges support by NSF (1108928; and 1109061), NASA (10-ATP10-0053, 
10-ADAP10-0073, and NNX12AH73G), and STScI (HST-AR-12125.01, 
GO-12560, and HST-GO-12309).
MC acknowledges support by NASA (HST-AR-14286.001-A).
FG acknowledges support by NSF grant 1412155.
PvH was funded by the Belgian Science
Policy Office through the ESA Prodex program
and contract no.\ BR/154/PI/MOLPLAN.
FPK acknowledges support from the UK Science and Technology Facilities Council. 

\section*{DEDICATION}

We dedicate this paper to the many friends of the \Cloudy{} project across Mexico,
and to the Mexican people for making our many visits to their beautiful country so wonderful.

\appendix

\section{Stout Data}
\label{sec:AppendixStoutData}

Table~\ref{table:stout-refs} lists the species included in
the Stout database.
Species names followed by the symbol ``$\ddagger$'' are used
by default by \Cloudy{}, as they are listed in Stout.ini masterlist.
For each species, the references for the energy, transition
probability, and collisional data are given.
For the latter, the designation ``baseline''  has the same
meaning as in \citet{LykinsStout15}, namely that no collisional
data are available, and that the ``g-bar'' approximation is
used instead.

The table was prepared with the aid of the Perl scripts
\verb+cloudy/scripts/db-ref-bib2json.pl+
and \verb+cloudy/scripts/db-ref-json2tex.pl+ in the \Cloudy{} distribution.
The former walks through the Stout database, gathers bibliographical references,
and stores them in an external file in JSON format.
The latter script operates on the data in that file to create
the table itself, in \TeX{} format.

\section{Atomic and Ionic Species Treated in \Cloudy{}}
\label{sec:AppendixSpecies}

Table~\ref{table:species-db} lists the species treated by \Cloudy{},
followed by the database from which their energy and transition,
radiative and collisional, data are drawn.

The table was prepared with the aid of the \verb+cloudy/scripts/db-species-tex.pl+
Perl script in the \Cloudy{} distribution, which operates on the output of the
\verb+save species labels+ command.

\bibliography{LocalBibliography,bibliography2}

%
%
\begin{table*}
	{\small
	\centering
	\caption
	{
		Species data bases.
	}
	\label{table:species-db}
\begin{tabular}{|l|l|p{0.2cm}|l|l|p{0.2cm}|l|l|p{0.2cm}|l|l|}
		\hline
		\hline
		Species	&
		Database	&
			&
		Species	&
		Database	&
			&
		Species	&
		Database	&
			&
		Species	&
		Database
		\\
		\hline
			H	&
			H-like 	&
				&
			O$^{7+}$	&
			H-like 	&
				&
			Mg$^{4+}$	&
			Chianti	&
				&
			P	&
			Stout
		\\
		\hline
			He	&
			He-like	&
				&
			F	&
			\ldots	&
				&
			Mg$^{5+}$	&
			Chianti	&
				&
			P$^{+}$	&
			Stout
		\\
		\hline
			He$^{+}$	&
			H-like 	&
				&
			F$^{+}$	&
			Stout	&
				&
			Mg$^{6+}$	&
			Chianti	&
				&
			P$^{2+}$	&
			Stout
		\\
		\hline
			Li	&
			Stout	&
				&
			F$^{2+}$	&
			Stout	&
				&
			Mg$^{7+}$	&
			Chianti	&
				&
			P$^{3+}$	&
			Stout
		\\
		\hline
			Li$^{+}$	&
			He-like	&
				&
			F$^{3+}$	&
			Stout	&
				&
			Mg$^{8+}$	&
			Chianti	&
				&
			P$^{4+}$	&
			Chianti
		\\
		\hline
			Li$^{2+}$	&
			H-like 	&
				&
			F$^{4+}$	&
			Stout	&
				&
			Mg$^{9+}$	&
			Chianti	&
				&
			P$^{5+}$	&
			Stout
		\\
		\hline
			Be	&
			Stout	&
				&
			F$^{5+}$	&
			Stout	&
				&
			Mg$^{10+}$	&
			He-like	&
				&
			P$^{6+}$	&
			Chianti
		\\
\hline
	\end{tabular}
	}
\end{table*}
\begin{table*}
	{\small
	\addtocounter{table}{-1}
	\centering
	\caption
	{
		{\it (Cont.{})}
	}
	\begin{tabular}{|l|l|p{0.2cm}|l|l|p{0.2cm}|l|l|p{0.2cm}|l|l|}
		\hline
		\hline
		Species	&
		Database	&
			&
		Species	&
		Database	&
			&
		Species	&
		Database	&
			&
		Species	&
		Database
		\\
		\hline
			Cl$^{4+}$	&
			Stout	&
				&
			K$^{4+}$	&
			Chianti	&
				&
			Sc	&
			Stout	&
				&
			Ti$^{14+}$	&
			Chianti
		\\
		\hline
			Cl$^{5+}$	&
			Stout	&
				&
			K$^{5+}$	&
			Chianti	&
				&
			Sc$^{+}$	&
			Stout	&
				&
			Ti$^{15+}$	&
			Chianti
		\\
		\hline
			Cl$^{6+}$	&
			Stout	&
				&
			K$^{6+}$	&
			Stout	&
				&
			Sc$^{2+}$	&
			Stout	&
				&
			Ti$^{16+}$	&
			Chianti
		\\
		\hline
			Cl$^{7+}$	&
			Stout	&
				&
			K$^{7+}$	&
			Stout	&
				&
			Sc$^{3+}$	&
			Stout	&
				&
			Ti$^{17+}$	&
			Chianti
		\\
		\hline
			Cl$^{8+}$	&
			Stout	&
				&
			K$^{8+}$	&
			Chianti	&
				&
			Sc$^{4+}$	&
			Stout	&
				&
			Ti$^{18+}$	&
			Chianti
		\\
		\hline
			Cl$^{9+}$	&
			Chianti	&
				&
			K$^{9+}$	&
			Stout	&
				&
			Sc$^{5+}$	&
			Stout	&
				&
			Ti$^{19+}$	&
			Chianti
		\\
		\hline
			Cl$^{10+}$	&
			Chianti	&
				&
			K$^{10+}$	&
			Chianti	&
				&
			Sc$^{6+}$	&
			Stout	&
				&
			Ti$^{20+}$	&
			He-like
		\\
\hline
	\end{tabular}
	}
\end{table*}
\begin{table*}
	{\small
	\addtocounter{table}{-1}
	\centering
	\caption
	{
		{\it (Cont.{})}
	}
	\begin{tabular}{|l|l|p{0.2cm}|l|l|p{0.2cm}|l|l|p{0.2cm}|l|l|}
		\hline
		\hline
		Species	&
		Database	&
			&
		Species	&
		Database	&
			&
		Species	&
		Database	&
			&
		Species	&
		Database
		\\
		\hline
			Cr$^{4+}$	&
			Stout	&
				&
			Mn$^{15+}$	&
			Stout	&
				&
			Fe$^{25+}$	&
			H-like 	&
				&
			Ni$^{6+}$	&
			Stout
		\\
		\hline
			Cr$^{5+}$	&
			\ldots	&
				&
			Mn$^{16+}$	&
			Chianti	&
				&
			Co	&
			\ldots	&
				&
			Ni$^{7+}$	&
			\ldots
		\\
		\hline
			Cr$^{6+}$	&
			Chianti	&
				&
			Mn$^{17+}$	&
			Chianti	&
				&
			Co$^{+}$	&
			\ldots	&
				&
			Ni$^{8+}$	&
			Stout
		\\
		\hline
			Cr$^{7+}$	&
			Chianti	&
				&
			Mn$^{18+}$	&
			Chianti	&
				&
			Co$^{+}$	&
			Stout	&
				&
			Ni$^{9+}$	&
			\ldots
		\\
		\hline
			Cr$^{8+}$	&
			Chianti	&
				&
			Mn$^{19+}$	&
			Chianti	&
				&
			Co$^{2+}$	&
			Stout	&
				&
			Ni$^{10+}$	&
			Stout
		\\
		\hline
			Cr$^{9+}$	&
			Stout	&
				&
			Mn$^{20+}$	&
			Chianti	&
				&
			Co$^{3+}$	&
			\ldots	&
				&
			Ni$^{11+}$	&
			Chianti
		\\
		\hline
			Cr$^{10+}$	&
			Stout	&
				&
			Mn$^{21+}$	&
			Chianti	&
				&
			Co$^{4+}$	&
			\ldots	&
				&
			Ni$^{12+}$	&
			Chianti
		\\
\hline
	\end{tabular}
	}
\end{table*}
\begin{table*}
	{\small
	\addtocounter{table}{-1}
	\centering
	\caption
	{
		{\it (Cont.{})}
	}
	\begin{tabular}{|l|l|p{0.2cm}|l|l|p{0.2cm}|l|l|p{0.2cm}|l|l|}
		\hline
		\hline
		Species	&
		Database	&
			&
		Species	&
		Database	&
			&
		Species	&
		Database	&
			&
		Species	&
		Database
		\\
		\hline
			Cu$^{13+}$	&
			Stout	&
				&
			Zn$^{19+}$	&
			Chianti	&
				&
			CN	&
			LAMDA	&
				&
			HTiS	&
			\ldots
		\\
		\hline
			Cu$^{14+}$	&
			Stout	&
				&
			Zn$^{20+}$	&
			\ldots	&
				&
			CN$^{+}$	&
			\ldots	&
				&
			HTiS$^{+}$	&
			\ldots
		\\
		\hline
			Cu$^{15+}$	&
			Stout	&
				&
			Zn$^{21+}$	&
			\ldots	&
				&
			CO	&
			LAMDA	&
				&
			HeH$^{+}$	&
			\ldots
		\\
		\hline
			Cu$^{16+}$	&
			Stout	&
				&
			Zn$^{22+}$	&
			Chianti	&
				&
			CS	&
			LAMDA	&
				&
			Li$^{-}$	&
			\ldots
		\\
		\hline
			Cu$^{17+}$	&
			Stout	&
				&
			Zn$^{23+}$	&
			Chianti	&
				&
			CS$^{+}$	&
			\ldots	&
				&
			LiH	&
			\ldots
		\\
		\hline
			Cu$^{18+}$	&
			\ldots	&
				&
			Zn$^{24+}$	&
			Chianti	&
				&
			ClO	&
			\ldots	&
				&
			LiH$^{+}$	&
			\ldots
		\\
		\hline
			Cu$^{19+}$	&
			\ldots	&
				&
			Zn$^{25+}$	&
			\ldots	&
				&
			ClO$^{+}$	&
			\ldots	&
				&
			N$_{2}$	&
			\ldots
		\\
\hline
	\end{tabular}
	}
\end{table*}
\begin{table*}
	{\small
	\addtocounter{table}{-1}
	\centering
	\caption
	{
		{\it (Cont.{})}
	}
	\begin{tabular}{|l|l|p{0.2cm}|l|l|p{0.2cm}|l|l|p{0.2cm}|l|l|}
		\hline
		\hline
		Species	&
		Database	&
			&
		Species	&
		Database	&
			&
		Species	&
		Database	&
			&
		Species	&
		Database
		\\
		\hline
			SO$^{+}$	&
			\ldots	&
				&
			SiO$^{+}$	&
			\ldots	&
				&
			TiH$^{+}$	&
			\ldots	&
				&
			TiO$^{+}$	&
			\ldots
		\\
		\hline
			SO$_{2}$	&
			LAMDA	&
				&
			SiOH$^{+}$	&
			\ldots	&
				&
			TiH$_{2}$	&
			\ldots	&
				&
			TiO$_{2}$	&
			\ldots
		\\
		\hline
			SiC$_{2}$	&
			LAMDA	&
				&
			TiC	&
			\ldots	&
				&
			TiH$_{2}$$^{+}$	&
			\ldots	&
				&
			TiOH$^{+}$	&
			\ldots
		\\
		\hline
			SiH	&
			\ldots	&
				&
			TiC$^{+}$	&
			\ldots	&
				&
			TiN	&
			\ldots	&
				&
			TiS	&
			\ldots
		\\
		\hline
			SiH$_{2}$$^{+}$	&
			\ldots	&
				&
			TiC$_{2}$	&
			\ldots	&
				&
			TiN$^{+}$	&
			\ldots	&
				&
			TiS$^{+}$	&
			\ldots
		\\
		\hline
			SiN	&
			\ldots	&
				&
			TiC$_{2}$$^{+}$	&
			\ldots	&
				&
			TiNC	&
			\ldots	&
				&
			$^{13}$CO	&
			LAMDA
		\\
		\hline
			SiN$^{+}$	&
			\ldots	&
				&
			TiF$^{+}$	&
			\ldots	&
				&
			TiNC$^{+}$	&
			\ldots	&
				&
				&
		\\
		\hline
			SiO	&
			LAMDA	&
				&
			TiH	&
			\ldots	&
				&
			TiO	&
			\ldots	&
				&
				&
		\\
		\hline
	\end{tabular}
	}
\end{table*}

%
%
\begin{table*}
	{\small
	\centering
	\caption
	{
		References for Stout database.
		Species marked with $\ddagger$ are listed in Stout.ini, the default masterlist.
	}
	\label{table:stout-refs}
	\begin{tabular}{|l|p{5cm}|p{5cm}|p{5cm}|}
		\hline
		\hline
			Species &
			Energy &
			Transition &
			Collision
		\\
		\hline
			Li	$\ddagger$	&
			NIST  2013-11-08	&
			NIST  2013-10-18	&
			baseline
		\\
		\hline
			Be	$\ddagger$	&
			NIST  2014-09-16	&
			NIST  2014-09-16	&
			baseline
		\\
		\hline
			Be$^{+}$	$\ddagger$	&
			NIST  2013-11-08	&
			NIST  2013-10-18	&
			baseline
		\\
		\hline
			B	$\ddagger$	&
			NIST  2013-10-18	&
			NIST  2013-10-18	&
			baseline
		\\
		\hline
			B$^{+}$	$\ddagger$	&
			NIST  2014-09-16	&
			NIST  2014-09-16	&
			baseline
		\\
		\hline
			B$^{2+}$	$\ddagger$	&
			NIST  2014-09-16	&
			NIST  2014-09-16	&
			baseline
		\\
		\hline
			C	$\ddagger$	&
			NIST  2013-10-01	&
			NIST  2013-10-01	&
			\citet{Johnson1987}
		\\
\hline
	\end{tabular}
	}
\end{table*}
\begin{table*}
	{\small
	\addtocounter{table}{-1}
	\centering
	\caption
	{
		{\it (Cont.{})}
	}
	\begin{tabular}{|l|p{5cm}|p{5cm}|p{5cm}|}
		\hline
		\hline
			Species &
			Energy &
			Transition &
			Collision
		\\
		\hline
			O$^{+}$	$\ddagger$	&
			NIST  2014-01-26	&
			\citet{FroeseFischer2004}	&
			\citet{Kisielius2009}
		\\
					&
				&
			NIST  2014-01-26	&
			
		\\
		\hline
			O$^{2+}$	$\ddagger$	&
			NIST  2014-09-16	&
			NIST	&
			\citet{Aggarwal1983}
		\\
					&
				&
			\citet{Galavis1997}	&
			\citet{Aggarwal1985}
		\\
					&
				&
			\citet{Storey2000}	&
			\citet{Bhatia1993}
		\\
					&
				&
			\citet{Tachiev2001}	&
			\citet{Lennon1994}
		\\
					&
				&
				&
			\citet{Storey2014}
		\\
\hline
	\end{tabular}
	}
\end{table*}
\begin{table*}
	{\small
	\addtocounter{table}{-1}
	\centering
	\caption
	{
		{\it (Cont.{})}
	}
	\begin{tabular}{|l|p{5cm}|p{5cm}|p{5cm}|}
		\hline
		\hline
			Species &
			Energy &
			Transition &
			Collision
		\\
		\hline
			Na$^{8+}$	&
			NIST  2014-09-16	&
			NIST  2014-09-16	&
			baseline
		\\
		\hline
			Mg	$\ddagger$	&
			NIST  2013-10-01	&
			NIST  2013-10-01	&
			\citet{Barklem2012}
		\\
					&
				&
				&
			\citet{Osorio2015}
		\\
		\hline
			Mg$^{+}$	&
			NIST  2014-09-16	&
			NIST  2014-09-16	&
			baseline
		\\
		\hline
			Mg$^{2+}$	$\ddagger$	&
			NIST  2013-10-01	&
			NIST  2013-10-01	&
			\citet{Liang2010}
		\\
					&
				&
				&
			NIST 2013-09-07
		\\
		\hline
			Mg$^{3+}$	&
			NIST  2014-09-16	&
			NIST  2014-09-16	&
			baseline
		\\
\hline
	\end{tabular}
	}
\end{table*}
\begin{table*}
	{\small
	\addtocounter{table}{-1}
	\centering
	\caption
	{
		{\it (Cont.{})}
	}
	\begin{tabular}{|l|p{5cm}|p{5cm}|p{5cm}|}
		\hline
		\hline
			Species &
			Energy &
			Transition &
			Collision
		\\
		\hline
			Si$^{6+}$	$\ddagger$	&
			NIST  2014-05-24	&
			NIST  2014-05-24	&
			\citet{Butler1994}
		\\
		\hline
			Si$^{7+}$	&
			NIST  2014-09-16	&
			NIST  2014-09-16	&
			baseline
		\\
		\hline
			Si$^{8+}$	$\ddagger$	&
			NIST  2013-10-01	&
			NIST  2013-10-01	&
			\citet{Lennon1994}
		\\
		\hline
			Si$^{9+}$	&
			NIST  2014-09-16	&
			NIST  2014-09-16	&
			baseline
		\\
		\hline
			Si$^{10+}$	&
			NIST  2014-09-16	&
			NIST  2014-09-16	&
			baseline
		\\
		\hline
			Si$^{11+}$	&
			NIST  2014-09-16	&
			NIST  2014-09-16	&
			baseline
		\\
		\hline
			P	$\ddagger$	&
			NIST  2013-10-18	&
			NIST  2013-10-18	&
			baseline
		\\
\hline
	\end{tabular}
	}
\end{table*}
\begin{table*}
	{\small
	\addtocounter{table}{-1}
	\centering
	\caption
	{
		{\it (Cont.{})}
	}
	\begin{tabular}{|l|p{5cm}|p{5cm}|p{5cm}|}
		\hline
		\hline
			Species &
			Energy &
			Transition &
			Collision
		\\
		\hline
			S$^{13+}$	&
			NIST  2014-09-16	&
			NIST  2014-09-16	&
			baseline
		\\
		\hline
			Cl	$\ddagger$	&
			NIST  2013-10-01	&
			NIST  2013-10-01	&
			\citet{Hollenbach1989}
		\\
		\hline
			Cl$^{+}$	&
			NIST  2014-09-16	&
			NIST  2014-09-16	&
			baseline
		\\
		\hline
			Cl$^{2+}$	&
			NIST  2014-09-16	&
			NIST  2014-09-16	&
			baseline
		\\
		\hline
			Cl$^{3+}$	&
			NIST  2014-09-16	&
			NIST  2014-09-16	&
			baseline
		\\
		\hline
			Cl$^{4+}$	$\ddagger$	&
			NIST  2014-09-16	&
			NIST  2014-09-16	&
			baseline
		\\
		\hline
			Cl$^{5+}$	$\ddagger$	&
			NIST	&
			\citet{FernandezMenchero2014Mg}	&
			\citet{FernandezMenchero2014Mg}
		\\
\hline
	\end{tabular}
	}
\end{table*}
\begin{table*}
	{\small
	\addtocounter{table}{-1}
	\centering
	\caption
	{
		{\it (Cont.{})}
	}
	\begin{tabular}{|l|p{5cm}|p{5cm}|p{5cm}|}
		\hline
		\hline
			Species &
			Energy &
			Transition &
			Collision
		\\
		\hline
			K$^{7+}$	$\ddagger$	&
			NIST	&
			\citet{FernandezMenchero2014Mg}	&
			\citet{FernandezMenchero2014Mg}
		\\
					&
			\citet{FernandezMenchero2014Mg}	&
				&
			
		\\
		\hline
			K$^{8+}$	&
			NIST  2014-09-16	&
			NIST  2014-09-16	&
			baseline
		\\
		\hline
			K$^{9+}$	$\ddagger$	&
			NIST  2014-09-16	&
			NIST  2014-09-16	&
			baseline
		\\
		\hline
			K$^{10+}$	&
			NIST  2014-09-16	&
			NIST  2014-09-16	&
			baseline
		\\
		\hline
			K$^{11+}$	&
			NIST  2014-09-16	&
			NIST  2014-09-16	&
			baseline
		\\
		\hline
			K$^{12+}$	&
			NIST  2014-09-16	&
			NIST  2014-09-16	&
			baseline
		\\
\hline
	\end{tabular}
	}
\end{table*}
\begin{table*}
	{\small
	\addtocounter{table}{-1}
	\centering
	\caption
	{
		{\it (Cont.{})}
	}
	\begin{tabular}{|l|p{5cm}|p{5cm}|p{5cm}|}
		\hline
		\hline
			Species &
			Energy &
			Transition &
			Collision
		\\
		\hline
			Sc$^{9+}$	$\ddagger$	&
			NIST	&
			\citet{FernandezMenchero2014Mg}	&
			\citet{FernandezMenchero2014Mg}
		\\
					&
			\citet{FernandezMenchero2014Mg}	&
				&
			
		\\
		\hline
			Sc$^{10+}$	$\ddagger$	&
			NIST  2014-09-16	&
			NIST  2014-09-16	&
			baseline
		\\
		\hline
			Sc$^{11+}$	$\ddagger$	&
			NIST  2014-09-16	&
			NIST  2014-09-16	&
			baseline
		\\
		\hline
			Sc$^{12+}$	$\ddagger$	&
			NIST  2013-10-01	&
			NIST  2013-10-01	&
			\citet{Saraph1994}
		\\
		\hline
			Sc$^{13+}$	$\ddagger$	&
			NIST  2014-09-16	&
			NIST  2014-09-16	&
			baseline
		\\
		\hline
			Sc$^{14+}$	$\ddagger$	&
			NIST  2014-09-16	&
			NIST  2014-09-16	&
			baseline
		\\
\hline
	\end{tabular}
	}
\end{table*}
\begin{table*}
	{\small
	\addtocounter{table}{-1}
	\centering
	\caption
	{
		{\it (Cont.{})}
	}
	\begin{tabular}{|l|p{5cm}|p{5cm}|p{5cm}|}
		\hline
		\hline
			Species &
			Energy &
			Transition &
			Collision
		\\
		\hline
			V$^{10+}$	$\ddagger$	&
			NIST  2014-09-16	&
			NIST  2014-09-16	&
			baseline
		\\
		\hline
			V$^{11+}$	$\ddagger$	&
			NIST	&
			\citet{FernandezMenchero2014Mg}	&
			\citet{FernandezMenchero2014Mg}
		\\
		\hline
			V$^{12+}$	$\ddagger$	&
			NIST  2014-09-16	&
			NIST  2014-09-16	&
			baseline
		\\
		\hline
			V$^{14+}$	$\ddagger$	&
			NIST	&
			NIST	&
			\citet{Berrington1998}
		\\
					&
			NIST	&
			NIST	&
			
		\\
		\hline
			V$^{15+}$	$\ddagger$	&
			NIST  2014-09-16	&
			NIST  2014-09-16	&
			baseline
		\\
		\hline
			V$^{16+}$	$\ddagger$	&
			NIST  2014-09-16	&
			NIST  2014-09-16	&
			baseline
		\\
\hline
	\end{tabular}
	}
\end{table*}
\begin{table*}
	{\small
	\addtocounter{table}{-1}
	\centering
	\caption
	{
		{\it (Cont.{})}
	}
	\begin{tabular}{|l|p{5cm}|p{5cm}|p{5cm}|}
		\hline
		\hline
			Species &
			Energy &
			Transition &
			Collision
		\\
		\hline
			Mn$^{8+}$	&
			NIST  2014-09-16	&
			NIST  2014-09-16	&
			baseline
		\\
		\hline
			Mn$^{9+}$	&
			NIST  2014-09-16	&
			NIST  2014-09-16	&
			baseline
		\\
		\hline
			Mn$^{10+}$	$\ddagger$	&
			NIST  2014-09-16	&
			NIST  2014-09-16	&
			baseline
		\\
		\hline
			Mn$^{11+}$	$\ddagger$	&
			NIST  2014-09-16	&
			NIST  2014-09-16	&
			baseline
		\\
		\hline
			Mn$^{12+}$	$\ddagger$	&
			NIST  2014-09-16	&
			NIST  2014-09-16	&
			baseline
		\\
		\hline
			Mn$^{13+}$	$\ddagger$	&
			NIST	&
			\citet{FernandezMenchero2014Mg}	&
			\citet{FernandezMenchero2014Mg}
		\\
					&
			\citet{FernandezMenchero2014Mg}	&
				&
			
		\\
\hline
	\end{tabular}
	}
\end{table*}
\begin{table*}
	{\small
	\addtocounter{table}{-1}
	\centering
	\caption
	{
		{\it (Cont.{})}
	}
	\begin{tabular}{|l|p{5cm}|p{5cm}|p{5cm}|}
		\hline
		\hline
			Species &
			Energy &
			Transition &
			Collision
		\\
		\hline
			Fe$^{18+}$	&
			NIST  2014-09-16	&
			NIST  2014-09-16	&
			baseline
		\\
		\hline
			Fe$^{19+}$	&
			NIST  2014-09-16	&
			NIST  2014-09-16	&
			baseline
		\\
		\hline
			Fe$^{20+}$	&
			NIST  2014-09-16	&
			NIST  2014-09-16	&
			baseline
		\\
		\hline
			Fe$^{21+}$	&
			NIST  2014-09-16	&
			NIST  2014-09-16	&
			baseline
		\\
		\hline
			Fe$^{22+}$	&
			NIST  2014-09-16	&
			NIST  2014-09-16	&
			baseline
		\\
		\hline
			Fe$^{23+}$	&
			NIST  2014-09-16	&
			NIST  2014-09-16	&
			baseline
		\\
		\hline
			Co$^{+}$	$\ddagger$	&
			NIST  2014-09-16	&
			NIST  2014-09-16	&
			\citet{Storey2015}
		\\
\hline
	\end{tabular}
	}
\end{table*}
\begin{table*}
	{\small
	\addtocounter{table}{-1}
	\centering
	\caption
	{
		{\it (Cont.{})}
	}
	\begin{tabular}{|l|p{5cm}|p{5cm}|p{5cm}|}
		\hline
		\hline
			Species &
			Energy &
			Transition &
			Collision
		\\
		\hline
			Ni$^{8+}$	$\ddagger$	&
			NIST  2014-09-16	&
			NIST  2014-09-16	&
			baseline
		\\
		\hline
			Ni$^{10+}$	&
			NIST  2014-09-16	&
			NIST  2014-09-16	&
			\citet{DelZanna2014b}
		\\
					&
			\citet{DelZanna2014b}	&
			\citet{DelZanna2014b}	&
			
		\\
		\hline
			Ni$^{11+}$	&
			NIST  2014-09-16	&
			NIST  2014-09-16	&
			baseline
		\\
		\hline
			Ni$^{12+}$	&
			NIST  2014-09-16	&
			NIST  2014-09-16	&
			baseline
		\\
		\hline
			Ni$^{13+}$	&
			NIST  2014-09-16	&
			NIST  2014-09-16	&
			baseline
		\\
		\hline
			Ni$^{14+}$	&
			\citet{DelZanna2014}	&
			\citet{DelZanna2014}	&
			\citet{DelZanna2014}
		\\
\hline
	\end{tabular}
	}
\end{table*}

\end{document}